\RequirePackage{rotating}
\documentclass[useAMS,usenatbib,uselongtable]{mn2e}

\usepackage{times} \usepackage{graphicx} \usepackage{xspace} \usepackage{epsfig}
\usepackage{natbib} 
\usepackage{rotating} 
\usepackage{dcolumn}
\usepackage{amssymb} \usepackage{amsmath}
\usepackage{caption}

\title[The rotation-activity relation of M dwarfs with K2 mission, X-ray and UV data]{A path towards understanding the rotation-activity relation of M dwarfs 
with K2 mission, X-ray and UV data}
\author[]{B. Stelzer$^{1}$\thanks{E-mail:stelzer@astropa.inaf.it}, M. Damasso$^2$, 
A. Scholz$^3$, S.P. Matt$^4$ \\ 
$^{1}$ INAF - Osservatorio Astronomico di Palermo, Piazza del Parlamento 1, 90134 Palermo,
Italy \\ 
$^{2}$ INAF - Osservatorio Astrofisico di Torino, via Osservatorio 20, 10025 Pino Torinese, Italy \\
$^{3}$ SUPA School of Physics \& Astronomy, University of St. Andrews, North Haugh, 
St.Andrews KY\,16\,9SS, United Kingdom \\
$^{4}$ Department of Physics and Astronomy, University of Exeter, Physics Building, 
Stocker Road, Exeter, EX4 4QL, United Kingdom}
\begin{document}

\date{Accepted 2016 August 2. Received 2016 August 1; in original form 2016 March 22}

\pagerange{\pageref{firstpage}--\pageref{lastpage}} \pubyear{2013}

\maketitle

\label{firstpage}

\begin{abstract} 
We study the relation between stellar rotation and magnetic activity for a sample
of $134$ bright, nearby M dwarfs observed in the Kepler Two-Wheel (K2) mission during campaigns 
C0 to C4. The K2 lightcurves yield photometrically derived rotation periods for $97$
stars ($79$ of which without previous period measurement), as well as various
measures for activity related to cool spots and flares. We find a clear difference between 
fast and slow rotators with a dividing line at a period of $\sim 10$\,d
at which the activity level changes abruptly. All photometric diagnostics of activity
(spot cycle amplitude, flare peak amplitude and residual variability after subtraction
of spot and flare variations) display the same dichotomy, pointing to a quick transition
between a high-activity mode for fast rotators and a low-activity mode for slow rotators. 
This unexplained behavior is reminiscent of a dynamo mode-change seen in numerical
simulations that separates a dipolar from a multipolar regime. 
A substantial number of the fast rotators are visual binaries. A tentative explanation
is accelerated disk evolution in binaries leading to higher initial rotation rates on the
main-sequence  
and associated longer spin-down and activity lifetimes. 
We combine the K2 rotation periods with archival X-ray and UV data. 
X-ray, FUV and NUV detections are found for $26$, $41$, and $11$ stars from our
sample, respectively.
Separating the fast from the slow rotators, we determine for the first time 
the X-ray saturation level 
separately for early- and for mid-M stars. 
\end{abstract}


\section{Introduction}\label{sect:intro}

Together with convection,
rotation is the main driver of stellar dynamos and ensuing magnetic activity
phenomena (e.g. Kosovichev et al. 2013). In a feedback mechanism, magnetic fields are
responsible for the spin-evolution of stars: during part of the pre-main sequence
phase the magnetic field couples the star to its accretion disk dictating angular
momentum transfer \citep{Bouvier14.0} and during the main-sequence phase magnetized
winds remove angular momentum leading to spin-down \citep{Kawaler88.0, Matt15.0}.
Rotation and magnetic fields are, therefore, intimately linked and play 
a fundamental role in stellar evolution. 

Magnetic field strength and topology can be measured through Zeeman broadening
and polarization, respectively. Collecting the required optical high-resolution (polarimetric) 
spectroscopic observations is time-consuming, and each of these techniques can be applied only 
to stars with a limited range of rotation rates \citep[e.g.][]{Donati09.0, Vidotto14.1}. 
However, how the stellar
dynamo and the spin-evolution are linked can be addressed by measuring both magnetic
activity and rotation rate across evolutionary timescales. While the activity-age
relation is a proxy for the evolution of the stellar dynamo, the rotation-age
relation discriminates between models of angular momentum evolution. 

In a seminal work by \cite{Skumanich72.1} the age decay of both
activity and rotation of solar-type stars was established by extrapolating between
the age of the oldest known open cluster (600Myr) and the Sun (4.5Gyr). 
Unfortunately, stellar ages are notoriously difficult to assess. Therefore,
the direct relation between rotation and activity -– observed first some decades ago 
\citep[e.g.][]{Pallavicini81.1, Vilhu84.1} –- has widely substituted studies which involve
age-estimates. The early works cited above have used spectroscopic measurements as  
measure for stellar rotation ($v \sin{i}$), and carry intrinsic ambiguities
related to the unknown inclination angle of the stars. 
Stellar rotation rates are best derived from the periodic brightness
variations induced by cool star spots moving across the line-of-sight, which can be
directly associated with the rotation period. 
In more recent studies of the rotation-activity connection, 
photometrically measured rotation periods have proven more useful than $v \sin{i}$
\citep[e.g.][]{Pizzolato03.1, Wright11.0}. Especially, for M dwarfs `saturation' sets
in at relatively small values of $v \sin{i}$ due to their small radii and it is hard to
probe the slow-rotator regime with spectroscopic rotation measurements. 

Theory predicts a qualitative change of the dynamo mechanism 
at the transition into the fully convective regime 
\citep[spectral type $\sim$\,M3;][]{Stassun11.0}. Fully convective stars lack the
tachocline in which solar-like $\alpha\Omega$-dynamos originate. Alternative field
generation mechanisms must be at work: 
a turbulent dynamo was proposed by \cite{Durney93.1} but it is expected to generate only
small-scale fields, in contrast to recent results from Zeeman Doppler Imaging (ZDI) 
which have shown evidence
for large-scale dipolar fields in some fully convective stars \citep{Morin08.0, Morin10.0}. 
Current studies of field generation in the fully convetive regime 
are, therefore, concentrating on $\alpha^2$-dynamos \citep{Chabrier06.1}. 
While rotation has no influence on turbulent 
dynamo action, it is considered an important ingredient of mean-field $\alpha^2$-dynamos. 
This attributes studies of the rotation dependence of magnetic activity across the M spectral
type range a crucial meaning for understanding fully convective dynamos. 
Moreover, while improved spin-down models based on stellar wind simulations have been
developed for solar-type stars \citep{Gallet13.0}, angular momentum evolution
models of M stars are still controversial \citep{Reiners12.3}. 
Therefore, for the most abundant type of stars in our Galaxy, 
both the characteristics of the stellar dynamo and the angular momentum evolution are
still widely elusive. 

Rotation-activity studies have been presented with different diagnostics for activity, 
the most frequently used ones being H$\alpha$ and X-ray emission.
While H$\alpha$ measurements are available
for larger samples, especially thanks to surveys such as the {\it Sloan Digital Sky Survey} 
\citep[e.g.][]{West04.1}, X-ray emission was shown
to be more sensitive to low activity levels in M dwarfs \citep{Stelzer13.0}.
The samples for the most comprehensive 
rotation-activity studies involving X-ray data have been assembled
from a literature compilation, providing a large number of stars, at the expense of
homogeneity. \cite{Wright11.0} discuss a sample of more than $800$ late-type stars
(spectral type FGKM). However, the rotation-activity relation is not studied
separately for M stars, possibly due to a strong bias towards X-ray luminous stars
which affects especially the M stars as seen from their Fig.~5. 
Overall, the lack of unbiased overlapping samples with known rotation period and X-ray 
activity level has left the X-ray - rotation relation of M stars
nearly unconstrained \cite[see bottom right panels of Fig.5 and~6 in][]{Pizzolato03.1}. 
Studies with optical emission lines (H$\alpha$, Ca\,{\sc ii}\,H\&K) 
as activity indicator have for convenience mostly
been coupled with $v \sin{i}$ as rotation measure because both parameters can be obtained
from the same set of spectra \citep{Browning10.0,Reiners12.1}. Only lately has it
become possible to combine H$\alpha$
data with photometrically measured M star rotation periods, since a larger sample of
periods have become
available from ground-based planet transit search programs \citep{West15.0}.

M dwarfs have not yet reached a common rotational sequence even at Gyr-ages, suggesting
weaker winds and longer spin-down timescales as compared to solar-like stars 
\citep{Irwin11.0}. The old and slowly rotating M dwarfs generally have  
low variability amplitudes resulting
from reduced spot coverage and long rotation periods (up to months). 
From the ground, significant numbers of field M dwarf rotation periods have recently 
been measured \citep{Newton16.0}.  
However, the sample of their study comprises only very low-mass stars ($R_* \leq 0.33\,R_\odot$) 
and seems to be incomplete in terms of the period detection efficiency 
\citep{Irwin11.0}. The Kepler mission \citep{Borucki10.0} 
with its ability to provide high-precision, long and
uninterrupted photometric lightcurves has led to the detection of rotation periods in
$> 2000$ field M dwarfs \citep{Nielsen13.0, McQuillan13.0, McQuillan14.0}, 
a multiple of the number known before.
Interesting findings of this Kepler-study are (i) the evidence for a bimodal
distribution of rotation periods for M dwarfs with $P_{\rm rot} = 0.4$...$70$\,d 
and (ii) the
fact that the envelope for the slowest observed rotation periods shifts towards
progressively larger periods for stars with mass below $\sim 0.5\,M_\odot$. How these
features in the rotational distribution are connected to stellar activity has not yet
been examined. Most of the Kepler stars are too distant for detailed characterization 
in terms of magnetic activity diagnostics. However, the Kepler Two-Wheel (K2) mission 
is ideally suited to study both rotation and activity for nearby M stars. 

Since March 2014, with its two remaining reaction wheels, the Kepler spacecraft is 
restricted to observations in the ecliptic plane changing the pointing direction every 
$\sim 80$\,d \citep{Howell14.0}. With special data processing correcting for the 
spacecraft's pointing drift, the photometric precision of K2 is similar to that achieved
by the preceding fully functional Kepler mission \citep{Vanderburg14.0}. A great number
of field M dwarfs have been selected as K2 targets with the goal of detecting planet
transits. Several lists of planet candidates have already been published
\citep[e.g.][]{ForemanMackey15.0, Montet15.0, Vanderburg16.0}, and some interesting planet 
systems have already been validated, including objects from the target list of
this study (see Sect~\ref{subsect:results_planethosts}). 

In our program to study the M star rotation-activity connection we limit the sample to 
nearby, bright M stars which provide the largest signal-to-noise in the K2 lightcurves
and are most likely to be detectable at the high energies that are the best probes of 
magnetic activity. 
We derive from the K2 mission data both rotation periods and various diagnostics of
magnetic activity, and we combine this with X-ray and UV 
activity from past and present space missions ({\em ROSAT}, {\em XMM-Newton}, {\em GALEX}). 
As mentioned above, X-ray wavelengths have proven
to be more sensitive to low activity levels in M dwarfs than optical emission lines.  
Moreover, both X-rays and UV photons are known to have a strong impact on 
close-in planets, providing another motivation for characterizing the high-energy emission 
of these stars. Given the high occurrence rate of terrestrial planets
in the habitable zone of M dwarfs \citep[$\sim 50$\,\% according to][]{Kopparapu13.0}, 
a substantial number of the stars
we survey may soon be found to host potentially habitable worlds. 

The importance of stellar magnetic activity for exoplanet studies is twofold. 
First, star spots and chromospheric structures 
introduce noise in measured radial velocity curves, so-called
RV `jitter', which depends strongly on the properties of the spots \citep{Andersen15.0}. 
The spectra collected to perform radial velocity measurements can also be used to model
starspots \citep[see e.g.][]{Donati15.0}. However, since 
it is an impossible task to measure the spot distribution for every potentially
interesting star, relations between star spot characteristics and other activity diagnostics 
such as UV or X-ray emission -- if applied to statistical samples -- can provide useful 
estimates of the expected RV noise.
Secondly, as mentioned above, the stellar X-ray and UV emission is crucial for the evolution and the 
photochemistry of planet atmospheres. While the magnetic activity of the star may erode
the atmospheres of planets formed in close orbits \citep[e.g.][]{Penz08.0}, it may by the same
effect remove the gaseous envelopes of planets migrated inward from beyond the snow line 
and render them habitable \citep{Luger15.0}. 
Until recently, photochemical models for planets around M dwarfs
relied exclusively on the observed UV properties of a single strongly active 
star, AD\,Leo 
\citep{Segura05.0}. Lately, \cite{Rugheimer15.0} have modeled the effect of an M star 
radiation field on exoplanet atmospheres based on the Hubble Space Telescope (HST) UV 
spectra of six exoplanet host stars \citep{France13.0}. These stars are apparently only
weakly active, as none of them displays
H$\alpha$ emission. Yet their HST spectra show hot emission lines proving the presence
of a chromosphere and transition region.  
The lower limit of the chromospheric UV flux and its dependence
on stellar parameters has not been constrained so far. 
Similarly, on the high end of the activity range, with exceptions such as AD\,Leo
\citep[e.g.][]{SanzForcada02.1, Crespo-Chacon06.0}, 
the frequencies and luminosities of flares on M dwarfs are still largely unknown. 

There has been significant recent progress in studies of M dwarf flares 
based on data from the main Kepler mission 
\citep{Ramsay13.0, Hawley14.0, Davenport14.0, Lurie15.0}. The time
resolution of $1$\,min obtained in the Kepler short-cadence data proves essential for catching
small events, adding to the completeness of the observed flare distributions and 
enabling the examination of flare morphology. The drawback is that 
these results are limited to individual objects or a very small group of bright stars.  
The K2 mission gives access to much larger samples of bright M dwarfs, for which we can
examine the relation between flaring and rotation in a statistical way, albeit 
at lower cadence. In this work we establish, to our knowledge for the first time, 
a direct connection between white-light flaring and stellar rotation rate. 

As described above, the sample selection is the key to success in constraining the
rotation-activity relation of M dwarfs. We present our sample in Sect.~\ref{sect:sample}.
In Sect.~\ref{sect:stepar} we derive the stellar parameters.
This is necessary in order to investigate the dependence of rotation and activity 
on effective temperature ($T_{\rm eff}$) and mass ($M_*$), and to compute commonly used
activity indices which consist of normalizing the magnetically-induced 
emission (X-ray, UV, etc.) to the bolometric luminosity. We then 
describe the analysis of K2 data involving the detection of flares and rotation periods 
(Sect.~\ref{sect:k2_analysis}), 
archival X-ray (Sect.~\ref{sect:xray_analysis}) and UV (Sect.~\ref{sect:uv_analysis}) data.
We present our results in Sect.~\ref{sect:results}. The implications are discussed
in Sect.~\ref{sect:discussion}, and we provide a summary in Sect.~\ref{sect:summary}.

\section{Sample}\label{sect:sample}

This work is based on all bright and nearby M dwarfs
from the Superblink proper motion catalog by \citet[][henceforth LG11]{Lepine11.0} 
observed within the K2 mission's campaigns C0....C4. The Superblink catalog
comprises an All-Sky list of 8889 M dwarfs (spectral type K7 to M7) brighter than
$J=10$, within a few tens of parsec. Many other programs focusing on M stars are carried
out within the K2 mission, and rotation periods have been measured for more than
a thousand M stars during the main Kepler mission \citep{McQuillan13.0}. 
However, a careful sample selection comprising stars with already known
or easily accessible magnetic activity characteristics is mandatory to nail down the
rotation-activity relation. 
The majority of the Kepler stars are too distant ($> 200$\,pc) and, therefore, too 
faint for the {\em ROSAT} All-Sky Survey, the main source for X-ray
studies of widely dispersed samples. The proper-motion-selected M stars of the 
LG11 catalog are much closer and consequently brighter, facilitating
the detection of both rotation periods and X-ray and UV emission. 

A total of $134$ Superblink M dwarfs have been observed in K2 campaigns C0...C4. 
Henceforth we will refer to these objects as the ``K2 Superblink M star sample". 
The target list is given in Table~\ref{tab:targettable}. We list the identifier from
the EPIC catalog, the campaign in which the object was observed, the designation
from the {\it Third Catalog of Nearby Stars} \citep[CNS\,3; ][]{Gliese91.1}, 
magnitudes in the Kepler band and further parameters,  
the calculation of which is described in the next section.

\section{Fundamental stellar parameters}\label{sect:stepar}

We derive physical parameters of the K2 Superblink M stars (effective temperature, mass, 
radius, and bolometric luminosity) by adopting empirical and semi-empirical calibrations 
from \cite{Mann15.0}, which are based on the color indices \textit{V$-$J} 
and \textit{J$-$H}, and on the absolute magnitude in the 2\,MASS $K$ band,
\textit{M}$\rm_{K_{S}}$. The calibrations of \cite{Mann15.0} are valid for
dwarf stars, and can be expected to hold for the K2 Superblink M star 
sample which has been cleaned by LG\,11 from contaminating giants.   
Due to a press error some wrong values appeared in the tables of \cite{Mann15.0}. We 
use here the correct values reported in the 
erratum\footnote{http://iopscience.iop.org/article/10.3847/0004-637X/819/1/87/meta}. 
Stellar magnitudes and their uncertainties 
are obtained from the UCAC4 catalog \citep{Zacharias13.0} which provides 2MASS near-IR 
photometry and $V$ band magnitudes from 
{\it The AAVSO Photometric All Sky Survey (APASS)}; \cite{Henden14.0}.
These latter ones are more accurate and have significantly
better precision than the $V$ band magnitudes given in LG\,11. For the $6$ stars
with no $V$ magnitude in UCAC4, we found measurements in Data Release\,9 of the
APASS catalog\footnote{https://www.aavso.org/apass}.  

We derive an empirical linear calibration to calculate \textit{M}$\rm_{K_{S}}$ for 
our sample, using a list of $1,078$ M dwarfs with apparent \textit{K}$\rm_{S}$ magnitude 
from UCAC\,4 and trigonometric parallax in LG\,11. 
This allows us to obtain estimates for $M_{\rm K_s}$ independent on the trigonometric parallax
which is available for only $27$ stars in our sample. 
The best linear least-squares fit to the data is obtained through a Monte-Carlo analysis.
This approach provides more realistic errors than simple least-squares fitting 
because the uncertainties are derived from posterior distributions 
of the parameters and take into account all the errors affecting the 
measurements. 

\begin{figure}
  \resizebox{\hsize}{!}{\includegraphics{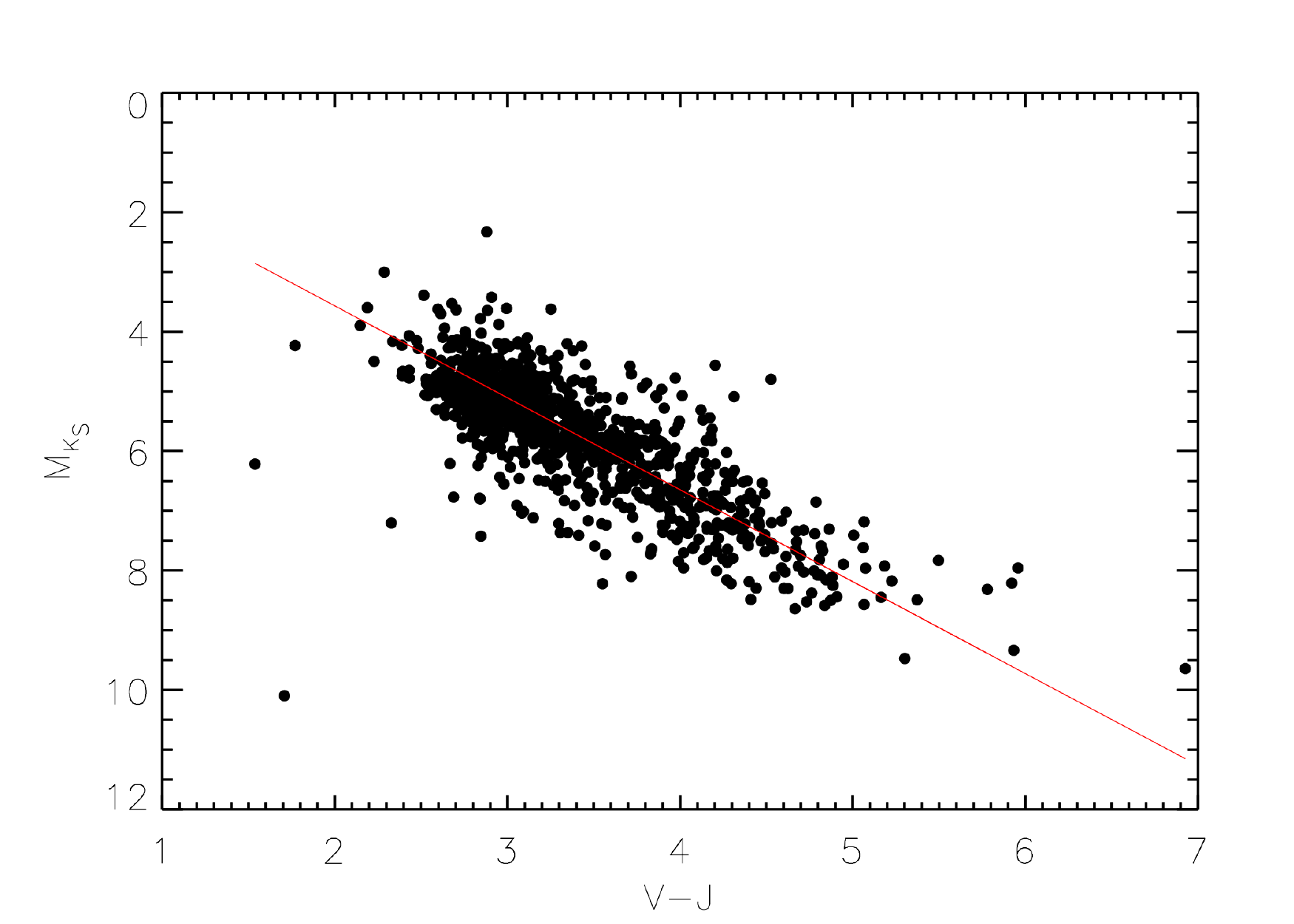}}
 \caption{Calibration of the linear relation between absolute $K_{\rm S}$ magnitude, 
 $M_{\rm K_S}$, and $V-J$ color 
 obtained from a sample of $1,078$ M dwarfs with measured trigonometric parallax 
 in LG11 
 (black circles). Our best-fit model is represented by a red solid line. The 
 residuals have a scatter of $0.56$\,mag.}
 \label{fig:pxphoto_calibration}
\end{figure}

Specifically, we generate $10,000$ synthetic samples (each composed of $1,078$ stars) 
drawing $V-J$ and $M_{\rm K_S}$ randomly from 2D normal distributions with mean equal to 
the observed values and standard deviation (henceforth STD) 
equal to the uncertainties. We then fit 
to each of the $10,000$ representations a straight line 
with the IDL\footnote{IDL is a product of the Exelis Visual Information 
Solutions, Inc.}  \texttt{FITEXY} routine, assuming for each simulated point the
original errors in both variables. The best-fit relation is then defined by the
median values and standard deviations of the {\sl a posteriori} Monte-Carlo distribution 
for the coefficients in the linear fit, given by   
\begin{equation}
M_{\rm K_S} = 0.49(\pm 0.02) + 1.539(\pm 0.006) \cdot (V-J)
\end{equation}
The residuals of this solution, which is applicable in the 
range $1.54 < V-J < 6.93$, show a rms of $0.56$\,mag. 
In Fig.~\ref{fig:pxphoto_calibration} we show this relation overplotted on the observed 
data. 

All other stellar parameters and their uncertainties are calculated in the same manner
through a Monte-Carlo analysis. 
In particular, the stellar effective temperatures ($T_{\rm eff}$) are obtained from 
the calibration relation which uses $V-J$ and $J-H$  
(Eq.~$7$ in \citealt{Mann15.0}), while radii ($R_*$) and masses ($M_*$) are calculated 
from relations with \textit{M}$\rm_{K_{S}}$ (Eqs.~$4$ and~$10$ in \citealt{Mann15.0}, 
respectively), and the bolometric correction $BC_{\rm K}$ is derived through a third-degree 
polynomial with $V-J$ as independent variable \citep[presented in Table~3 of ][]{Mann15.0}.   

Thus, we first generate for each star a sample of 
$10,000$ synthetic $V-J$, $J-H$, and $M_{\rm K_S}$ datasets drawn from normal 
distributions with mean and sigma equal to the observed value and its error. 
Then we apply to each star 
the above-mentioned calibrations from \cite{Mann15.0}. The best estimate of each parameter 
($T_{\rm eff}$, $R_*$, $M_*$ and $BC_{\rm K}$) is 
then obtained as the median value of the corresponding {\sl a posteriori} distribution, 
with its standard deviation assumed as the uncertainty. 

To provide conservative estimates of the stellar parameters, 
the uncertainties representing the scatter of the relations of 
\citet[][see Tab. 1, 2, and 3 therein]{Mann15.0}
are propagated into the Monte-Carlo process. 
Specifically, for $T_{\rm eff}$ we consider the scatter in the difference 
between the predicted and the spectroscopically observed temperature  
($48$\,K), and the typical uncertainty on the spectroscopic 
value of $T_{\rm eff}$ ($60$\,K) adding both in quadrature, 
while for $BC_{\rm K}$ we consider the uncertainty of $0.036$\,mag. 
These additional uncertainties are taken into account in the Monte-Carlo analysis 
when drawing randomly the samples. 
For radius and mass, \cite{Mann15.0} provide relative uncertainties of $2.89$\,\% and 
$1.8$\%, respectively. These values are calculated from the median values of our
posterior distributions for $R_*$ and $M_*$ and are then added in quadrature to their  
standard deviations. 

\cite{Mann15.0} argue that some of the above-mentioned relations for the stellar 
parameters can be improved by including an additional term involving metallicity 
([Fe/H]).
We found [Fe/H] measurements in the literature \citep{Newton14.0} for only $6$ stars from 
the K2 Superblink M star 
sample, and we verified for these objects that the radii and temperatures derived by 
taking account of [Fe/H] (Eqs.~5 and~6 in \citealt{Mann15.0}) are compatible
with our estimates described above. 

From $BC_{\rm K}$ and 
\textit{M}$\rm_{K_{S}}$ we calculate the absolute bolometric magnitudes of our sample, 
which are then converted into luminosities assuming 
the absolute bolometric magnitude of the Sun is $M_{\rm bol,\odot}=4.7554$. 
We note, that the distances we infer from our $M_{\rm K_s}$ values and the observed 
$K_s$ magnitudes are systematically larger, on average by about $\sim 25\,\%$, 
than the photometric distances presented by LG11 for the same stars. 
For the $27$ stars with trigonometric parallax in the literature 
\citep[LG11,][]{Dittmann14.0} our newly derived photometric distances are in 
excellent agreement with the trigonometric distances. 
In the near future, {\em Gaia} 
measurements will provide the ultimate and accurate distances for all 
K2 Superblink M stars. In the meantime, as corroborated by the comparison to 
trigonometric parallaxes, our estimates, which are based on the most accurate
photometry available to date, can be considered as a fairly reliable 
guess on the distances. 

\begin{figure}
  \resizebox{\hsize}{!}{\includegraphics{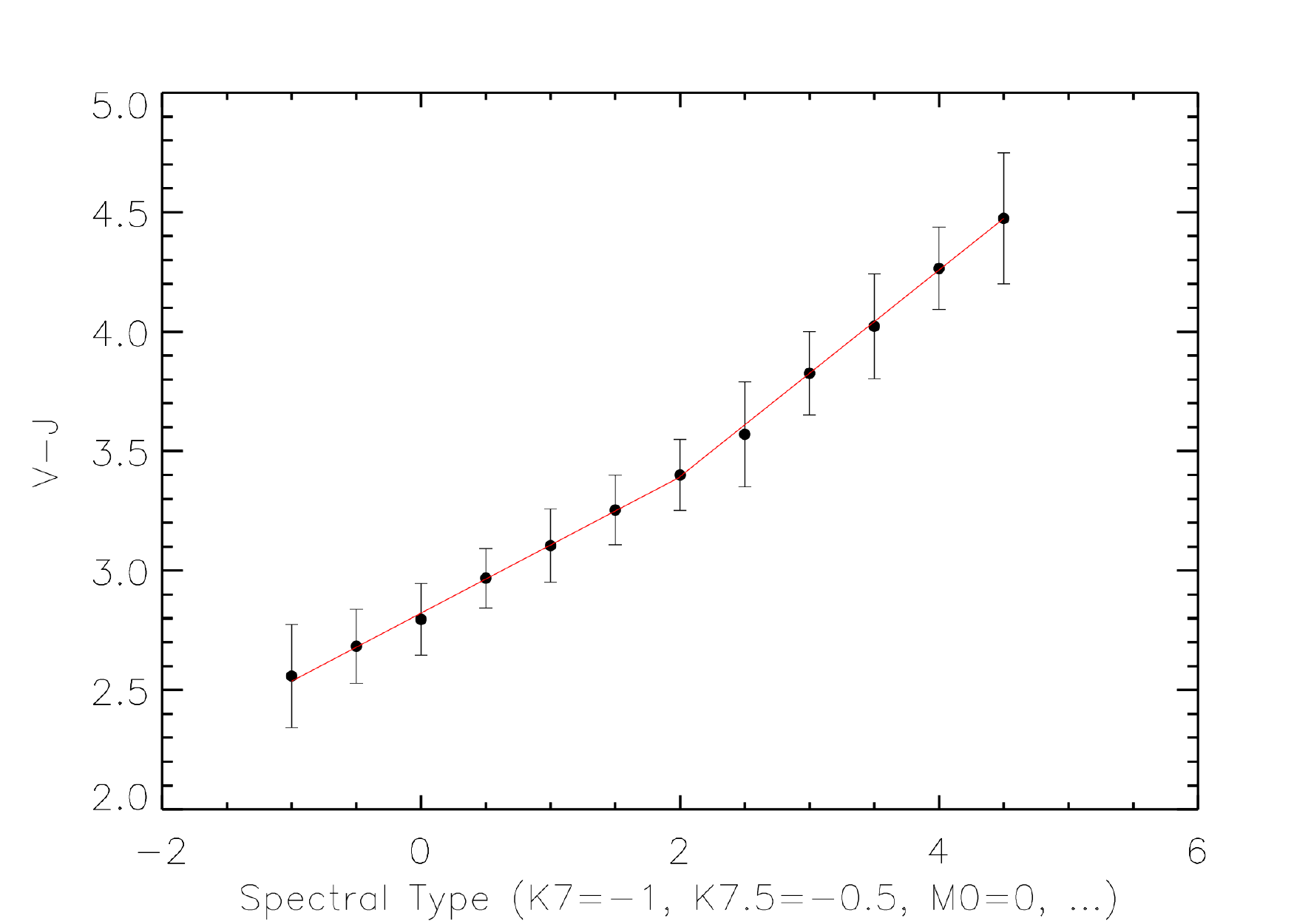}}
  \caption{Calibration of the relation between spectral type and $V-J$ obtained from a 
  sample of $1,173$ M dwarfs with spectroscopically determined spectral type 
  in \protect\cite{Lepine13.0}. 
  We fit the data with two 
  straight lines (red solid lines), one for stars with sub-types earlier than M2 and the other one for those 
  with sub-types later than M2. 
  }
  \label{fig:sptypephoto_calibration}
\end{figure}

All stars in the K2 Superblink M star sample have a photometric estimate of the 
spectral type in LG\,11, based on an empirical relation of spectral type 
with $V-J$ color which was calibrated with SDSS spectra. 
Since we use here the higher-precision UCAC\,4 $V$ band magnitudes, for consistency with
our calculation of the other stellar parameters, we derive an analogous
relation between $V-J$ and spectral type. To this end, we make use of $1,173$ stars 
classified as K7 or M-type dwarfs by \cite{Lepine13.0} based on spectroscopy.
We group the stars in bins of $0.5$ spectral subclasses,  
with K7 corresponding to $-1$, M0 to $0$, and so on until M$4.5$, which is 
the last sub-type for which we have enough stars in the calibration sample for a useful fit. 
We calculate the mean and standard deviation of $V-J$ for each spectral type bin, 
and notice that the data can be fitted with a combination of two straight lines 
for the ranges [K7,M2] and [M2,M4.5] (see Fig.~\ref{fig:sptypephoto_calibration}). 
Our fit, performed through a Monte-Carlo procedure as described above, results in the 
relations 
\begin{eqnarray}
V - J = 2.822( \pm 0.067) + 0.285( \pm 0.061) \cdot SpT 
\label{eq:vminj_spt_early} \\
V - J = 2.53( \pm 0.29) + 0.432( \pm 0.093) \cdot SpT
\label{eq:vminj_spt_late} 
\end{eqnarray}
which are valid for $2.5 \leqslant V-J \leqslant 3.4$ 
and $3.4 \leqslant V-J \leqslant 4.5$, 
for the hotter (Eq.~\ref{eq:vminj_spt_early}) and cooler (Eq.~\ref{eq:vminj_spt_late}) 
spectral types respectively. 
We use this calibration to classify the K2 Superblink M star sample, 
by rounding the results of the linear relations to the closest spectral sub-type. 
Nine K2 Superblink M stars 
have $V-J$ colors slightly beyond the boundaries for which we calibrated
Eqs.~\ref{eq:vminj_spt_early} and~\ref{eq:vminj_spt_late} 
and we extrapolate the relations at the ends to spectral types
K5 and M5, respectively. No star deviates by more than $0.5$ spectral subclasses from
Eqs.~\ref{eq:vminj_spt_early} and~\ref{eq:vminj_spt_late}.
The spectroscopically determined spectral types from the literature, which are available
for roughly three dozens of the K2 Superblink M stars, are in excellent agreement with 
our values \citep[see][]{Reid04.0, Reiners12.1, Lepine13.0}.  

In Table~\ref{tab:targettable} we provide the photometry 
(Kepler magnitude $K_{\rm p}$, $V$, $J$, and $K_{\rm s}$), 
the distances obtained from the absolute $K$ band magnitude,  
the fundamental parameters ($M_*$, $R_*$, $\log{L_{\rm bol}}$ and
$T_{\rm eff}$)
 and the spectral type (SpT) derived as described above.
 The few stars with $M_{\rm K_s}$ slightly more than $3\,\sigma$ smaller than the lower 
 boundary of the calibrated range ($4.6 < M_{\rm K_s} < 9.8$) are flagged with an asterisk in
 Table~\ref{tab:targettable}. 
 
 Stars for which the K2 photometry -- and in some cases also the optical/IR photometry
 used by us to calculate the stellar parameters -- comprises a potential
 contribution from a close binary companion are discussed in detail in the 
 Appendix~\ref{sect:appendix_bin}. These stars are 
 also highlighted in Table~\ref{tab:targettable} and flagged in all figures where
 relevant. The Gl\,852\,AB binary is represented
 in our target list by two objects (EPIC\,206262223 and EPIC\,206262336) but they are not 
 resolved in the K2 aperture\footnote{Our analysis relies on the data reduction performed
 by A.Vanderburg; see Sect.~\ref{sect:k2_analysis}.}, 
 i.e. only the combined lightcurve of both stars is at our disposition. We compute the
 stellar parameters for both components in the binary using the individual $V$ magnitudes
 from \cite{Reid04.0}; then we assign the rotational parameters and the X-ray/UV emission
 to the brighter, more massive star (EPIC\,206262336) and we do not consider 
 the secondary (EPIC\,206262223) any further. 

\begin{figure*}
\begin{center}
\parbox{18cm}{
\parbox{9cm}{
\includegraphics[width=8.5cm]{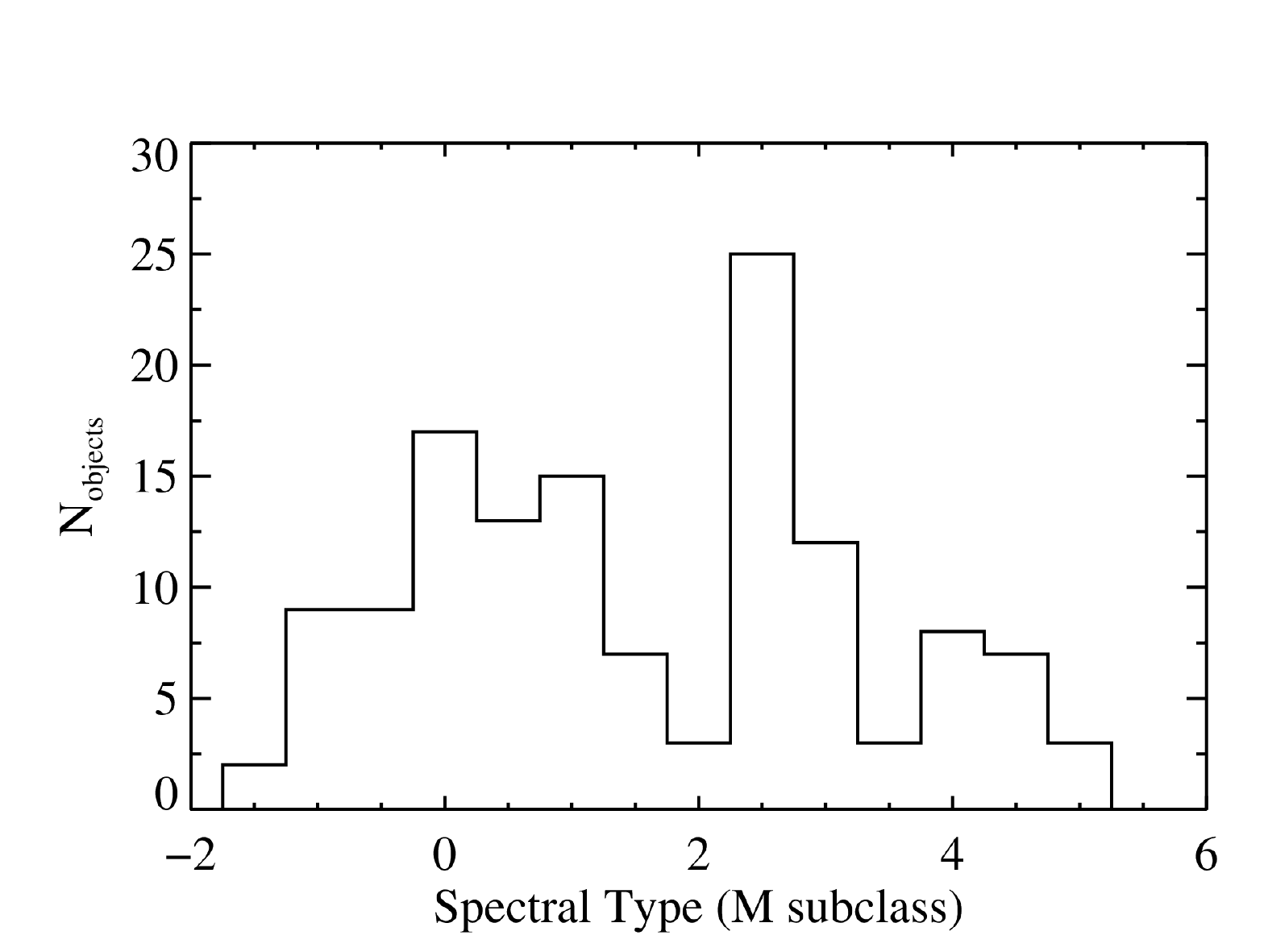}
}
\parbox{9cm}{
\includegraphics[width=8.5cm]{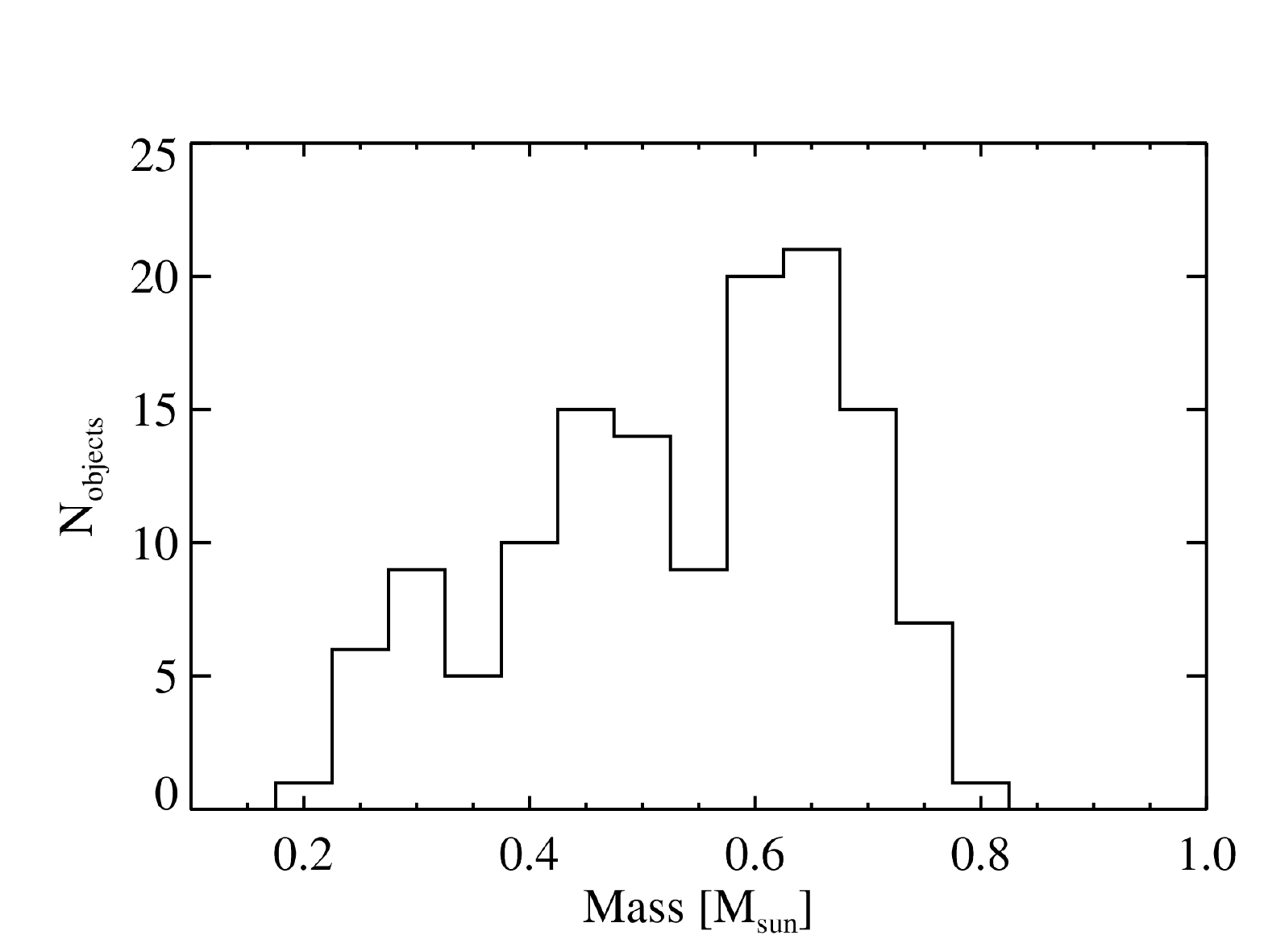}
}
}
\caption{Distribution of spectral types and masses for the K2 Superblink M star sample 
observed in campaigns C0 to C4. 
Negative indices denote spectral types earlier than M, where the value
$-1$ stands for K7. See text in Sect.~\ref{sect:stepar} for details.}
\label{fig:histo_spt_mass}
\end{center}
\end{figure*}

The distributions of spectral type and mass for the K2 Superblink M star sample are
shown in Fig.~\ref{fig:histo_spt_mass}. Covering spectral type K5 to M5 (masses between
about $0.2$ and $0.9\,M_\odot$), 
this is an excellent database for investigating the
connection between rotation and activity across the fully convective boundary 
(SpT $\sim$M3/M4).
%
%


\begin{sidewaystable*}
\caption{Target list for K2 Campaign C0...C4 with stellar parameters.}
\label{tab:targettable}
\begin{tabular}{lclrrrrrrrrrc} \hline
EPIC ID & Campaign & CNS-name & $K_{\rm p}$ & \multicolumn{1}{c}{$V$}   & \multicolumn{1}{c}{$J$}   & \multicolumn{1}{c}{$K_s$} & \multicolumn{1}{c}{$d$} & \multicolumn{1}{c}{$M_*$}       & \multicolumn{1}{c}{$R_*$}       & \multicolumn{1}{c}{$\log{L_{\rm bol}}$} & \multicolumn{1}{c}{$T_{\rm eff}$} & SpT  \\
        &          &          & [mag]       & \multicolumn{1}{c}{[mag]} & \multicolumn{1}{c}{[mag]} & \multicolumn{1}{c}{[mag]} & \multicolumn{1}{c}{[pc]} & \multicolumn{1}{c}{[$M_\odot$]} & \multicolumn{1}{c}{[$R_\odot$]} & \multicolumn{1}{c}{[$L_\odot$]}   &  \multicolumn{1}{c}{[K]} & \\ \hline
       202059188$^B$ & C0 &                &  14.70 & $ 14.32 \pm   0.05$ & $  9.88 \pm   0.02$ & $  9.04 \pm   0.02$ & $  22.0$ & $  0.27 \pm   0.01$ & $  0.27 \pm   0.01$ & $ -2.13 \pm   0.04$ & $  3181 \pm  80 $ & M4.5   \\
           202059192 & C0 &                &  13.10 & $ 13.09 \pm   0.02$ & $  9.99 \pm   0.02$ & $  9.15 \pm   0.02$ & $  59.9$ & $  0.59 \pm   0.01$ & $  0.56 \pm   0.02$ & $ -1.23 \pm   0.03$ & $  3712 \pm  80 $ & M1.0   \\
           202059193 & C0 &                &  12.50 & $ 12.42 \pm   0.03$ & $  9.54 \pm   0.02$ & $  8.71 \pm   0.02$ & $  57.2$ & $  0.65 \pm   0.02$ & $  0.63 \pm   0.02$ & $ -1.08 \pm   0.03$ & $  3835 \pm  82 $ & M0.0   \\
           202059195 & C0 &      G 103-029 &  14.70 & $ 14.18 \pm   0.06$ & $  9.95 \pm   0.02$ & $  9.07 \pm   0.02$ & $  25.9$ & $  0.31 \pm   0.01$ & $  0.31 \pm   0.01$ & $ -1.99 \pm   0.05$ & $  3306 \pm  81 $ & M4.0   \\
           202059198 & C0 &                &  11.60 & $ 11.62 \pm   0.02$ & $  8.48 \pm   0.06$ & $  7.65 \pm   0.02$ & $  29.2$ & $  0.58 \pm   0.02$ & $  0.55 \pm   0.02$ & $ -1.26 \pm   0.05$ & $  3724 \pm  99 $ & M1.0   \\
           202059199 & C0 &       LP 420-6 &  12.60 & $ 12.60 \pm   0.02$ & $  9.06 \pm   0.02$ & $  8.22 \pm   0.03$ & $  28.6$ & $  0.47 \pm   0.01$ & $  0.45 \pm   0.02$ & $ -1.53 \pm   0.03$ & $  3533 \pm  83 $ & M2.5   \\
           202059203 & C0 &      LP 362-257 &  14.10 & $ 13.64 \pm   0.03$ & $  9.71 \pm   0.02$ & $  8.81 \pm   0.02$ & $  28.5$ & $  0.38 \pm   0.01$ & $  0.37 \pm   0.01$ & $ -1.79 \pm   0.03$ & $  3414 \pm  79 $ & M3.0   \\
\hline
\multicolumn{13}{l}{$^B$ Potential contamination by companion star in binary (see Appendix~\ref{sect:appendix_bin}).} \\
\multicolumn{13}{l}{$^*$ $M_{\rm Ks}$ slightly outside the range of calibrated values (see text in Sect.~\ref{sect:stepar}).} \\
\multicolumn{13}{l}{$^a$ This star (Gl\,852B) is not considered further because we assign the rotation and activity to the primary (Gl\,852A) in the unresolved binary.} \\
\multicolumn{13}{l}{The full table is available in the electronic edition of the journal.} \\
\end{tabular}
\end{sidewaystable*}

%
%
%
%
%
%

\section{K2 data analysis}\label{sect:k2_analysis}

We base our analysis of K2 time-series mostly on the lightcurves made publicly
available by A.Vanderburg 
\citep[see][ and Sect.~\ref{subsect:k2_analysis_prep}]{Vanderburg14.0}. 
We use the ``corrected" fluxes in which the features and trends 
resulting from the satellite pointing instability have been eliminated. 
All stars of the K2 Superblink M star sample have been observed in long-cadence (LC) 
mode with time-resolution of $\Delta t_{\rm LC} = 29.4$\,min. Nine stars have in addition 
short-cadence (SC) data available ($\Delta t_{\rm SC} = 1$\,min). In the following, where 
not explicitly stated, we refer to the LC data. 

Our analysis comprises both the measurement of rotation periods
and an assessment of photometric activity indicators. 
In particular, the identification of flares is of prime value both for activity studies
and for obtaining a ``cleaned" lightcurve allowing to perform more accurate 
diagnostics on the rotation cycle, e.g. its amplitude. 
The main limitation of the LC data is the difficulty in identifying short-duration 
flares, as a result of poor temporal resolution combined with 
the presence of some residual artefacts from instrumental effects in the lightcurves that 
have not been removed in the K2 data reduction pipeline. 
However, in this work we aim at elaborating 
trends between activity and rotation, and for this purpose completeness 
of the flare sample is less important than having 
statistically meaningful numbers of stars. 

Rotation and activity diagnostics are determined  
with an iterative process in which we identify ``outliers"  
in the K2 lightcurves. This involves removing any slowly varying signal by 
subtracting a smoothed lightcurve from the original data. The appropriate 
width of the boxcar in the smoothing process 
depends on the time-scale of the variation to be approximated, i.e. on the length of
the rotational cycle. Therefore, we start
the analysis with a first-guess period search on the original, corrected lightcurve. 
We use three methods to determine rotation periods which are laid out in 
Sect.~\ref{subsect:k2_analysis_period}. 
Before presenting the details of our period search 
we describe how we prepare the lightcurves and how we extract the flares and ``clean" 
the corrected lightcurves further, thus removing both astrophysical flare events and 
residual noise from the data reduction.

\subsection{Data preparation}\label{subsect:k2_analysis_prep}

We download the lightcurves reduced and made publicly available by
A. Vanderburg\footnote{The reduced K2 lightcurves
were downloaded from https://www.cfa.harvard.edu/$\sim$avanderb/k2.html}. 
The data reduction steps are described by \cite{Vanderburg14.0}.
In short, the authors extracted raw photometry from K2 images by aperture photometry.
The variability in the resulting lightcurves is dominated by a zigzag-like pattern
introduced by the instability of the satellite pointing and its correction
with help of spacecraft thruster fires taking place approximately every $6$\,h.
This artificial variability can be removed by a ``self-flat fielding" process
described in detail by \cite{Vanderburg14.0}. We base our analysis on these
``corrected" or ``detrended" lightcurves to which we apply some additional corrections
described below. 

Upon visual inspection of each individual corrected lightcurve we notice some flux jumps. 
As explained by A.Vanderburg in his data release notes\footnote{The technical reports on the detrending
process carried out by A.Vanderburg are accessible at 
https://www.cfa.harvard.edu/$\sim$avanderb/k2.html}  
such offsets can arise due
to the fact that he divides the lightcurves in pieces and performs the data reduction 
separately on each individual section.  
In stars with long-term variations these offsets are 
clearly seen to be an artefact of the data reduction, 
and we remove them by applying a vertical shift to the lightcurve rightwards of the 
feature. Note that, since the absolute fluxes are irrelevant for our
analysis it does not matter which side is used as the baseline for the normalization. 
While such flux jumps are evident in lightcurves with slow variations, for stars
with short periods it is much more difficult to identify such systematic offsets
and even if they are identified it is impossible to perform the normalization without
{\sl a priori} knowledge of the (periodic) variation pattern. However, since such short-period
lightcurves comprise many rotational cycles, the period search is much less sensitive
to such residual artifacts than it is for long-period lightcurves.

In a second step, we remove all cadences in which the satellite thrusters were on 
(and the telescope was moving). The thruster fires are shorter than the cadence
of observations in LC mode such that each of the corresponding gaps regards a single data 
point. Several lightcurves have spikes and decrements produced by incomplete background
removal or individual null values among the fluxes. 
We identify such obvious artefacts by visual inspection of each individual lightcurve
and remove the respective data points. 
We then fill all gaps in the K2 lightcurves, i.e. all data points separated by 
multiples of $\Delta t_{\rm LC}$, by interpolation on the neighboring data points. 
Evenly spaced data is required for the auto-correlation function, one of the
methods we use for the period search (see Sect.~\ref{subsubsect:k2_analysis_period_detrended}). 
We add Gaussian noise to the interpolated data points. To avoid that 
the width of the distribution from which the errors are drawn 
is dominated by the rotational variation we use only the nearest data points
to the right and to the left of the gap to define mean and sigma of the Gaussian.

\subsection{Identification of flares}\label{subsect:k2_analysis_flares}

Then we start the iterative flare search and cleaning process. Our approach is
similar to the methods presented in previous systematic Kepler flare studies 
\citep[][and subsequent papers of that series]{Hawley14.0}. 
Specifically, our procedure  
 consists in 
(i) boxcar smoothing of the lightcurve, 
(ii) subtraction of the smoothed from the original lightcurve (i.e. removal 
of the rotational signal), 
and (iii) flagging and removal of all data points which deviate by more than 
a chosen threshold from the subtracted curve. 
We repeat this procedure three times with successively smaller width
of the boxcar. Subsequently, the removed cadences are 
regenerated by interpolation and addition of white noise as described above. 
This provides a lightcurve that is free from flares (henceforth referred to as the
``cleaned" lightcurve). When subtracted from the original corrected data, 
the result is a flat lightcurve (henceforth referred to as the ``flattened" lightcurve) 
in which the rotational variation has been removed and the dominating variations are 
flares, eclipses and artefacts. 

\begin{figure*}
\begin{center}
\parbox{18cm}{
\parbox{9cm}{
\includegraphics[width=9.0cm]{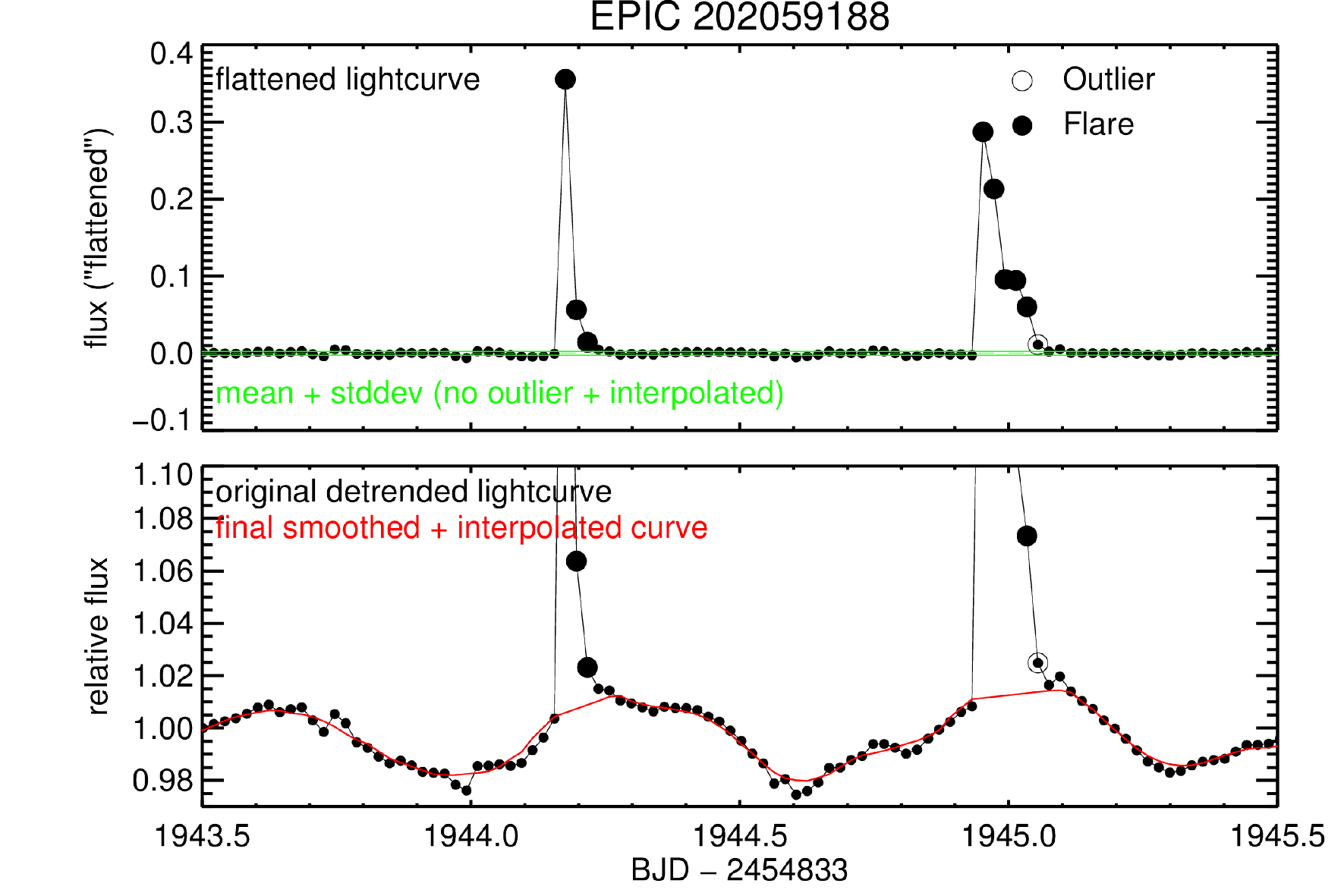}
}
\parbox{9cm}{
\includegraphics[width=9.0cm]{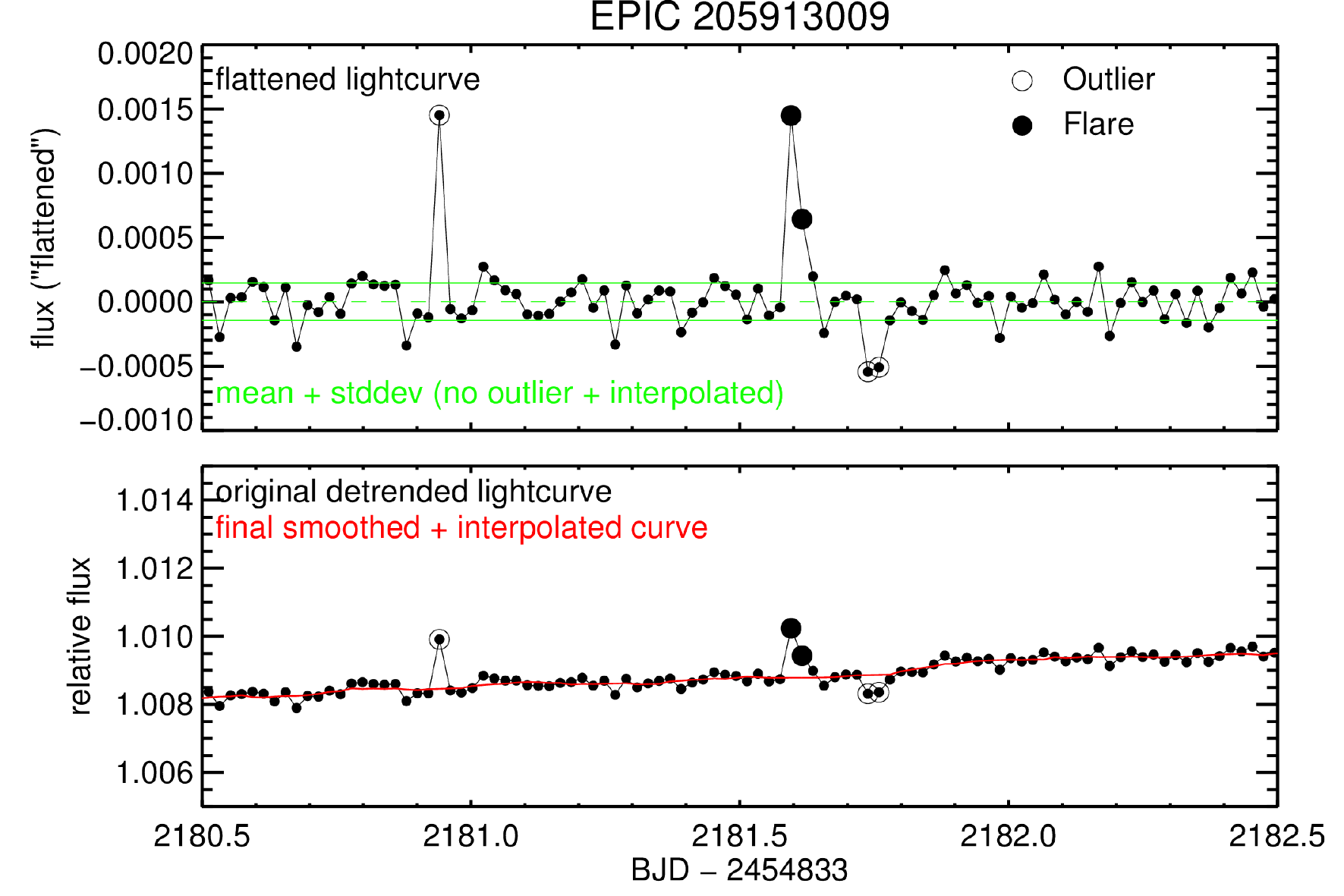}
}
}
\caption{Examples of lightcurves illustrating the procedure applied to identify flares.
Data points identified as flares are indicated, as well as other outlying points, 
according to the procedure described in Sect.~\ref{subsect:k2_analysis_flares}.} 
\label{fig:flares}
\end{center}
\end{figure*}

%
%
\begin{figure*}
\begin{center}
\parbox{18cm}{
\parbox{9cm}{
\includegraphics[width=8.5cm]{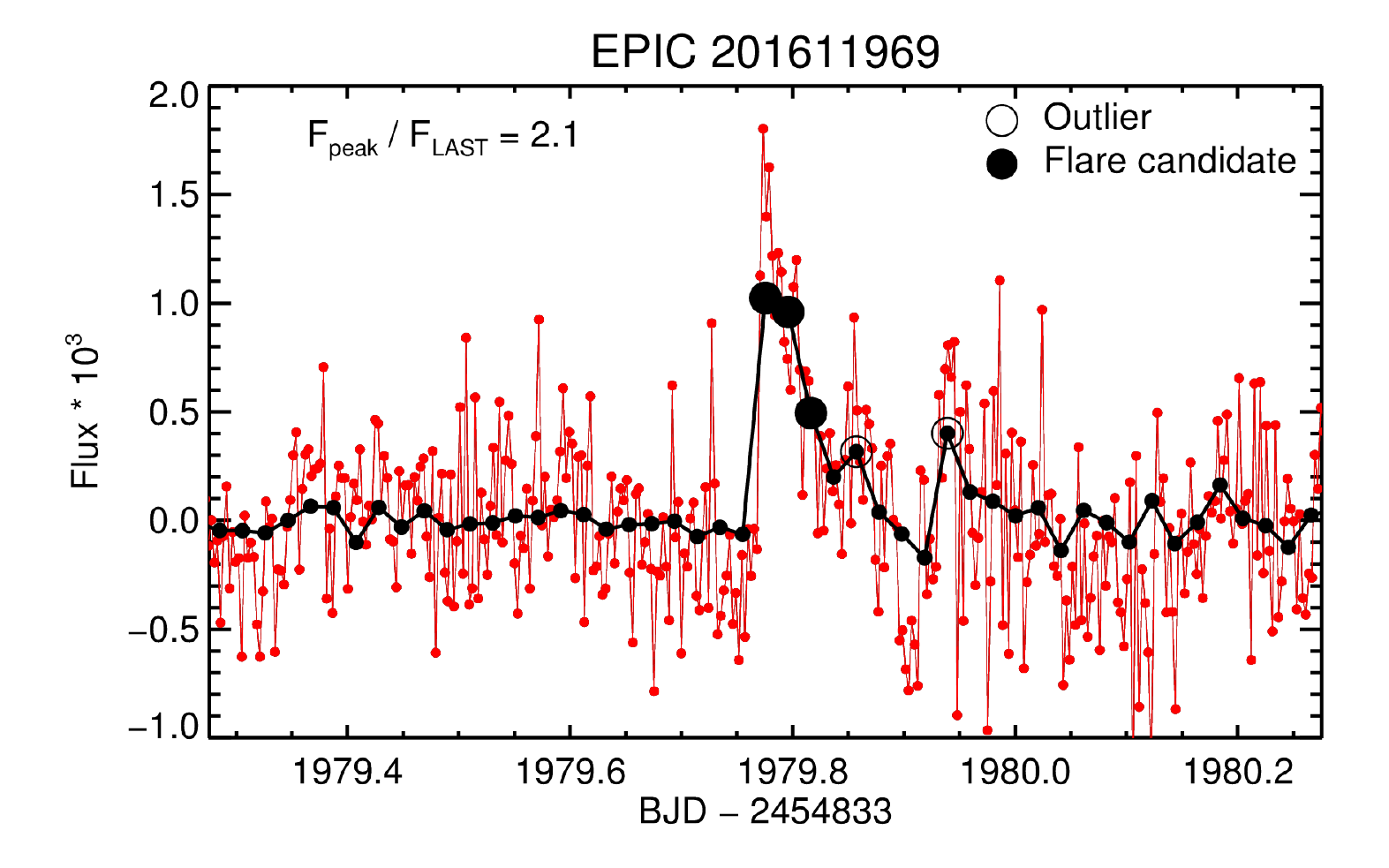}
}
\parbox{9cm}{
\includegraphics[width=8.5cm]{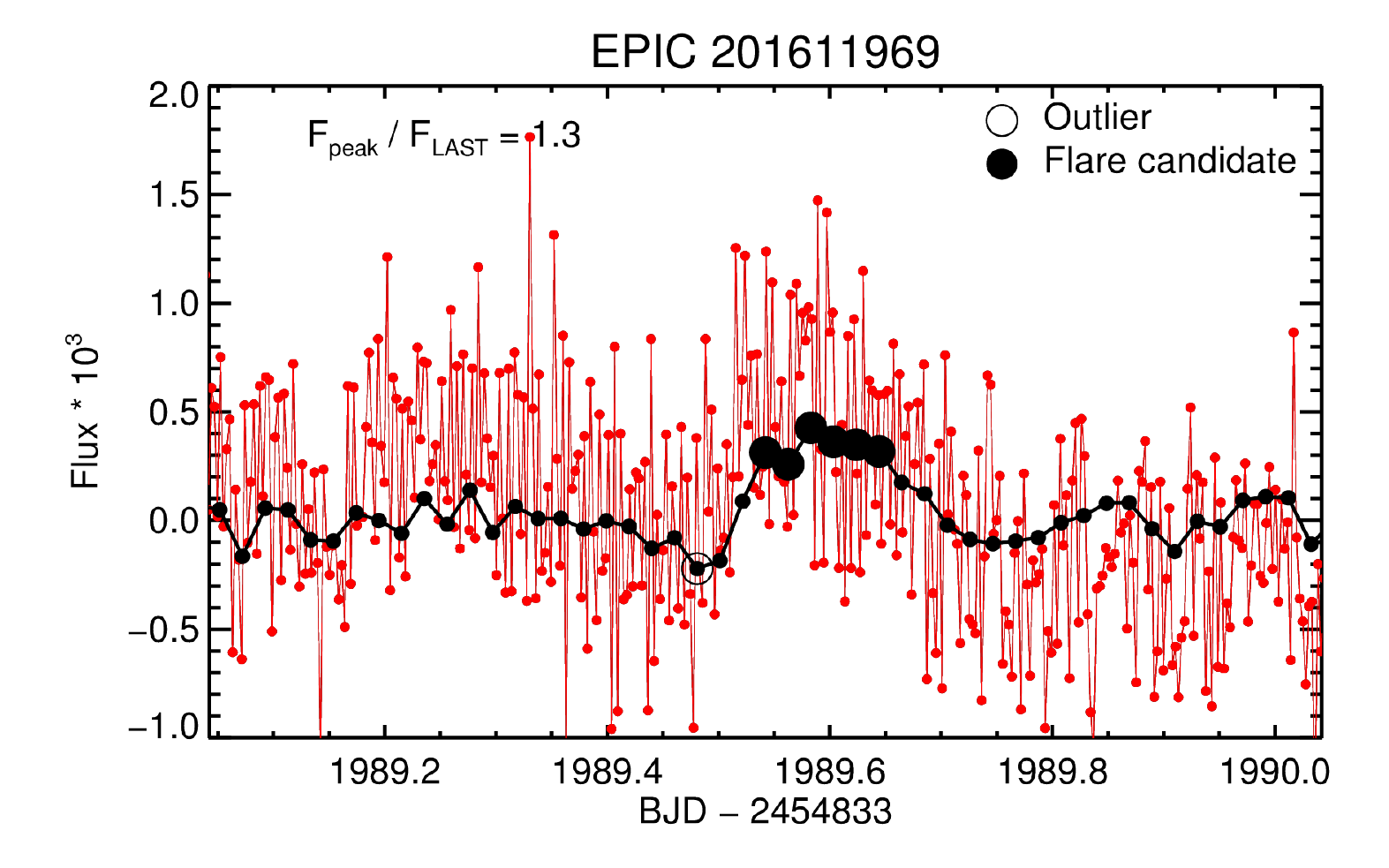}
}
}
\caption{Portion of the long-cadence (black) and short-cadence (red) K2 lightcurve for 
EPIC\,201611969. Flare candidates in LC data according to criteria (i) - (iii) described in 
Sect.~\ref{subsect:k2_analysis_flares} are marked with large filled circles. 
On the left a bona-fide flare, on the right a flare candidate which we discard
on the basis of its shape ($F_{\rm peak}/F_{\rm last} < 2$). 
The SC lightcurve has been binned to a cadence 
of $4\,s$; any vertical offset is the result of the different analysis for SC and LC
data and is irrelevant for our purpose of comparison the shape of flare candidates.}
\label{fig:lc_sc_lcs_flares}
\end{center}
\end{figure*}

A significant fraction of the data points that have been removed in the above 
$\sigma$-clipping process are isolated cadences. 
Such events are found both as up- and downward excursions in the flattened 
lightcurves. 
The number of upward outliers is for most lightcurves much larger
than the number of downward outliers, 
suggesting that many of these events are genuine flares. 
However, we assume here a conservative approach aimed at avoiding counting 
spurious events as flare. Therefore, we select all groups of at least two consecutive 
upwards deviating data points as flare candidates. In practice, this means that 
the minimum duration of the recognized flares is $\sim 1$\,hr 
(two times the cadence of $29.4$\,min).
Note, that as a result of the sigma-clipping, all flare peaks ($F_{\rm peak}$)
have a minimum significance of $3\,\sigma$, as measured with respect to the 
mean and standard deviation of the 
flattened lightcurve from which outliers have been removed, which is 
defined and further discussed in Sect.~\ref{subsect:k2_analysis_noise}. 
Finally, we require that $F_{\rm peak}$ must be at least twice the
flux of the last of the data points defining the flare ($F_{\rm last}$). 
As shown below, this last criterion removes
``flat-topped" events from our list of bona-fide flares which we trust less than ``fast-decay"
events given the possibility of residual artifacts from the data acquisition and reduction. 

A zoom into two examples of LC lightcurves with flares is shown in Fig.~\ref{fig:flares} and
illustrates our flare search algorithm. 
The lower panel shows the original, detrended lightcurve
and overlaid (in red) the smoothed lightcurve. The upper panel shows the result from 
the subtraction of these two curves, i. e. the flattened lightcurve. 
We highlight 
data points identified as outliers (open circles),
and data points that belong to bona-fide flares (filled circles).
The example on the right demonstrates the inability of 
recognizing short flares with our detection procedure. 
Short-cadence data from the main Kepler mission have shown that many flares on active 
M stars are, in fact, significantly shorter than one hour 
\citep[see e.g.][]{Hawley14.0}. 
SC lightcurves are available for $9$ stars from the K2 Superblink M star sample. 
The analysis of SC lightcurves will be described elsewhere. In this work we use
the SC data only as a cross-check on the quality of our flare search criteria applied
to the LC data (see below and Fig.~\ref{fig:lc_sc_lcs_flares}).
We recall that we aim at a conservative approach, avoiding at best possible 
spurious events in the flare sample, because our aim is to study trends with rotation. 

To summarize, the parameters of our flare search algorithm are 
(i) the width of the boxcar [adapted individually according to the first-guess period], 
(ii) the threshold for outliers identified in the $\sigma$-clipping process [adopted
to be $3\,\sigma$],
(iii) the minimum number of consecutive data points defining a flare [$2$],  
and (iv) the flux ratio between the flare peak bin and the last flare bin 
[$F_{\rm peak}/F_{\rm last} \geq 2$]. The values for these parameters have been chosen
by testing various combinations of criteria (i) - (iv) with different parameter values
and comparing the results to a by-eye inspection of the ``flattened" lightcurves.
In particular, criterion (iv) is introduced after a comparison of LC and SC lightcurves
which shows that, generally, the LC flare candidates correspond to analogous features in the
SC data but in some cases the features are very different from the canonical
flare shape (characterized by fast rise and exponential decay). 
Fig.~\ref{fig:lc_sc_lcs_flares} demonstrates that with criterion (iv) we
de-select such broad events from the list of bona-fide flares: Two flare candidates according
to criterion (i) - (iii) are shown; the event on the left panel is a bona-fide flare
according to criterion (iv) while the event on the right does not fullfill 
$F_{\rm peak}/F_{\rm last} \geq 2$.

\subsection{Residual variability and photometric noise}\label{subsect:k2_analysis_noise}

\begin{figure}
\begin{center}
\includegraphics[width=8.5cm]{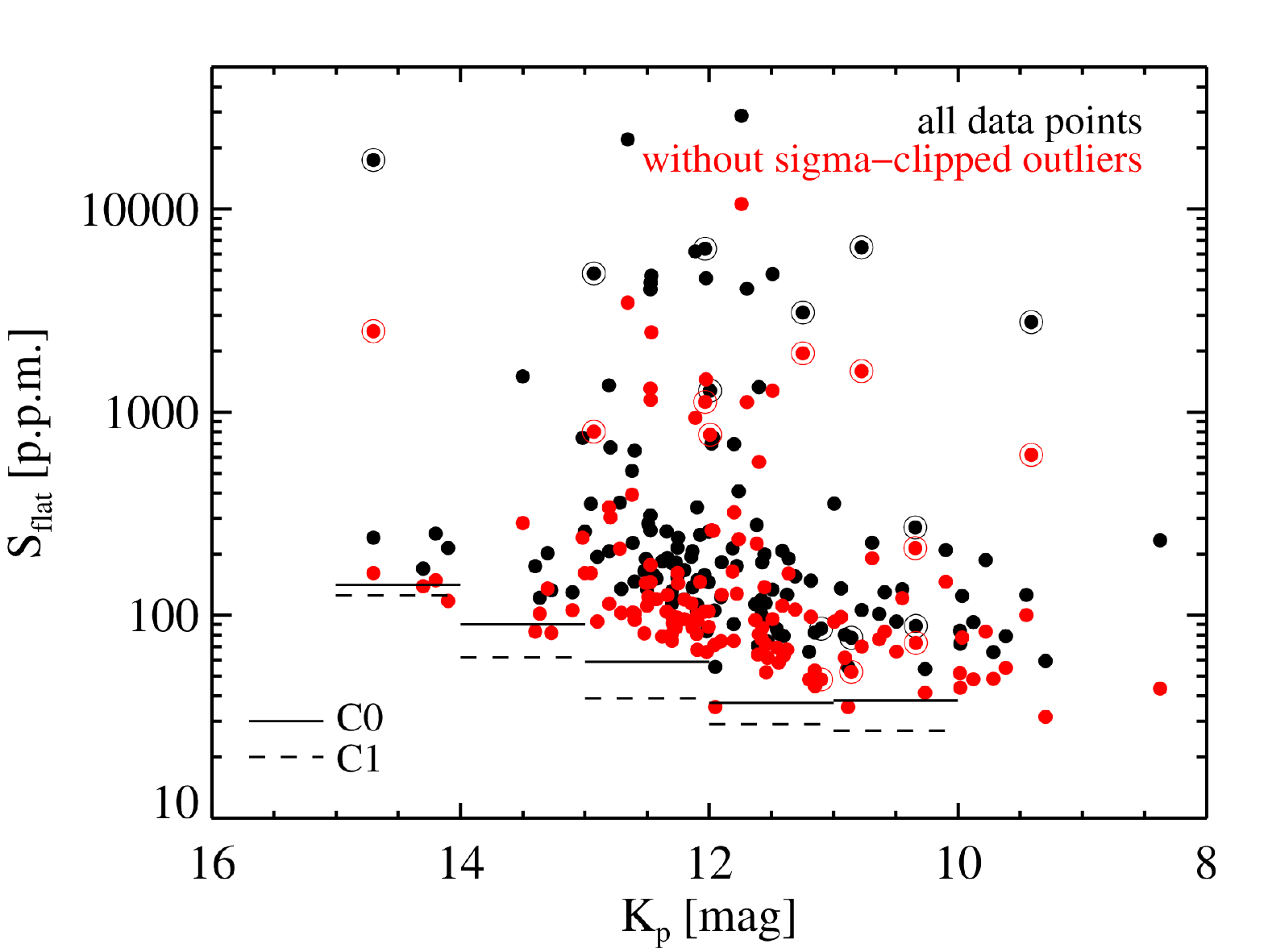}
\caption{Standard deviation for the lightcurves of the Superblink stars 
after ``flattening" by removal of the rotational signal
as described in Sect.~\ref{subsect:k2_analysis_flares}. $K_{\rm p}$ is the 
magnitude in the Kepler band. 
$S_{\rm flat}$ is calculated for two data sets: 
the full lightcurve (black circles) 
and the lightcurve without all data points that were 
identified as outliers during the clipping process (red circles). 
Horizontal lines represent the $6$\,hr-precisions for
C0 and C1 calculated by 
A.Vanderburg (see footnote to Sect.\ref{subsect:k2_analysis_prep}) 
for a sample of cool dwarfs drawn from different K2 
Guest Observer programs.
Stars with a possible contribution in the K2 photometry from an unresolved 
binary companion are high-lighted with large annuli.} 
\label{fig:kp_stddev}
\end{center}
\end{figure}

In Fig.~\ref{fig:kp_stddev} we show the standard deviations
of the flattened lightcurves, $S_{\rm flat}$, for two cases: 
including and excluding the data points identified as outliers. 
The `outliers' comprise flares, transits or eclipses, and artefacts 
from the data reduction. Therefore, for the case without outliers (red circles) the
standard deviation is 
calculated on the residual lightcurve from which the known astrophysical sources of 
variability have been removed, and it can be expected to represent the noise level in
our data. 

In Fig.~\ref{fig:kp_stddev} we compare our standard deviations $S_{\rm flat}$
to the estimated precision of K2 lightcurves provided for 
campaigns C0 and C1 in 
the data release notes of A.Vanderburg (see footnote to Sect.~\ref{subsect:k2_analysis_prep}). 
That estimate represents the $6$\,hr-precision based on a sample of cool dwarfs that is
not clearly specified. Our $S_{\rm flat}$ measurements suggest a somewhat
lower precision for the K2 Superblink M star sample. 
This might be due to differences in the definitions. 
Vanderburgs's $6$\,hr-precisions are medians for their sample 
and the scatter among their stars is much larger than the factor two difference with our 
$S_{\rm flat}$ values. 
Also, we measure the standard deviation 
on the full lightcurve while Vanderburg's precisions are based on a
running $6$\,hr mean. 
An alternative explanation for the apparently different photometric precisions 
could lie in different 
activity levels of the two samples, implying residual fluctuations of astrophysical
origin in our ``noise". 
In fact, in Sect.~\ref{subsect:results_noise} we present evidence 
that $S_{\rm flat}$ comprises an astrophysical signal.
Overall, our analysis presented in Fig.~\ref{fig:kp_stddev} confirms the high
precision achieved in K2 lightcurves with the detrending method applied by A.Vanderburg. 

\subsection{Period search}\label{subsect:k2_analysis_period}

We explore multiple approaches to measure rotation periods on the K2 data.

\subsubsection{Period search on detrended lightcurves}\label{subsubsect:k2_analysis_period_detrended}

We apply standard time-series analysis techniques, the Lomb Scargle (LS) periodogram 
and the auto-correlation function (ACF), to the detrended K2 lightcurves made publicly
available \citep[see][]{Vanderburg14.0}. 
As mentioned in Sect.~\ref{sect:k2_analysis}, as a first step we perform the period search
directly on the corrected version of the downloaded lightcurves with the purpose of  
adapting the boxcar width in the course of the
search for flares. We then repeat the period search on the ``cleaned" lightcurves 
obtained after the $\sigma$-clipping process 
and the regeneration of the missing data points through interpolation, i.e. after removal
of the flares and other outliers. 
The analysis is carried out in the IDL environment using the {\sc scargle} and 
{\sc a\_correlate} routines. 

Periodograms and ACFs have already been used successfully to determine rotation periods in 
Kepler data \citep[e.g.][]{McQuillan13.0, Nielsen13.0, Rappaport14.0, McQuillan14.0}.
As a cross-check on our procedure, we have downloaded Kepler 
lightcurves from the {\em Mikulski Archive for Space Telescopes} 
(MAST)\footnote{We downloaded the Kepler lightcurves from the Target Search page at 
https://archive.stsci.edu/kepler/kepler\_fov/search.php}
for some M stars from the \cite{McQuillan13.0} sample and we have verified that we 
correctly reproduce the published periods. 

\begin{figure*}
\begin{center}
\parbox{18cm}{
\parbox{6cm}{
\includegraphics[width=6cm]{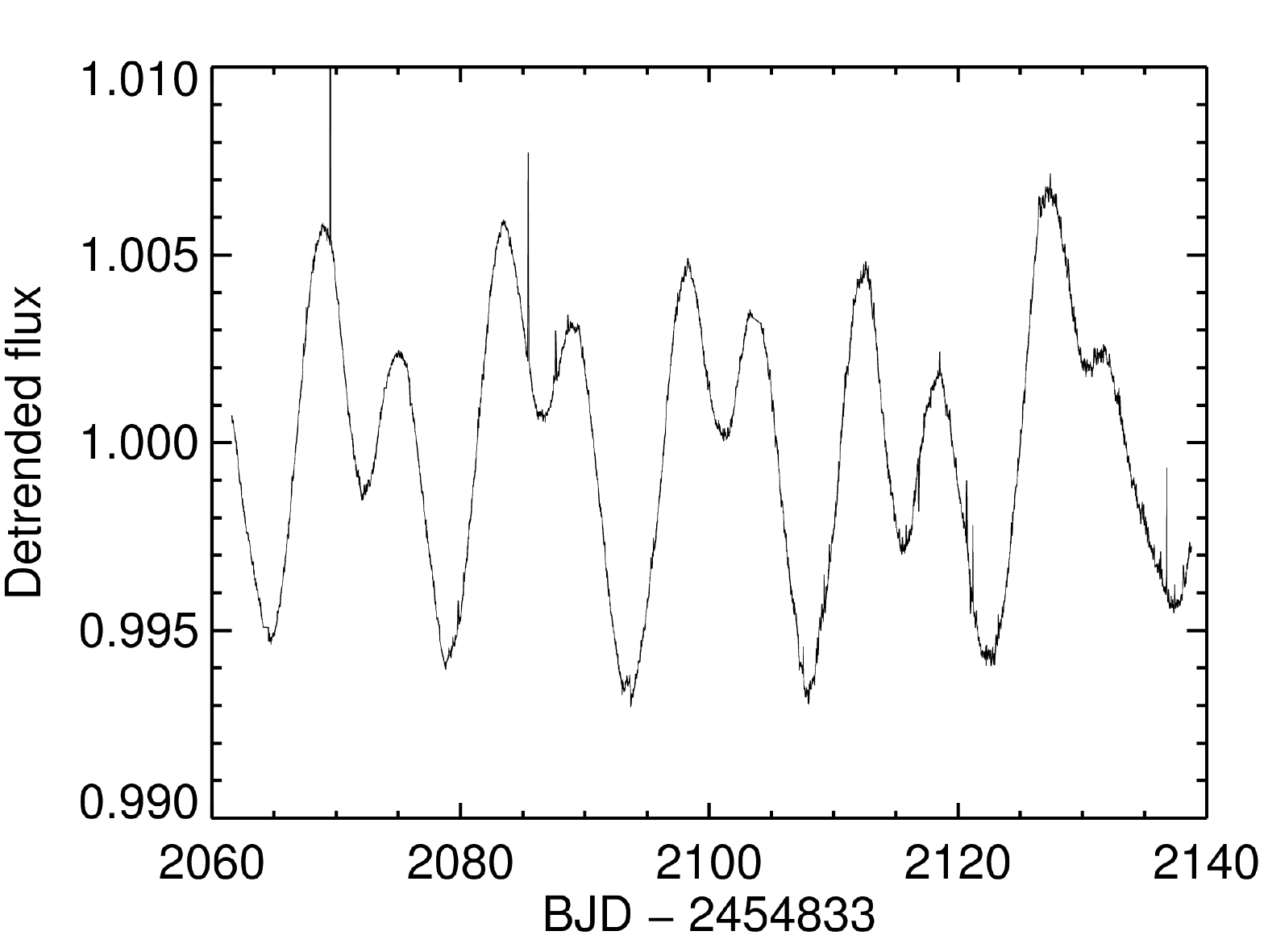}
}
\parbox{6cm}{
\includegraphics[width=6cm]{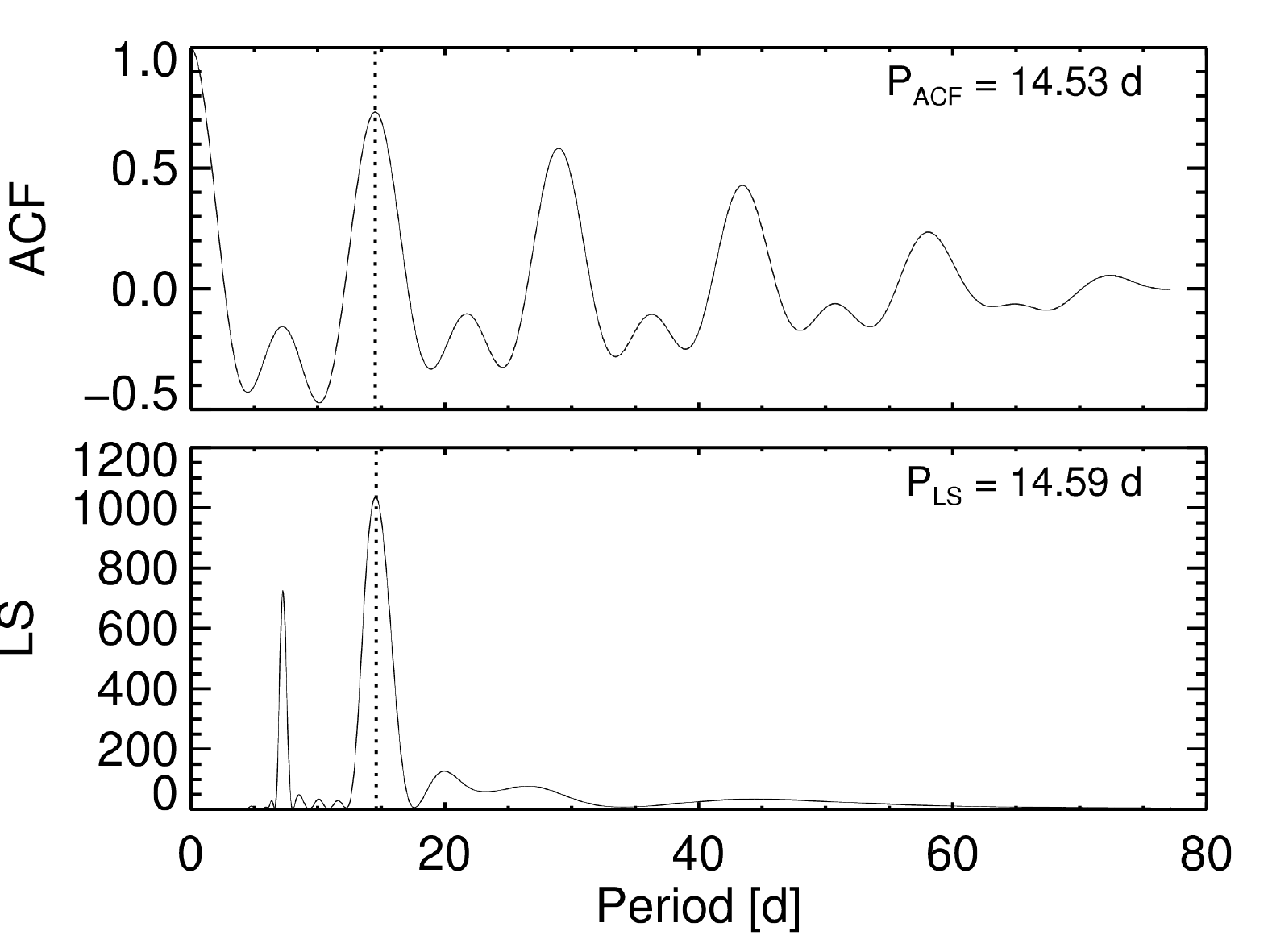}
}
\parbox{6cm}{
\includegraphics[width=6cm]{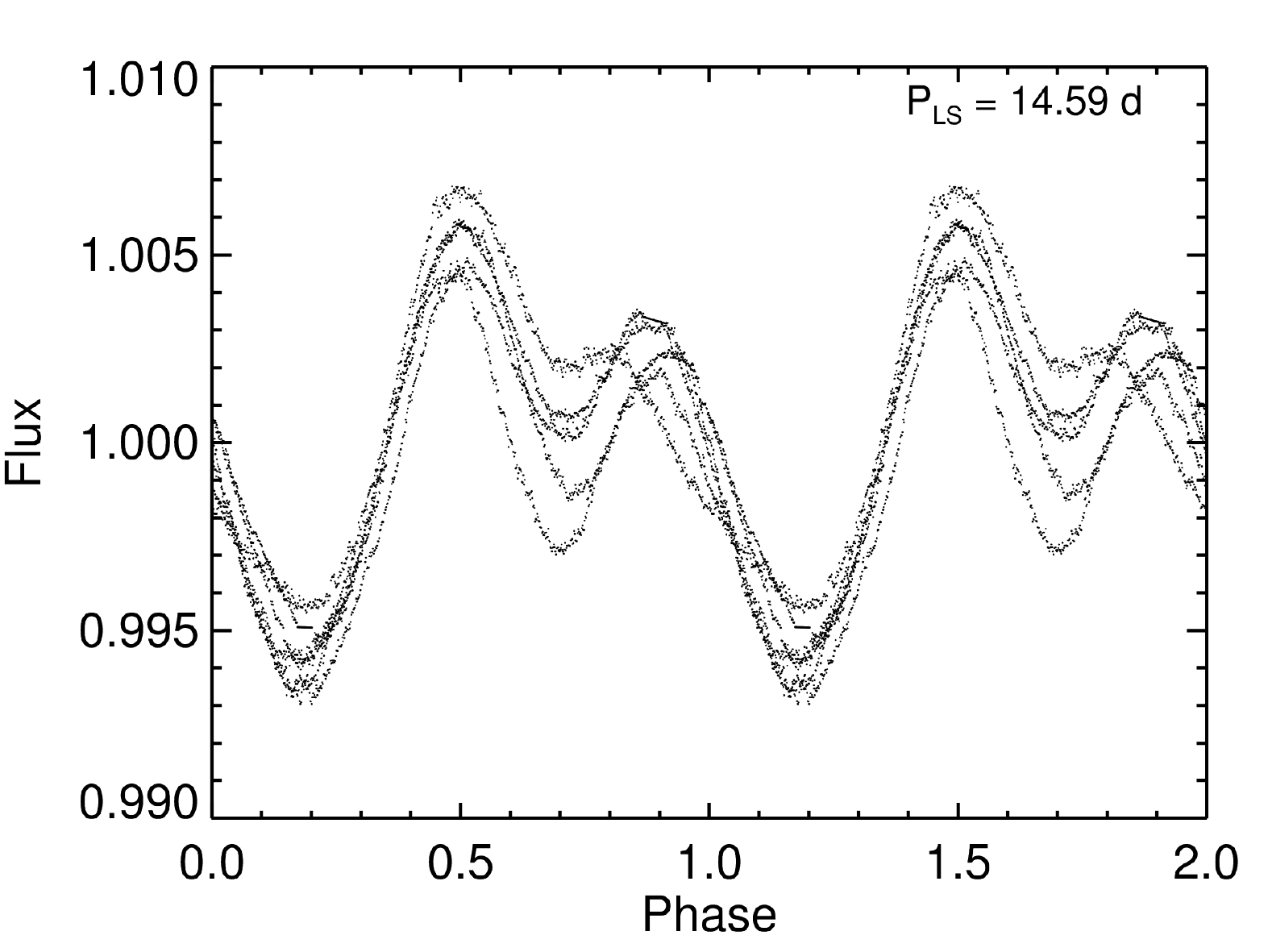}
}
}
\caption{Example of detrended K2 lightcurve (left panel), ACF and LS periodogram (middle panel), and lightcurve folded with the ACF period (right panel).\label{fig:prot_examples}
}
\end{center}
\end{figure*}

Following \cite{McQuillan13.0}, in our use of the ACF method we generally identify 
the rotation period as the time lag, $k \cdot \Delta t_{\rm LC}$ with integer number $k$, 
corresponding to the first peak in the ACF. Subsequent peaks are located at multiples of 
that period, resulting in the typical oscillatory behavior of the ACF. 
Exceptions are double-peaked lightcurves where the ACF presents two sequences 
of equidistant peaks  (see e.g. Fig.~\ref{fig:prot_examples}). 
Such lightcurves point to the presence of two dominant spots, 
and we choose the first peak of the sequence with higher ACF signal as representing the
rotation period. 
\cite{McQuillan13.0} have performed simulations that demonstrate the typical pattern of
the ACF for different effects in the lightcurve, such as changing phase and amplitude,
double peaks, and linear trends. All these features are also present in the K2 data,
although less pronounced than in the much longer main Kepler mission time-series examined
by \cite{McQuillan13.0}. 

The classical periodogram is based on a Fourier decomposition of the lightcurve.
In the form presented by \cite{Scargle82.1}, it can be applied to unevenly sampled data
and is essentially equivalent to least-squares fitting of sine-waves. 
Realistic time-series deviate from a sine-curve, and are subject to the effects described
above. This introduces features in the power spectrum. Since the dominating periodicity in
the K2 Superblink stars is reasonably given by the stellar rotation cycle, the highest 
peak of the periodogram can be interpreted as representing the rotation period.  
The LS-periodograms are computed here for a false-alarm probability of $0.01$ 
using the fast-algorithm of \cite{Press89.1}.  

In Fig.~\ref{fig:prot_examples} we show an example for a detrended K2
lightcurve, its LS-periodogram and ACF, and the lightcurve folded with the derived period. 
An atlas with the phase-folded lightcurves for all periodic stars is provided
in Appendix~\ref{sect:appendix_folded}.

\subsubsection{Period search on un-detrended lightcurves}\label{subsubsect:k2_analysis_period_undetrended}

As an independent check we derive the stellar 
rotation periods with the Systematics-Insensitive Periodogram (SIP) algorithm developed 
by \cite{Angus16.0}, that produces periodograms calculated from the analysis of the raw 
K2 photometric time series. For each observing campaign, these are modelled 
with a linear combination of a set of $150$ `eigen light curves' (ELC), or basis functions, 
that describe the systematic trends present in K2 data, plus a sum of sine and 
cosine functions over a range of frequencies\footnote{K2 raw and `eigen' light 
curves were downloaded from http://bbq.dfm.io/ketu/lightcurves/ and 
http://bbq.dfm.io/ketu/elcs/}. For each test frequency, the system of linear equations 
is solved through a least-square fit to the data. The periodogram power is determined 
as described in \cite{Angus16.0}, by calculating the squared 
signal-to-noise ratio \textit{(S/N)}$^{2}$ for each frequency. \textit{(S/N)}$^{2}$ is a 
function of the sine and cosine coefficients (i.e. the amplitudes), where 
the frequencies corresponding to amplitudes 
not well constrained by the fit are penalized. The stellar rotation period is 
finally calculated as the inverse of the frequency having the highest power.

\subsubsection{Sine-fitting of stars with long periods}\label{subsubsect:k2_analysis_period_sinefits}

The techniques described in Sects.~\ref{subsubsect:k2_analysis_period_detrended}
and~\ref{subsubsect:k2_analysis_period_undetrended} are limited to periods shorter than
the duration of the K2 campaigns 
($33$\,d for C0 and $70...80$\,d for the other campaigns).
However, by visual inspection of the lightcurves we identify $11$ stars with clearly 
sine-like variations that exceed the K2 monitoring time baseline. 
For these objects a least-squares fit allows us to constrain the rotation periods. 
The fitting was done with the routine {\sc curve\_fit} in the
Python package SciPy \citep{Jones01.0} 
As initial guesses for the parameters we used 
four times the standard deviation as amplitude, a period of $30$\,d, 
and a phase of $0.0$, but the results do not depend on this choice. 
For all $11$ lightcurves the routine converges on a unique solution 
independent of the choice of the initial guesses for the parameters.
In three cases the sinecurve provides only a crude approximation because the lightcurve is
not symmetric around maximum or minimum and shows signs of spot evolution; in these cases
the results are treated with caution.

\subsubsection{Comparison of results from different period search algorithms}\label{subsubsect:k2_analysis_rot_compare}

\begin{figure}
\begin{center}
\includegraphics[width=8.5cm]{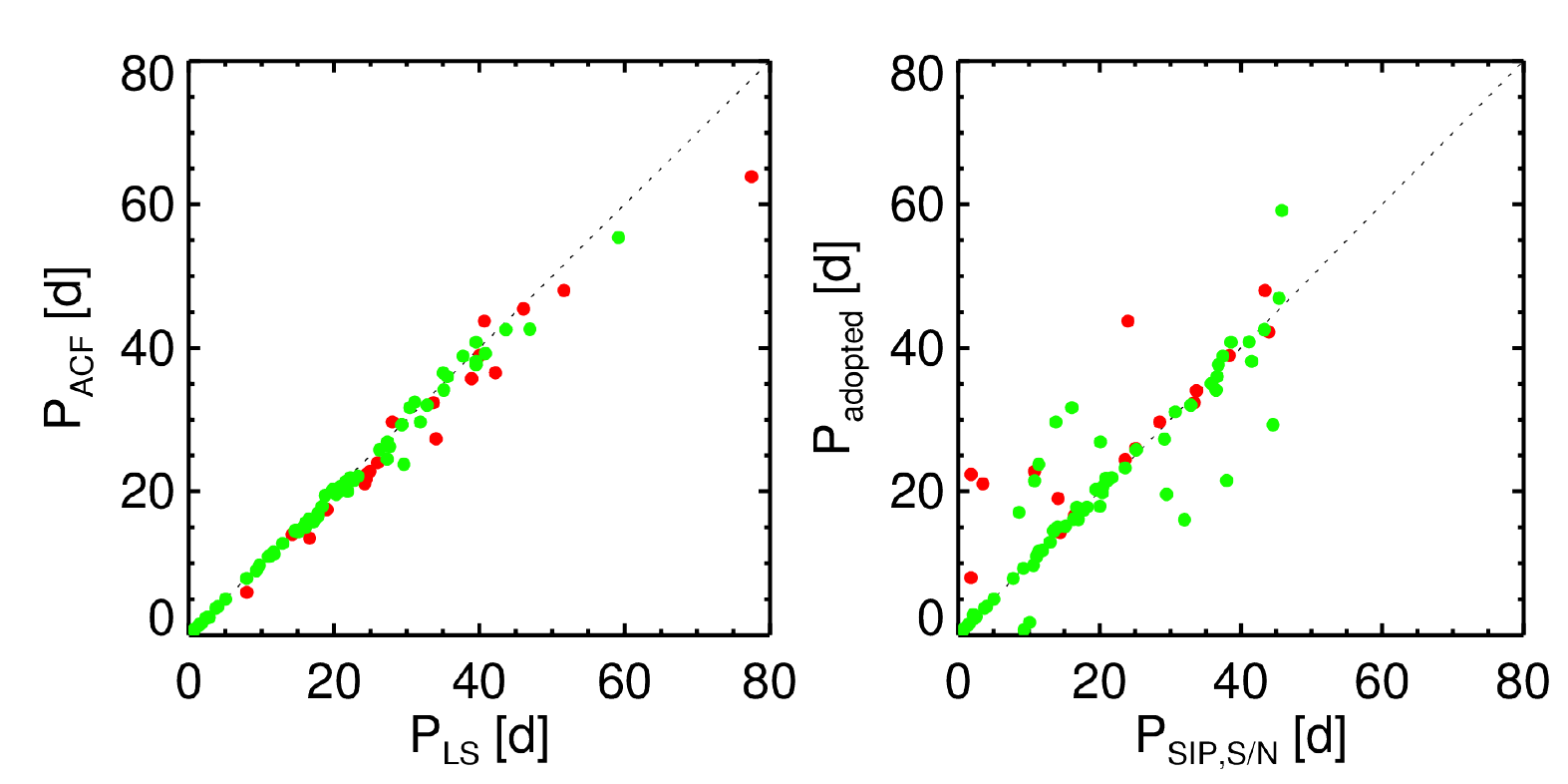}
\caption{Comparison of the periods derived with the different methods described in 
Sect.~\ref{subsect:k2_analysis_period}. 
Reliable periods (green; flag `y' in Table~\ref{tab:rot}), questionable periods 
(red; flag `?'). 
Periods equal to or longer than the data set according to the LS and ACF analysis
have been determined through sine-fitting and are not shown here.} 
\label{fig:period_comparison}
\end{center}
\end{figure}

The results of the different period search methods are compared 
in Fig.~\ref{fig:period_comparison}. 
Generally, the LS and the ACF periods are in excellent agreement. 
For further use, based on the agreement between the periods obtained with the two 
techniques and considering the appearance of the phase folded lightcurve, we adopt 
either the ACF or the LS period as rotation period. 
This selection is made independently by two 
members of the team (BS and AS),  
and the results deviate for only few stars. For those dubious cases
we make use of the SIP results as cross-check, and we adopt the period (either LS or
ACF) which is in better agreement with the SIP ``S/N" period. 
In addition, for the $11$ stars that have their highest peak in the ACF and LS at a 
period corresponding to the length of the data set ($T_{\rm tot}$) but for 
which visual inspection reveals a clear 
(sine-like) pattern indicating a spot-modulation with $P_{\rm rot} > T_{\rm tot}$
we use the periods from the sine-fitting.  
From a comparison of the values obtained with the ACF and with the LS periodogram
we estimate the typical error on our periods to be $\lesssim 3$\,\%. 
The final, adopted periods are given in Table~\ref{tab:rot} 
together with a quality flag and reference to the method with which it was derived. 
Flag `Y' stands for reliable periods, `?' for questionable period detections, 
and `N' for no period. 
These periods are obtained from the `cleaned' lightcurves, but 
due to the robustness of the detection techniques 
they are in agreement (within $< 5$\,\%) with the
periods found on the original, detrended lightcurves.


\begin{table}\begin{center}
\caption{Rotation and activity parameters derived from the K2 lightcurves. The full
table is available in the electronic version of the journal.}
\label{tab:rot}
\begin{tabular}{lrccrrr}\hline
EPIC ID & $P_{\rm rot}$ & method & flag & $R_{\rm 0}$ & $R_{\rm per}$ & $S_{\rm ph}$ \\
        &        [d]    &        &      &             &        [\%]   & [ppm]  \\ \hline
      202059188 &       0.69 &    LS & $Y$ &       0.01 &  2.754 &  10520.9  \\
      202059192 &      35.22 &  SINE & $Y$ &       0.78 &  0.559 &   1844.1  \\
      202059193 &      19.01 &    LS & $?$ &       0.42 &  0.469 &   1161.4  \\
      202059195 &      42.46 &  SINE & $Y$ &       0.63 &  1.814 &   6447.7  \\
      202059198 &      27.31 &    LS & $Y$ &       0.61 &  0.883 &   2802.3  \\
      202059199 &        ... &   --- & $N$ &        ... &  0.877 &   2508.0  \\
      202059203 &        ... &   --- & $N$ &        ... &  0.273 &    635.6  \\
      202059204 &       7.89 &   ACF & $Y$ &       0.18 &  2.760 &   8494.3  \\
\hline
\end{tabular}
\end{center}\end{table}

%
%
%
%

\section{X-ray emission}\label{sect:xray_analysis}

We perform a systematic archive search for X-ray observations of the K2
Superblink M stars. Specifically, we consult the 
{\em XMM-Newton} Serendipitous Source Catalogue \citep[3\,XMM-DR5;][]{Rosen16.0}, 
the {\em XMM-Newton} Slew Survey Source Catalogue \citep[XMMSL1\_Delta6;][]{Saxton08.0},  
the Second {\em ROSAT} Source Catalog of Pointed Observations (2RXP) 
and the {\em ROSAT} Bright and Faint Source
catalogs (BSC and FSC). Our procedure for cross-matching the K2 targets with these
catalogs and for deriving X-ray fluxes and luminosities follow those described by 
\cite{Stelzer13.0}. That work presented the X-ray and UV emission of the M dwarfs
within $10$\,pc of the Sun. Although that sample was drawn from the same catalog (LG11),
there are only two stars in common with our K2 study because most of the stars that
fall in the K2 fields have distances in the range $20...60$\,pc. 
We briefly summarize the individual analysis steps here and we refer to
\cite{Stelzer13.0} for details. 

First, in order to ensure that no matches are missed due to the high proper motion 
of the most nearby stars (stars at $< 10$\,pc have proper motions of
$\sim 1^{\prime\prime}$/yr), the cross-correlation
between the K2 target list and the X-ray catalogs is done after correcting the object
coordinates from the K2 catalog\footnote{Note, that the coordinates provided in the
target lists of the individual K2 campaigns at http://keplerscience.arc.nasa.gov/k2-approved-programs.html 
refer to epoch 2000, except for
campaign C0 where the coordinates seem to refer to the date of observation (Mar - May 2014).} 
to the date of the X-ray observation using the proper motions given by LG11. 
We then use the following match radii between the X-ray catalog positions and the K2
coordinates: $40^{\prime\prime}$ for RASS \citep{Neuhaeuser95.1}, 
$30^{\prime\prime}$ for XMMSL \citep{Saxton08.0}, 
$25^{\prime\prime}$ for 2\,RXP \citep{Pfeffermann03.0}
and $10^{\prime\prime}$ for 3XMM-DR5. With one exception all counterparts have much 
smaller separations than the respective cross-correlation radius (see Table~\ref{tab:x}).
The only doubtful X-ray counterpart is the XMMSL source associated with EPIC\,201917390. 
It has a separation close to the edge of our match circle which 
-- as stated by \cite{Saxton08.0} -- is a generous interpretation of the astrometric
uncertainty of the XMMSL. Since the same star is also clearly identified with a RASS source
at an X-ray luminosity within a factor of two of the XMMSL source, we decide to keep
the XMMSL counterpart. 
 For all but one of the X-ray detected stars the rotation 
period could be determined. The exception is EPIC\,210500368 for which hints for
pseudo-periodic variations in the K2 lightcurve can be seen by-eye but the ACF and
LS periodograms show no dominant peak.

After the identification of the X-ray counterparts we compute 
their $0.2-2$\,keV flux assuming a $0.3$\,keV thermal emission subject to an 
absorbing column of $N_{\rm H} = 10^{19}\,{\rm cm^{-2}}$. 
The {\em ROSAT} count-to-flux conversion factor is determined 
with PIMMS\footnote{The Portable Interactive Multi-Mission Simulator is accessible at 
http://cxc.harvard.edu/toolkit/pimms.jsp} to be 
$CF_{\rm ROSAT} = 2.03 \cdot 10^{11}\,{\rm cts\,erg^{-1}\,cm^{-2}}$ 
\citep[see also][]{Stelzer13.0} and we apply it to the
count rates given in the 2\,RXP catalog, the BSC and the FSC. 
For 3XMM-DR5 sources we use the tabulated EPIC/pn count rates in bands $1-3$
which represent energies of $0.2-0.5$, $0.5-1.0$, and $1.0-2.0$\,keV, respectively.
We sum the count rates in these bands, and perform the flux conversion for the combined
$0.2-2.0$\,keV band. All 3XMM-DR5 counterparts to K2 Superblink M stars were observed 
with the EPIC/pn medium filter, and for the $N_{\rm H}$ and $kT$ given above 
we find in PIMMS a count-to-flux conversion factor of 
$CF_{\rm 3XMM-DR5} = 9.22 \cdot 10^{11}\,{\rm cts\,erg^{-1}\,cm^{-2}}$. 
The XMMSL1 catalog has the three energy bands already combined in columns `B5'. 

A total of $26$ K2 Superblink stars have an X-ray counterpart in the archival 
databases that we have consulted. 
The X-ray fluxes obtained as described above
are converted to luminosities using the updated photometric distances of the stars
(see Sect.~\ref{sect:stepar}). The X-ray luminosities 
are given in Table~\ref{tab:x} together with the separation between X-ray and 
optical position and the respective X-ray catalog.
For stars with more than one epoch of X-ray detection 
the luminosities are in agreement within a factor of two and we provide
the mean of the two values. 
The errors of $L_{\rm x}$ comprise the uncertainties of the count rates and an assumed 
$20$\,\% error of the distances which yield roughly comparable contributions to the error
budget. Our assumption on the distance error is motivated by the distance spread between 
photometric and trigonometric distances for the subsample with both measurements
(described in Sect.~\ref{sect:stepar}).  

%
%
\begin{table}\begin{center}
\caption{X-ray parameters derived from archival data. For stars with multiple detections the mean X-ray luminosity is given. ${\rm Sep_{x,opt}}$ are the separations between X-ray and K2-EPIC position.}
\label{tab:x}
\begin{tabular}{lrrl}\hline
EPIC ID & \multicolumn{1}{c}{$\log{L_{\rm x}}$} & ${\rm Sep_{x,opt}}$ & Ref.cat  \\
        & \multicolumn{1}{c}{[erg/s]}                & \multicolumn{1}{c}{[$^{\prime\prime}$]} &  \\ \hline
      202059204  & $   28.6   \pm     0.2$ & $   6.2,   6.2$ &        BSC,       BSC  \\
      202059229  & $   29.2   \pm     0.2$ & $   2.9,   2.9$ &        BSC,       BSC  \\
      202059231  & $   28.2   \pm     0.3$ & $  13.3,  13.3$ &        FSC,       FSC  \\
      201482319  & $   28.2   \pm     0.2$ & $   7.6,  14.0$ &        BSC,      2RXP  \\
      201518346  & $   26.8   \pm     0.2$ & $   7.2,  10.8$ &       2RXP,       BSC  \\
      201675315  & $   27.1   \pm     0.2$ & $   1.4,   1.4$ &       3XMM,      3XMM  \\
      201806997  & $   29.4   \pm     0.2$ & $   6.8,   6.8$ &        BSC,       BSC  \\
      201917390  & $   28.5   \pm     0.2$ & $   8.4,  29.1$ &        BSC,     XMMSL  \\
      201842163  & $   28.5   \pm     0.2$ & $  22.8,  22.8$ &        BSC,       BSC  \\
      201909533  & $   28.8   \pm     0.2$ & $   2.8,   2.8$ &        BSC,       BSC  \\
      202571062  & $   29.5   \pm     0.2$ & $  13.0,   5.6$ &    ${\rm XMMSL_1}$, ${\rm XMMSL_2}$  \\
      204927969  & $   28.1   \pm     0.2$ & $   2.3,   2.3$ &        BSC,       BSC  \\
      204957517  & $   27.8   \pm     0.2$ & $   1.0,   1.0$ &       2RXP,      2RXP  \\
      205467732  & $   27.4   \pm     0.3$ & $   8.9,   8.9$ &        FSC,       FSC  \\
      205913009  & $   26.1   \pm     0.2$ & $   0.6,   0.6$ &       3XMM,      3XMM  \\
      206019392  & $   26.2   \pm     0.2$ & $   0.3,   9.0$ &       3XMM,      2RXP  \\
      206208968  & $   28.8   \pm     0.2$ & $  18.3,  18.3$ &        BSC,       BSC  \\
      206262336  & $   28.5   \pm     0.2$ & $  13.4,  13.4$ &        BSC,       BSC  \\
      206349327  & $   28.7   \pm     0.3$ & $   2.2,   2.2$ &        BSC,       BSC  \\
      210434976  & $   28.4   \pm     0.2$ & $  26.3,  26.3$ &        BSC,       BSC  \\
      210500368  & $   27.9   \pm     0.3$ & $   0.8,   0.8$ &       3XMM,      3XMM  \\
      210613397  & $   29.4   \pm     0.2$ & $  11.0,  11.0$ &        BSC,       BSC  \\
      210651981  & $   28.7   \pm     0.2$ & $   8.2,  12.5$ &      XMMSL,       BSC  \\
      210707811  & $   28.3   \pm     0.3$ & $  18.3,  18.3$ &        FSC,       FSC  \\
      210741091  & $   28.7   \pm     0.2$ & $   6.0,  10.7$ &     ${\rm 2RXP_2}$, ${\rm 2RXP_1}$  \\
      211111803  & $   28.7   \pm     0.3$ & $  17.8,  17.8$ &        FSC,       FSC  \\
\hline
\end{tabular}
\end{center}\end{table}

\section{Ultraviolet emission}\label{sect:uv_analysis}

To assess the UV activity of the K2 Superblink stars we cross-match our 
target list with the {\em GALEX-DR5 sources from AIS and MIS} \citep{Bianchi12.0}. 
GALEX performed imaging in two UV bands, far-UV 
(henceforth FUV; $\lambda_{\rm eff} = 1528$\,\AA, $\Delta \lambda = 1344-1786$\,\AA)
and 
near-UV (henceforth NUV; $\lambda_{\rm eff} =2271$\,\AA, $\Delta \lambda = 1771-2831$\,\AA). 
The All-Sky Survey (AIS) covered $\sim 85$\,\% of the high Galactic latitude ($\|b\|> 20^\circ$)
sky to m$_{AB} \sim 21$\,mag, and the Medium Imaging Survey (MIS) reached 
m$_{AB} \sim 23$\,mag on 1000 deg$^2$ \citep[e.g.][]{Bianchi09.0}. 

Analogous to our analysis of the X-ray data, we correct the coordinates from the K2
catalog to the date of the respective UV observation. 
We use a match radius of $10^{\prime\prime}$,
but none of the UV counterparts we identify is further than $3^{\prime\prime}$
from the proper motion corrected K2 position. The GALEX-DR5 catalog provides NUV
and FUV magnitudes which we convert to flux densities using the zero points 
given by \cite{Morrissey05.0}.

We isolate the chromospheric contribution to the UV emission from the photospheric part
with help of synthetic {\sc dusty} spectra of \cite{Allard01.1}, following the
procedure described by \cite{Stelzer13.0}. We adopt the model spectra with solar 
metallicity and $\log{g} = 4.5$, and we choose for each star that
model from the grid which has $T_{\rm eff}$ closest to the observationally determined 
value derived in Sect.~\ref{sect:stepar}. We then obtain the predicted photospheric UV flux density 
[$(f_{\rm UV_i,ph})_\lambda$] 
in the two GALEX bands ($i=NUV, FUV$) 
from the UV and $J$ band flux densities of the {\sc dusty} model 
(i.e. the synthetic $UV_i - J$ color) and the observed $J$ band flux density. 
The model flux densities in the FUV, NUV and $J$ bands are determined by convolving the
synthetic spectrum with the respective normalized filter transmission curve. 
Finally, the FUV and NUV fluxes are obtained by multiplying $(f_{\rm UV_i,ph})_\lambda$ 
with the effective band width of the respective GALEX filter 
 ($\delta \lambda_{\rm FUV} = 268$\,\AA;
$\delta \lambda_{\rm NUV} = 732$\,\AA); \cite{Morrissey07.0}. 
The expected photospheric fluxes ($f_{\rm UV_i,ph}$) are then subtracted from 
the observed ones to yield the chromospheric fluxes. We refer to these values as
`UV excess', $f_{\rm UV_i,exc}$. Finally, we define the UV activity index 
as $R^\prime_{UV_i} = \frac{f_{\rm UV_i,exc}}{f_{\rm bol}}$ where $f_{\rm bol}$ is the
bolometric flux. The superscript ($^\prime$) indicates, in the same manner as for the
well-known Ca\,{\sc ii}\,H\&K index, that the flux ratio has been
corrected for the photospheric contribution.  

We find NUV detections for $41$ stars from the K2 Superblink M star sample, i.e. roughly 
$30$\,\%, while only $11$ stars ($\sim 8$\,\%) are identified as FUV sources.
\cite{Stelzer13.0} have shown that the photospheric contribution to the FUV emission
of M stars is negligible while the fraction of the NUV emission emitted by the
photosphere can be significant. We confirm here for the K2 Superblink M stars with FUV 
detections that this emission is entirely emitted from the chromosphere, i.e. 
$f_{\rm FUV,ph}$ is orders of magnitude smaller than the observed FUV flux. 
The NUV emission of the K2 Superblink M stars is also only weakly affected by
photospheric contributions with $f_{\rm NUV,ph}$ less than $\sim 10$\,\% of the
observed flux for all stars. 
In Table~\ref{tab:uv} we provide the observed FUV and NUV magnitudes and the calculated 
chromospheric excess, $L_{\rm UV_i,exc}$, of all detected objects. 
The uncertainties for the UV luminosities comprise the magnitude errors and 
an assumed $20$\,\% error on the distances (as in Sect.~\ref{sect:xray_analysis}). 

%
\begin{table*}\begin{center}
\caption{UV parameters derived from archival GALEX data. Errors on the UV luminosities
comprise also an assumed uncertainty of $20$\,\% on the distance.}
\label{tab:uv}
\begin{tabular}{lrrrr}\hline
EPIC ID & \multicolumn{1}{c}{NUV}   & \multicolumn{1}{c}{FUV}   & \multicolumn{1}{c}{$\log{L_{\rm NUV}^\dagger}$} & \multicolumn{1}{c}{$\log{L_{\rm FUV}^\dagger}$}  \\
        & \multicolumn{1}{c}{[mag]} & \multicolumn{1}{c}{[mag]} & \multicolumn{1}{c}{[erg/s]}              & \multicolumn{1}{c}{[erg/s]}               \\ \hline
      201237257  & $ 19.95 \pm   0.13$ &                     & $   28.3 \pm     0.2$ &                        \\
      201460770  & $ 21.89 \pm   0.40$ &                     & $   27.8 \pm     0.3$ &                        \\
      201482319  & $ 20.17 \pm   0.16$ & $ 21.54 \pm   0.48$ & $   27.6 \pm     0.2$ & $   26.9 \pm     0.3$  \\
      201506253  & $ 21.20 \pm   0.19$ &                     & $   28.1 \pm     0.2$ &                        \\
      201518346  & $ 21.16 \pm   0.36$ &                     & $   26.0 \pm     0.2$ &                        \\
      201568682  & $ 21.07 \pm   0.22$ &                     & $   28.4 \pm     0.2$ &                        \\
      201611969  & $ 21.85 \pm   0.30$ &                     & $   27.4 \pm     0.2$ &                        \\
      201675315  & $ 20.95 \pm   0.27$ &                     & $   28.1 \pm     0.2$ &                        \\
      201719818  & $ 19.14 \pm   0.08$ & $ 22.05 \pm   0.47$ & $   28.5 \pm     0.2$ & $   27.3 \pm     0.3$  \\
      201917390  & $ 20.20 \pm   0.15$ & $ 21.47 \pm   0.38$ & $   27.9 \pm     0.2$ & $   27.3 \pm     0.2$  \\
      201367065  & $ 21.41 \pm   0.22$ &                     & $   28.2 \pm     0.2$ &                        \\
      201497866  & $ 21.28 \pm   0.34$ &                     & $   27.9 \pm     0.2$ &                        \\
      201842163  & $ 20.02 \pm   0.11$ & $ 21.39 \pm   0.29$ & $   28.2 \pm     0.2$ & $   27.6 \pm     0.2$  \\
      201909533  & $ 18.45 \pm   0.04$ & $ 20.06 \pm   0.12$ & $   28.6 \pm     0.2$ & $   27.9 \pm     0.2$  \\
      204963027  & $ 19.86 \pm   0.17$ &                     & $   28.3 \pm     0.2$ &                        \\
      204927969  & $ 20.18 \pm   0.20$ &                     & $   27.8 \pm     0.2$ &                        \\
      204994054  & $ 20.70 \pm   0.28$ &                     & $   28.2 \pm     0.2$ &                        \\
      206007536  & $ 19.99 \pm   0.09$ &                     & $   28.7 \pm     0.2$ &                        \\
      206019392  & $ 20.07 \pm   0.09$ & $ 22.63 \pm   0.42$ & $   26.4 \pm     0.2$ & $   25.3 \pm     0.2$  \\
      206054454  & $ 21.44 \pm   0.25$ & $ 22.12 \pm   0.46$ & $   27.7 \pm     0.2$ & $   27.4 \pm     0.3$  \\
      206055065  & $ 19.87 \pm   0.10$ &                     & $   29.0 \pm     0.2$ &                        \\
      206056832  & $ 21.45 \pm   0.25$ &                     & $   28.5 \pm     0.2$ &                        \\
      206107346  & $ 19.02 \pm   0.05$ & $ 21.86 \pm   0.30$ & $   28.2 \pm     0.2$ & $   27.0 \pm     0.2$  \\
      206208968  & $ 18.89 \pm   0.08$ & $ 20.14 \pm   0.21$ & $   28.2 \pm     0.2$ & $   27.6 \pm     0.2$  \\
      206368165  & $ 22.54 \pm   0.47$ &                     & $   27.5 \pm     0.3$ &                        \\
      206479389  & $ 21.57 \pm   0.33$ &                     & $   27.9 \pm     0.2$ &                        \\
      206490189  & $ 21.96 \pm   0.30$ &                     & $   27.3 \pm     0.2$ &                        \\
      210393283  & $ 21.57 \pm   0.41$ &                     & $   28.0 \pm     0.3$ &                        \\
      210434976  & $ 20.18 \pm   0.16$ &                     & $   27.7 \pm     0.2$ &                        \\
      210460280  & $ 20.76 \pm   0.20$ &                     & $   27.9 \pm     0.2$ &                        \\
      210500368  & $ 21.80 \pm   0.40$ &                     & $   27.9 \pm     0.3$ &                        \\
      210502828  & $ 20.49 \pm   0.18$ &                     & $   28.5 \pm     0.2$ &                        \\
      210535241  & $ 21.75 \pm   0.35$ &                     & $   28.1 \pm     0.2$ &                        \\
      210579749  & $ 19.58 \pm   0.09$ & $ 21.40 \pm   0.40$ & $   27.9 \pm     0.2$ & $   27.1 \pm     0.2$  \\
      210585703  & $ 21.97 \pm   0.40$ &                     & $   27.7 \pm     0.3$ &                        \\
      210592074  & $ 20.33 \pm   0.33$ &                     & $   28.8 \pm     0.2$ &                        \\
      210613397  & $ 19.69 \pm   0.09$ & $ 21.66 \pm   0.39$ & $   28.8 \pm     0.2$ & $   27.9 \pm     0.2$  \\
      210757663  & $ 21.65 \pm   0.39$ &                     & $   28.1 \pm     0.3$ &                        \\
      210778181  & $ 20.30 \pm   0.18$ &                     & $   28.5 \pm     0.2$ &                        \\
      211008819  & $ 18.81 \pm   0.08$ &                     & $   28.5 \pm     0.2$ &                        \\
      211036776  & $ 21.08 \pm   0.22$ &                     & $   27.9 \pm     0.2$ &                        \\
\hline
\multicolumn{5}{l}{$^\dagger$ Chromospheric excess luminosities after subtraction of the photo-} \\
\multicolumn{5}{l}{spheric contribution}\\
\end{tabular}
\end{center}\end{table*}

\section{Results}\label{sect:results}

\subsection{Period statistics and comparison with the literature}\label{subsect:results_stat_and_lit}

We could determine reliable periods for $75$ stars 
(flag `Y' in Table~\ref{tab:rot}), and periods with lower 
confidence are found for $22$ stars (flag `?'). 
Twelve stars of our sample have a previously reported period based on the same K2
data in \cite{Armstrong15.0}. In all but two cases those periods agree within $1-2$\,\%
with our values. The exceptions are EPIC-202059204 for which the lightcurves used by us
(and produced by A.Vanderburg) 
show no evidence for the $5.04$\,d period provided by \cite{Armstrong15.0}, and 
EPIC-201237257 for which our adopted period is twice the value of $16.2$\,d 
presented by \cite{Armstrong15.0}
based on the maximum peak in both our ACF and LS periodogram. 
Periods for a small number of K2 Superblink stars have been presented previously also 
in the following studies: Survey in the southern hemisphere using the 
All-Sky Automated Survey \cite[ASAS;][$6$\,stars]{Kiraga12.0},
HATnet survey in the Pleiades \citep[][$2$\,stars]{Hartman10.0},
SuperWASP survey in the Hyades and Pleiades \citep[][$1$\,star]{Delorme11.0}, 
and from the compilation of \cite[][$1$\,star]{Pizzolato03.1}. 
They are all in excellent agreement with our values derived from the K2 
lightcurves.
For the two stars we have in common with the HATNet survey of field stars presented by 
\cite{Hartman11.0}, however, we find strongly
discrepant values for the periods: 
$16.1$\,d vs $39.0$\,d in \cite{Hartman11.0} for EPIC-211107998 and
$12.9$\,d vs $0.86$\,d in \cite{Hartman11.0} for EPIC-211111803. We see no evidence in the K2
data for the period values determined by \cite{Hartman11.0}.   

\begin{figure}
\begin{center}
\includegraphics[width=8.5cm]{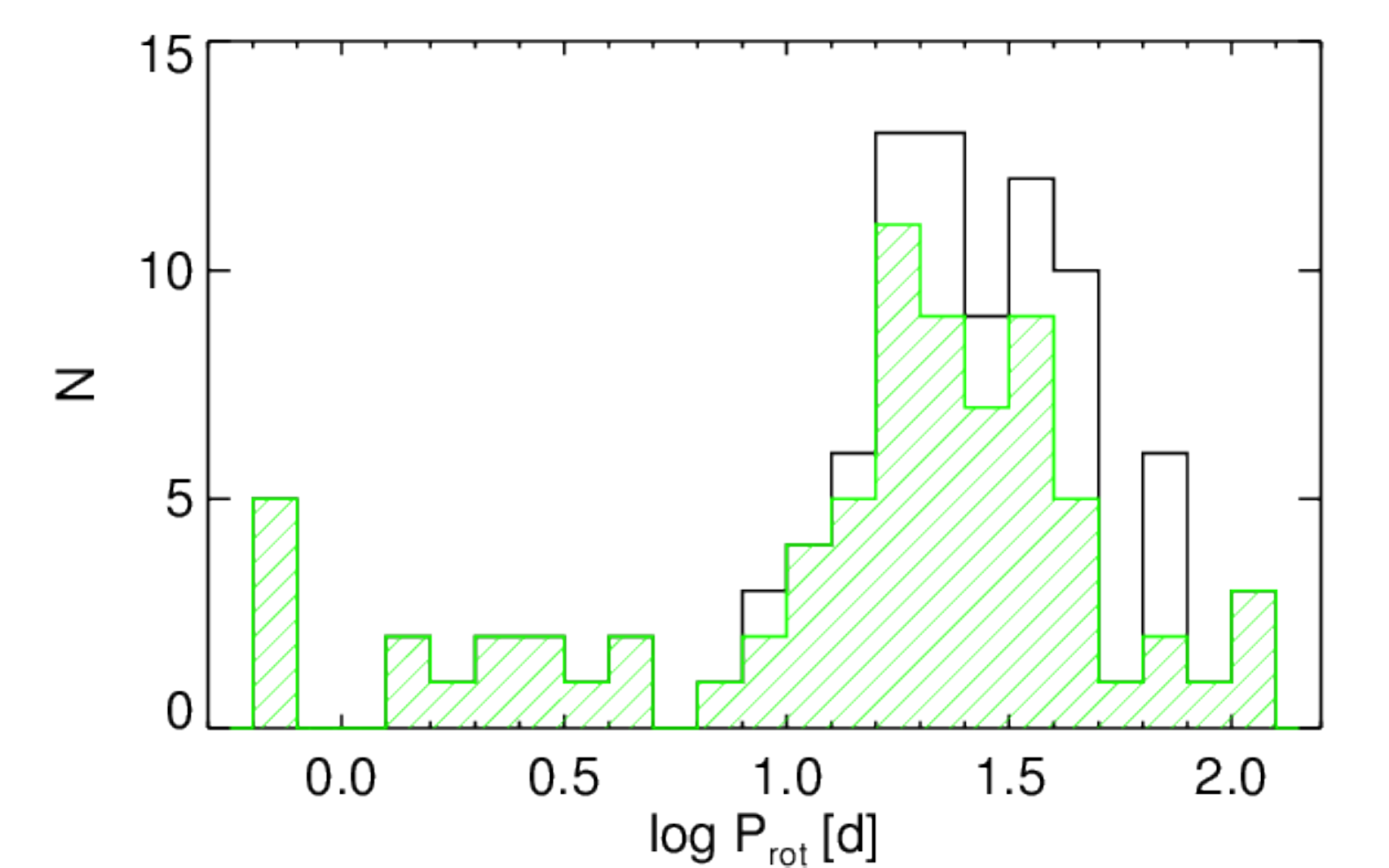}
\caption{Distribution of the $97$ rotation periods determined for the K2 Superblink 
M star sample. 
The black histogram represents the full sample of periods, and the overlaid
green histogram the subsample of reliable periods (flag `Y').} 
\label{fig:periods_histo}
\end{center}
\end{figure}

All in all, a $73$\,\% of the K2 Superblink sample shows  
periodic variability on timescales up to $\sim 100$\,d. 
Our period distribution is shown in Fig.~\ref{fig:periods_histo}. 
Studies of rotation of M stars in the main Kepler mission have come up with 
$63$\,\% \citep{McQuillan13.0} 
and $81$\,\% \citep{McQuillan14.0} of stars with detected periods. 
These differences may reflect the different data sets (each K2 campaign 
provides a lightcurve corresponding to the length of about one quarter of Kepler data) 
and detection methods (we use sine-fitting in addition to ACF and periodograms). 
In particular,  
we establish here in a relatively unbiased sample of M dwarfs periods of $\sim 100$\,d
and longer, in agreement with results from ground-based studies 
\citep{Irwin11.0, Newton16.0}. 
The period distribution of the Kepler sample from \cite{McQuillan13.0} shows
a cut at $\sim 65$\,d and \cite{McQuillan14.0} explicitly limit their sample to periods
$< 70$\,d. Note, that \cite{McQuillan13.0} have performed the period
search on individual Kepler Quarters which are of similar duration as the K2 campaigns. 
In fact, we are able to detect such long periods only thanks to the least-squares 
sine-fitting.  
We find that $\sim 10$\,\% of the periods are longer than $70$\,d. These 
would not have been detected by the methods of \cite{McQuillan13.0, McQuillan14.0}. 
An additional possible explanation for the absence of long-period variables in 
\cite{McQuillan13.0} -- related to photometric sensitivity -- is presented 
in Sect.~\ref{subsect:results_amplitude}.  

\subsection{Rotation period and stellar mass}\label{subsect:results_prot_mass}

\begin{figure}
\begin{center}
\includegraphics[width=8.5cm]{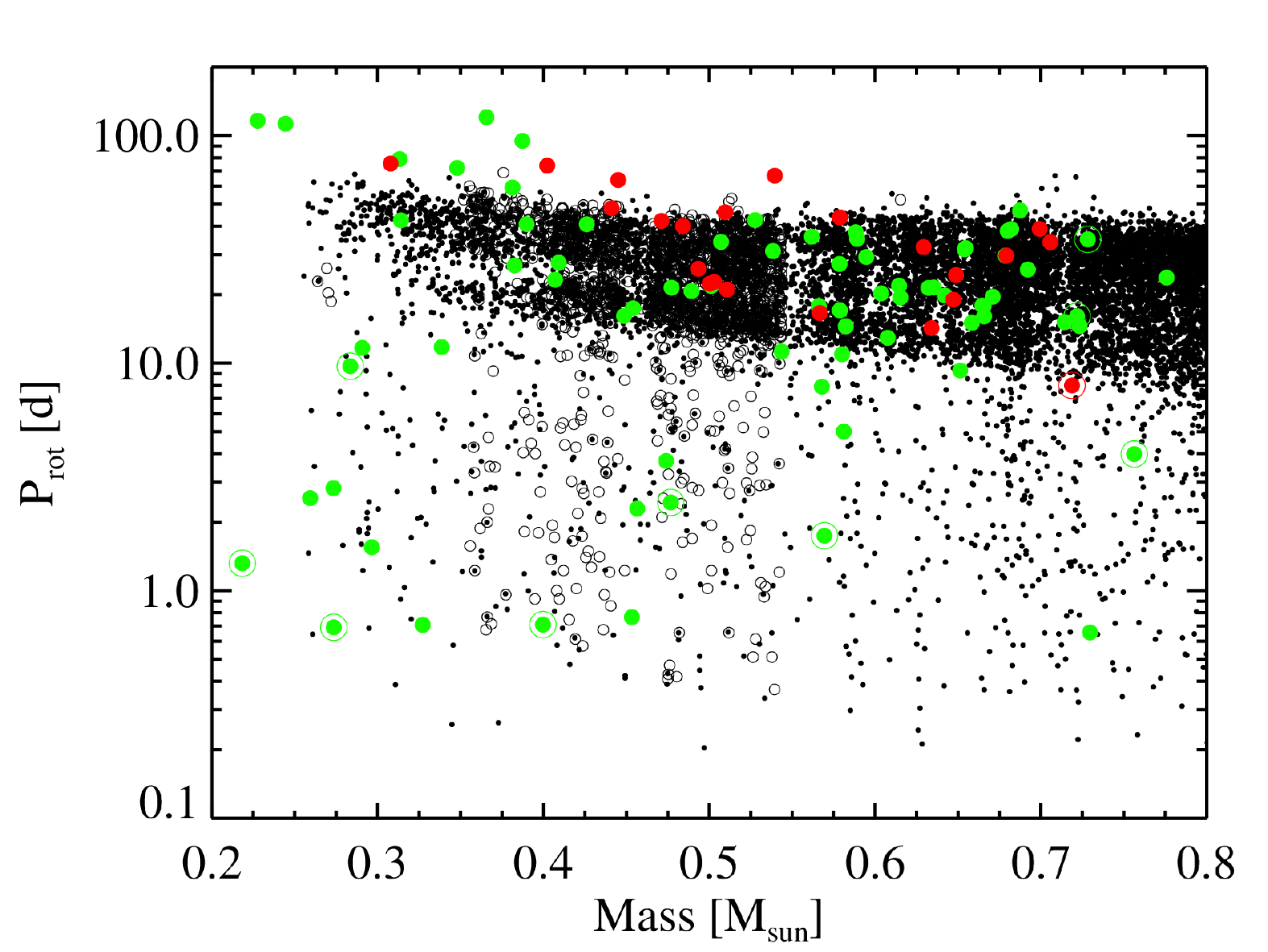}
\caption{Period versus mass for the K2 Superblink M star sample (green and red symbols for
periods flagged `Y' and `?', respectively). 
Binaries are marked with annuli (see Appendix~\ref{sect:appendix_bin}).
Data from Kepler studies are plotted as
open circles \citep{McQuillan13.0} and black dots \citep{McQuillan14.0}.} 
\label{fig:prot_mass}
\end{center}
\end{figure}

We present the newly derived rotation periods for the K2 Superblink M star sample in 
Fig.~\ref{fig:prot_mass} as a function of stellar mass together with results
for studies from the main Kepler mission. The sample of \cite{McQuillan13.0} (black 
open circles) covered stars in the mass range of $0.3 ... 0.55\,M_\odot$ 
selected based on the $T_{\rm eff}$ and $log{g}$ values from the Kepler input
catalog \citep{Brown11.0}. Subsequently, \cite{McQuillan14.0} (black dots) 
extended this study with similar selection criteria to all stars with $T_{\rm eff} < 6500$\,K. 
Among the most notable findings of these Kepler studies was a bimodal period distribution
for the lowest masses, and an increasing upper envelope of the period distribution 
for decreasing mass. While we have too few objects to identify the bimodality, we confirm
the upwards trend in the longest periods detected towards stars with lower mass. 
We are able to measure longer periods than \cite{McQuillan13.0} and \cite{McQuillan14.0}
because we add sine-fitting to the ACF and periodgram period search methods;
see Sect.~\ref{subsect:results_amplitude} for a more detailed comparison of the period
detection techniques and their implications. The fact that we measure periods in
excess of $\sim 100$\,d only in stars with very low mass ($M \leq 0.45\,M_\odot$) 
is interesting. If it is a real feature in the rotational distribution, 
it suggests a change of the spin-down efficiency 
at the low-mass end of the stellar sequence. Note, however, that the stellar masses
at which the upturn is seen to set in does not correspond to the fully convective transition
($\sim 0.35\,M_\odot$) where one might expect some kind of ``mode change" in the dynamo.  
Also, we can not exclude that there are detection biases,
e.g. the size and distribution of star spots and their lifetimes could be mass-dependent 
such that smaller and more quickly changing amplitudes are induced in higher-mass stars 
which would prevent us from detecting very long periods in them. 
A more detailed investigation of these features must be deferred to studies on a larger 
sample.

\subsection{Activity diagnostics from K2 rotation cycles}\label{subsect:results_amplitude}

We examine now various other diagnostics for rotation and activity derived from the K2 data. 
 These are listed together with the rotation periods in Table~\ref{tab:rot}. 

The Rossby number (in col.5) is defined as 
$R_{\rm 0} = P_{\rm rot}/\tau_{\rm conv}$, where $\tau_{\rm conv}$ is the convective
turnover time obtained from $T_{\rm eff}$ using Eq.~36 of \cite{Cranmer11.0} and
its extrapolation to $T_{\rm eff} < 3300$\,K. There is no consensus on the appropriate 
convective turnover times for M dwarfs beyond the fully convective boundary. 
As pointed out by \cite{Cranmer11.0}, the extrapolated values for late-M dwarfs
($\tau_{\rm conv} \sim 60...70$\,d) are in reasonable agreement with semi-empirical values
derived by \cite{Reiners09.3} but significantly lower than the predictions of
\cite{Barnes10.0}. 
The Rossby number is a crucial indicator of dynamo efficiency and is used 
in Sect.~\ref{subsect:results_rotact} for the description of the rotation - activity
relation. 
The parameters $R_{\rm per}$ (col.6) and $S_{\rm ph}$ (col.7) 
are measures for the variability in the
K2 lightcurve and are examined in this section.

\begin{figure}
\begin{center}
\includegraphics[width=8cm]{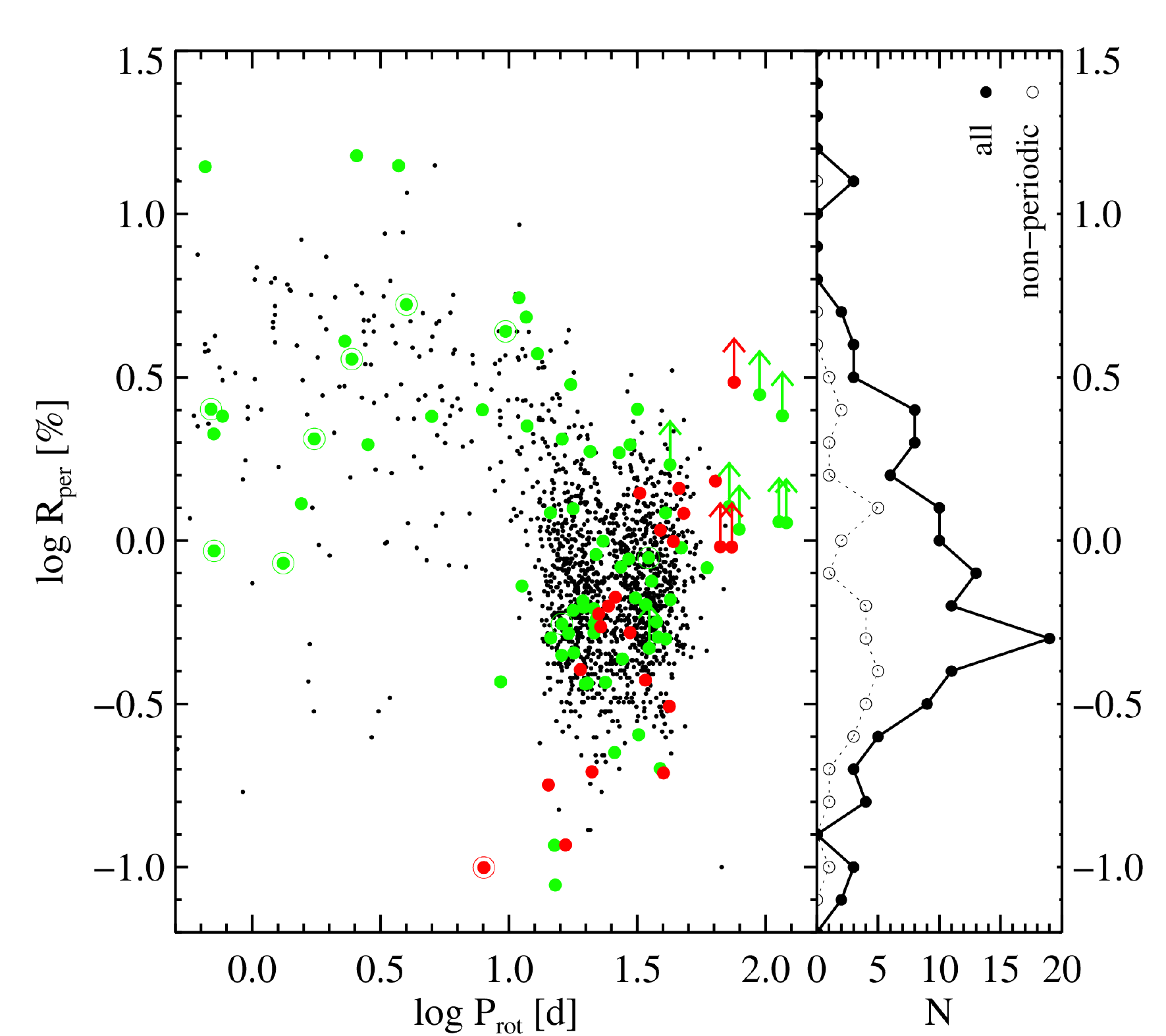}
\caption{Period versus amplitude for the K2 Superblink M dwarf sample (large, colored circles) 
compared to the Kepler field M dwarfs from \protect\cite{McQuillan13.0} (small, black dots).
For our K2 sample we distinguish reliable periods (green) and questionable periods (red). 
Banner on the right: histogram of $R_{\rm per}$ for the full
K2 Superblink star sample (solid line) and for the subsample classified as 
non-periodic (flag `N'). As in the other figures, binaries in our K2 Superblink sample 
are marked with large annuli.} 
\label{fig:ampl_vs_period}
\end{center}
\end{figure}
The amplitude of photometric variability associated with star spots is 
determined by the temperature contrast 
between spotted and unspotted photosphere and by the spot coverage, and may therefore, 
to first approximation, be considered a measure for magnetic activity. 
Various photometric activity indices characterizing the amplitude of Kepler lightcurves
have been used in the literature. 
\cite{Basri13.0} have introduced the range of variability
between the 5th and 95th percentile of the observed flux values, $R_{\rm var}$. 
This definition is meant to remove the influence of flares which are 
occasional events involving only a small fraction of a given rotational cycle. 
To further reduce the influence of outliers, we follow 
 the modified definition of \cite{McQuillan13.0}: $R_{\rm per}$ is 
the mean of the $R_{\rm var}$ values measured individually on all observed rotation 
cycles, expressed in percent. 

Since in the course of our flare analysis we produce
lightcurves where flares and other outliers have been eliminated we could use 
those ``cleaned" lightcurves for the analysis of the rotational variability. 
This way we could avoid cutting the top and bottom $5$\,\% of the data points. 
We compute the difference between the full amplitudes measured on the
``cleaned'' lightcurves and the $R_{\rm per}$ values measured on the original 
lightcurves and find them to differ by $\sim 0.05 \pm 0.05$\,dex in 
logarithmic space. This is negligible to the observed range of amplitudes. 
In order to enable a direct comparison to results from the literature, 
we prefer, therefore, 
to stick to the $R_{\rm per}$ values derived from the original lightcurves. 
In Fig.~\ref{fig:ampl_vs_period} we show the relation between $P_{\rm rot}$ and
$R_{\rm per}$ for the K2 Superblink M stars compared to the 
much larger Kepler M dwarf sample of \cite{McQuillan13.0}. 
The distribution of the two samples is in good agreement. In particular, 
there is a clear trend for stars with shorter
periods to have larger spot amplitudes. We examine this finding in more detail below. 

Fig.~\ref{fig:ampl_vs_period} also illustrates the difference between our results and
those of \cite{McQuillan13.0} for the longest periods. 
[Note that all K2 Superblink stars with periods inferred from sine-fitting 
have only a lower limit to the variability amplitude $R_{\rm per}$.] 
One reason for the absence of long-period stars in \cite{McQuillan13.0} 
could be the larger distance of the Kepler stars which results in lower
sensitivity for small amplitudes, suggesting that Kepler can find periods only in the more 
active stars likely to be rotating faster. 
However, remarkably, the long-period stars in the K2 Superblink M star 
sample seem to have 
larger spot amplitudes than stars with lower periods (from $\sim 15 ...50$\,d).
We recall again that we 
are able to detect such long periods only on stars with clear sine-like variation 
indicating the presence of a single dominating spot. Therefore, we can 
only speculate that stars with periods $\gtrsim 100$\,d and low spot amplitude may exist but
their more diffuse spot patterns or changes on time-scales shorter than the rotation period
yield a complex lightcurve. If so, one can expect these stars among the ones classified
as non-periodic (flag `N') by us. The bar on the right of Fig.~\ref{fig:ampl_vs_period} 
shows the distribution of $R_{\rm per}$ for all K2 Superblink M stars 
and for the subsample to which we could not assign a period. There is no clear preference of these
latter ones towards small amplitudes, and the above consideration does not allow us to
conclude on their periods. 
Constraining the range of spot amplitude of the slowest rotators should be a
prime goal of future studies on larger samples. As described in 
Sect.~\ref{subsect:results_prot_mass}
we find the longest periods exclusively in very low-mass stars. Therefore, the 
change in the distribution of the $R_{\rm per}$ values for the slowest rotators -- if 
truly existing -- might be a mass-dependent effect rather than related to rotation.

\begin{figure}
\begin{center}
\includegraphics[width=8.5cm]{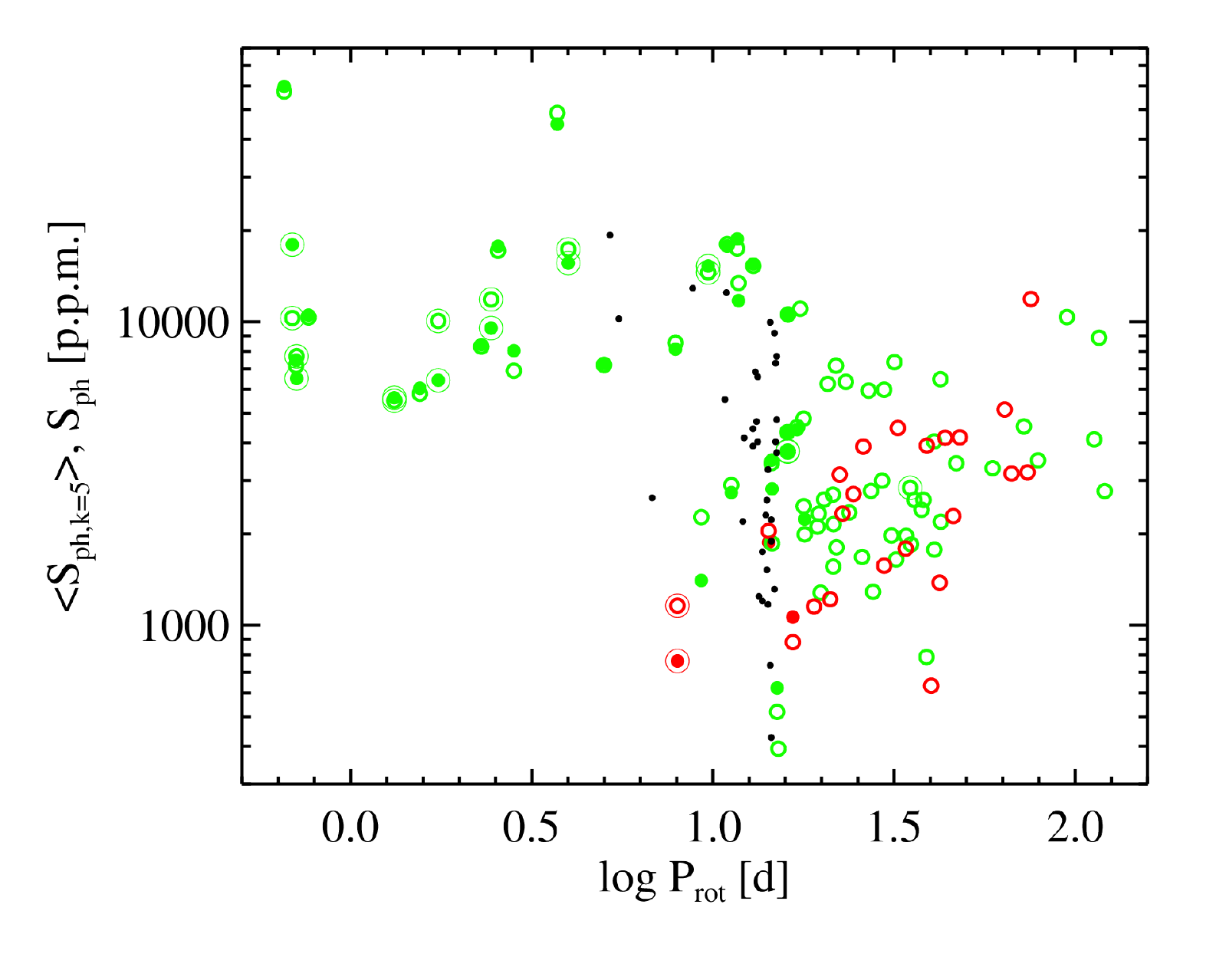}
\caption{Magnetic activity indices defined by \protect\cite{Mathur14.1} vs 
period for the K2 Superblink star sample, compared to the subsample of $34$ Kepler 
field M dwarfs studied by \protect\cite{Mathur14.1} (black dots). Green and red symbols
represent periods flagged `Y' and `?', respectively. Filled circles denote 
$\langle S_{\rm ph,k=5} \rangle$ values and open circles the $S_{\rm ph}$ values.
See text in Sect.~\ref{subsect:results_amplitude} for details
on the definition of these indices.
Large annuli mark binaries.}
\label{fig:sph_vs_period}
\end{center}
\end{figure}

\cite{Mathur14.1} defined the standard deviation
of the full lightcurve, $S_{\rm ph}$, and  $\langle S_{\rm ph,k} \rangle$,
the mean of the standard deviations computed for time intervals $k \cdot P_{\rm rot}$. 
They found that for increasing $k$ the index $\langle S_{\rm ph,k} \rangle$ 
approaches $S_{\rm ph}$. This way they were able to show that roughly after five rotation cycles 
($k = 5$) the full range of flux variation is reached, and they recommend 
$\langle S_{\rm ph,k=5} \rangle$ as measure of the global evolution of the 
variability. We compute 
$S_{\rm ph}$ and  $\langle S_{\rm ph,k=5} \rangle$ for the K2 Superblink M stars and show 
the results in Fig.~\ref{fig:sph_vs_period} versus the rotation periods;
filled circles represent $\langle S_{\rm ph,k=5} \rangle$  and open circles mark $S_{\rm ph}$. 
The sample studied by \cite{Mathur14.1} is also displayed (black dots for their
$\langle S_{\rm ph,k=5} \rangle$ values). That sample   
consists of $34$ Kepler M stars with $15$ Quarters of continuous observations and 
$P_{\rm rot} < 15$\,d from the Kepler study of \cite{McQuillan13.0}.
Our sample improves the period coverage especially for $P \lesssim 12$\,d. 
The fact that there are no very fast rotators
in the sample studied by \cite{Mathur14.1} is probably a bias related to their sample selection. 
We find that 
stars with short periods have systematically larger $\langle S_{\rm ph,k=5} \rangle$ 
index than stars with $P \gtrsim 10$\,d. 
The upper boundary of $15$\,d for the periods in the
\cite{Mathur14.1} sample is imposed by their requirement of covering at least $5$ cycles.
However, as explained above 
there is no dramatic difference between $\langle S_{\rm ph,k=5} \rangle$ and
$S_{\rm ph}$ for a given star. We verify this on the K2 Superblink M star sample
by showing as open circles their values $S_{\rm ph}$. 
The advantage of $S_{\rm ph}$ is that we can include in Fig.~\ref{fig:sph_vs_period}
the stars with $P > 1/5 \cdot \Delta t$. 
We can see that the pattern over the whole period range is very similar
to that of Fig.~\ref{fig:ampl_vs_period}, i.e. both spot amplitude and standard deviation
of the lightcurve show a dependence on rotation rate which seems to divide the stars
in two groups above and below $P \sim 10...12$\,d.   

\subsection{Activity diagnostics related to flares in K2 lightcurves}\label{subsect:results_flares}

\begin{figure}
\begin{center}
\includegraphics[width=8.5cm]{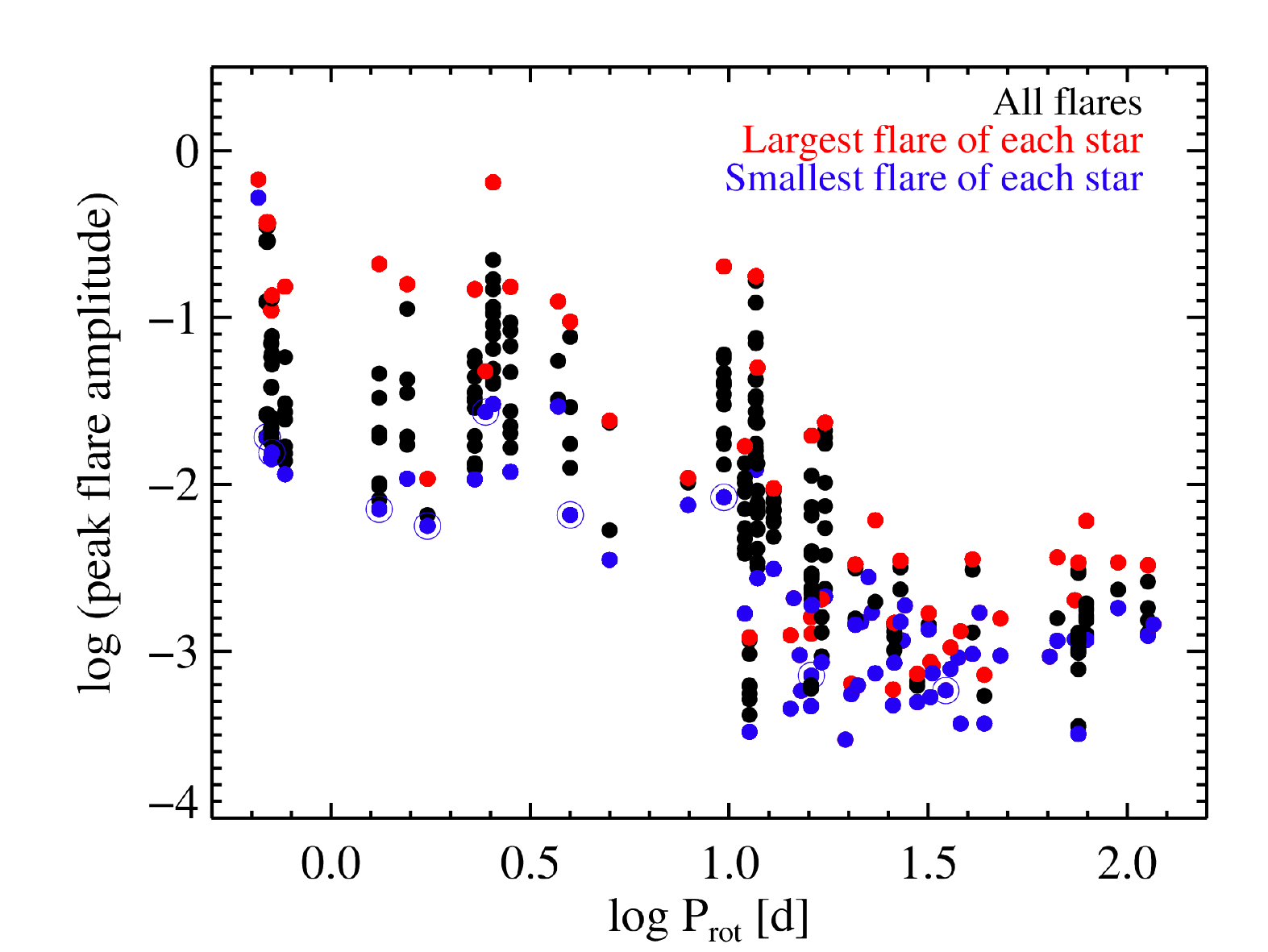}
\includegraphics[width=8.5cm]{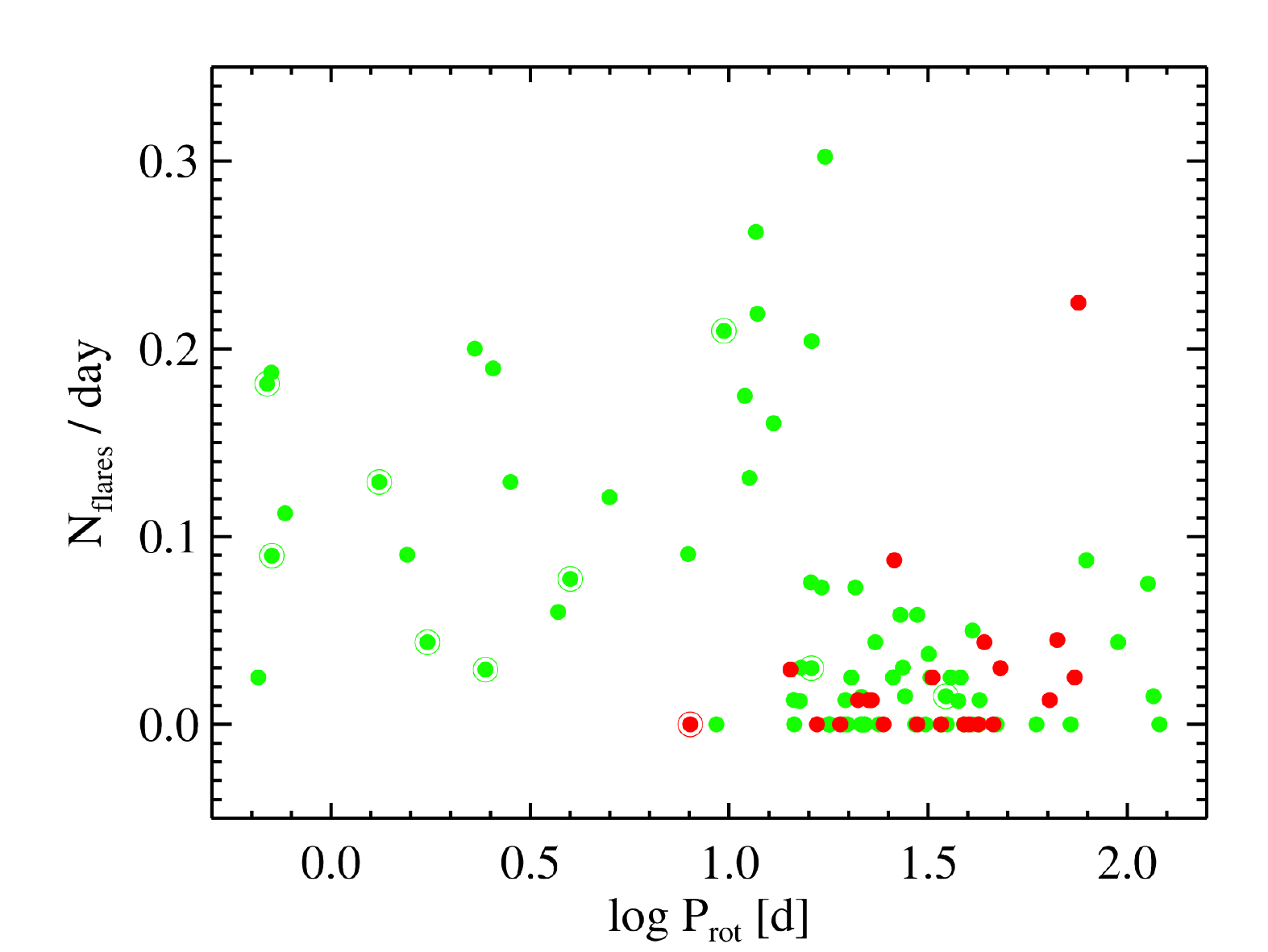}
\caption{{\em top} - Flare amplitude vs rotation period: All flares are shown and the
range of flare amplitudes for a given star is made evident by marking the largest and
smallest flare on each star with different colors, red and blue respectively. {\em bottom} -
Flare frequency vs rotation period: Each star is represented once. Binaries are
highlighted in both panels with annuli.} 
\label{fig:flares_vs_period}
\end{center}
\end{figure}

Our separate analysis of flares and rotation in the K2 lightcurves enables us to relate
flare activity to star spot activity. 
Fig.~\ref{fig:flares_vs_period} shows the peak amplitudes of all flares defined with respect
to the flattened K2 lightcurve (top panel) and the flare frequency of all stars (bottom panel) 
as function of the rotation period.
A clear transition takes place near $P_{\rm rot} \sim 10$\,d,
analogous to the case of the spot activity measures discussed in 
Sect.~\ref{subsect:results_amplitude}. While the absence of
small flares in fast rotators is determined by the noise level in the flattened 
lightcurve (see Fig.~\ref{fig:kp_stddev}), there is no bias against the detection of
large flares in slowly rotating stars. Note that our algorithm has lower flare detection 
sensitivity for events on fast-rotating stars because the presence of flares itself impacts 
on the quality of the smoothing process used to identify the flares. 
Therefore, especially for the stars with short periods, the 
number of flares observed per day ($N_{\rm flares}/{\rm day}$) 
may represent a lower limit to the actual flare frequency.

Considering the limitations of the K2 long-cadence data for flare statistics 
(see discussion in Sect.~\ref{subsect:k2_analysis_flares}) we do not put much
weight on the absolute numbers we derive for the flare rates. 
However, our results are in very good agreement with a dedicated M dwarf flare 
study based on short-cadence ($1$\,min) Kepler lightcurves. In particular, for the fast 
rotators the range we show 
in Fig.~\ref{fig:flares_vs_period} for the peak flare amplitudes ($\sim 0.01...0.5$) and for 
the number of flares per day ($\sim 0.05...0.2$), are similar to the numbers obtained
by \cite{Hawley14.0} for the active M star GJ\,1243 if only flares with duration of more
than one hour are considered from that work.  

\subsection{Residual activity in K2 lightcurves}\label{subsect:results_noise}

\begin{figure}
\begin{center}
\includegraphics[width=8.5cm]{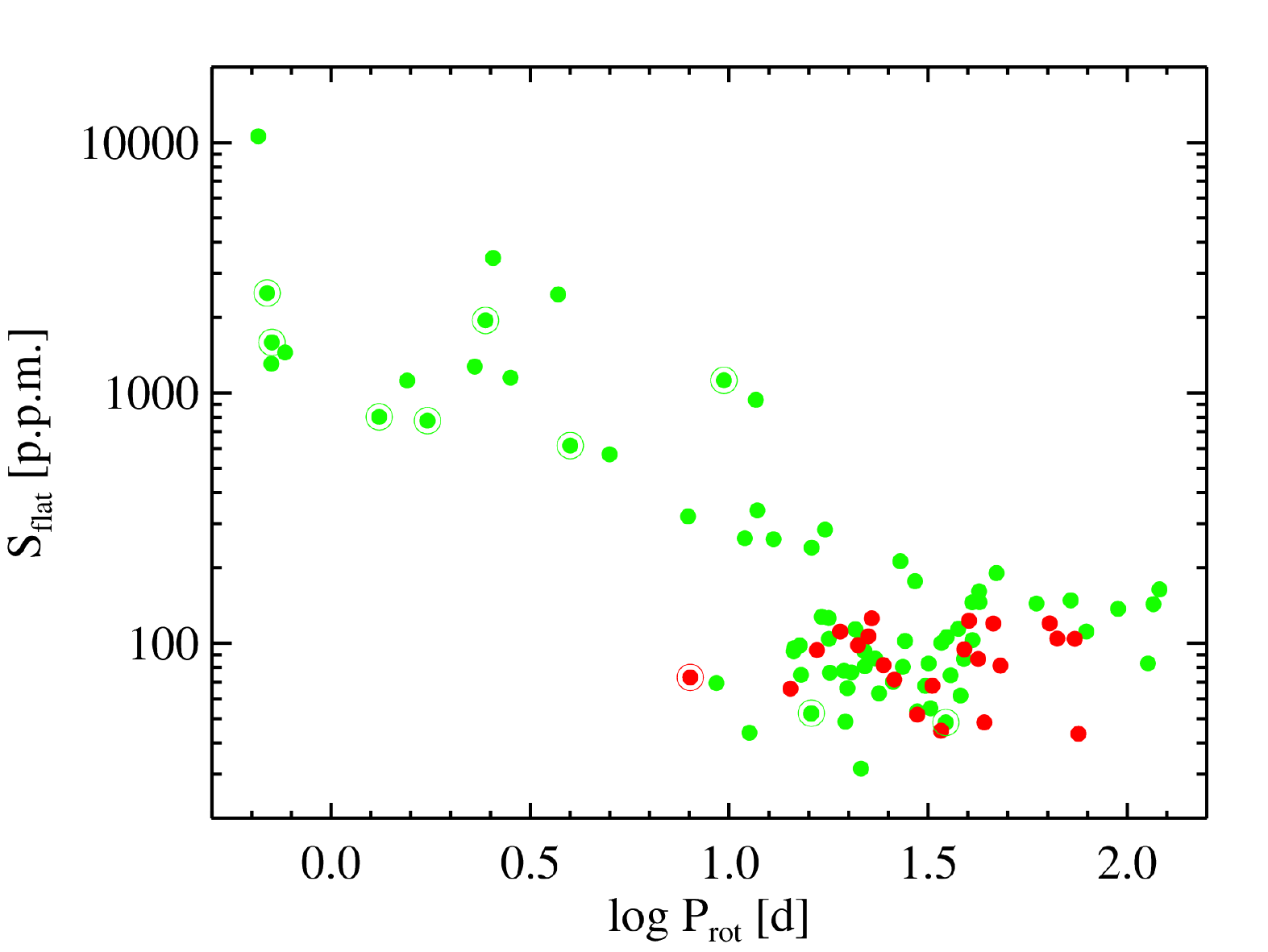}
\caption{Standard deviation of the flattened lightcurve excluding the outliers, 
as in Fig.~\ref{fig:kp_stddev}, shown here vs rotation period.
Green symbols (period flag `Y'), red symbols (period flag `?'). 
The clear transition between fast and slowly rotating stars indicates that for
fast rotators the origin
of this `noise' has an astrophysical component. Binaries are highlighted with annuli.}
\label{fig:period_stddev}
\end{center}
\end{figure}

Above we have shown that both the spot cycle amplitude and the flares display
a distinct behavior with rotation period. 
Here we examine the standard deviation of the ``flattened" lightcurves, $S_{\rm flat}$. 
As described in Sect.~\ref{subsect:k2_analysis_noise}, when measured without
considering the outliers, this parameter represents 
a measure for the noise after removal of the rotation cycle and of the flares.  
We notice a marked trend of $S_{\rm flat}$ with the rotation period 
(Fig.~\ref{fig:period_stddev}). 
A dependence of the noise level on the brightness of the star is expected and demonstrated
in Fig.~\ref{fig:kp_stddev}, where the lower envelope of the distribution increases
towards fainter $K_{\rm p}$ magnitude. However, the difference between the $S_{\rm flat}$ 
values seen for
slow and fast rotators in Fig.~\ref{fig:period_stddev} is clearly unrelated to this effect 
as there is no clustering of stars with large $S_{\rm flat}$ (and fast rotation) at
bright magnitudes in Fig.~\ref{fig:kp_stddev}. 
The evidently
bimodal distribution with rapid rotators showing larger values of $S_{\rm flat}$, therefore
suggests that there is a contribution to the `noise' in the K2 photometry that
is astrophysical in origin.

The similarity of the period dependence seen in $S_{\rm flat}$ 
(Fig.~\ref{fig:period_stddev}), the spot cycle (Fig.~\ref{fig:ampl_vs_period} 
and ~\ref{fig:sph_vs_period}) and the flares (Fig.~\ref{fig:flares_vs_period}) 
may indicate that the `noise' in the fastest rotators could be caused by 
unresolved spot or flare activity. 
Many small flares, so-called nano-flares, as well as 
many small and/or rapidly evolving spots can produce a seemingly stochastic signal. 
This astrophysical noise sources seem to be limited to fast rotators, while for
slow rotators the spot contrast drops below a constant minimum level of the variability   
which might be identified as the photometric precision 
(see Sect.~\ref{subsect:k2_analysis_noise}). 

\subsection{Photometric activity and binarity}\label{subsect:results_binaries}

In the relations between various activity indicators and rotation period 
presented in the previous sections, binary stars that have a possible contribution to the
rotational signal from the unresolved companion star are highlighted. Strikingly, the
binaries are mostly associated with rotation periods below the transition between
fast and slow regimes that we have identified. In Fig.~\ref{fig:period_stddev} this
could be taken as evidence that the presence of a companion increases the noise in the
K2 lightcurve. On the other hand, we have argued above that the coincidence of the
bimodality in $S_{\rm flat}$, spot and flare signatures with $P_{\rm rot}$ 
points at a fundamental transition taking place in these stars. We may speculate that
binarity is responsible for the observed dichotomy, e.g. by spinning up the star
through tidal interaction or by reducing angular momentum loss. 
The binary fraction ($BF$) for the fast rotators ($P_{\rm rot} < 10$\,d) is $8 / 19$,
i.e. $8$ stars out of $19$ are known binaries. For slow rotators ($P_{\rm rot} > 10$\,d)
the binary fraction is $2 / 78$. 
We calculate the $95$\,\% confidence levels for a binomial distribution and find 
the two samples to be significantly different: 
$BF_{\rm fast} = 0.42^{+0.67}_{-0.20}$ and $BF_{\rm slow} = 0.03^{+0.09}_{-0.00}$.
That said, we caution that no systematic and homogeneous search for multiplicity was done
for these stars and our literature compilation (Sect.~\ref{sect:appendix_bin}) 
may be incomplete.

\subsection{The X-ray and UV activity -- rotation relation}\label{subsect:results_rotact}

\begin{figure*}
\begin{center}
\parbox{18cm}{
\parbox{9cm}{
\includegraphics[width=8.5cm]{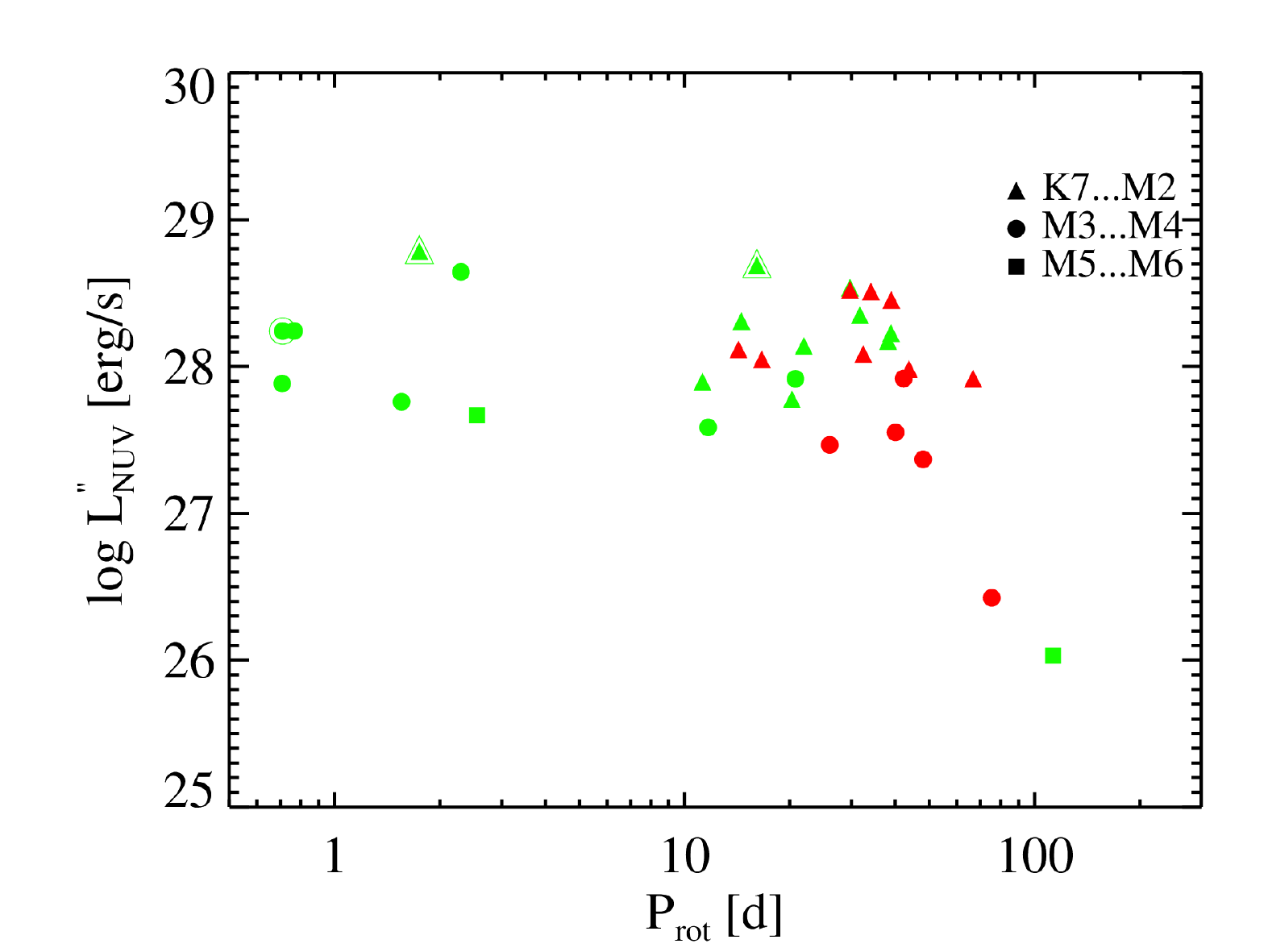}
}
\parbox{9cm}{
\includegraphics[width=8.5cm]{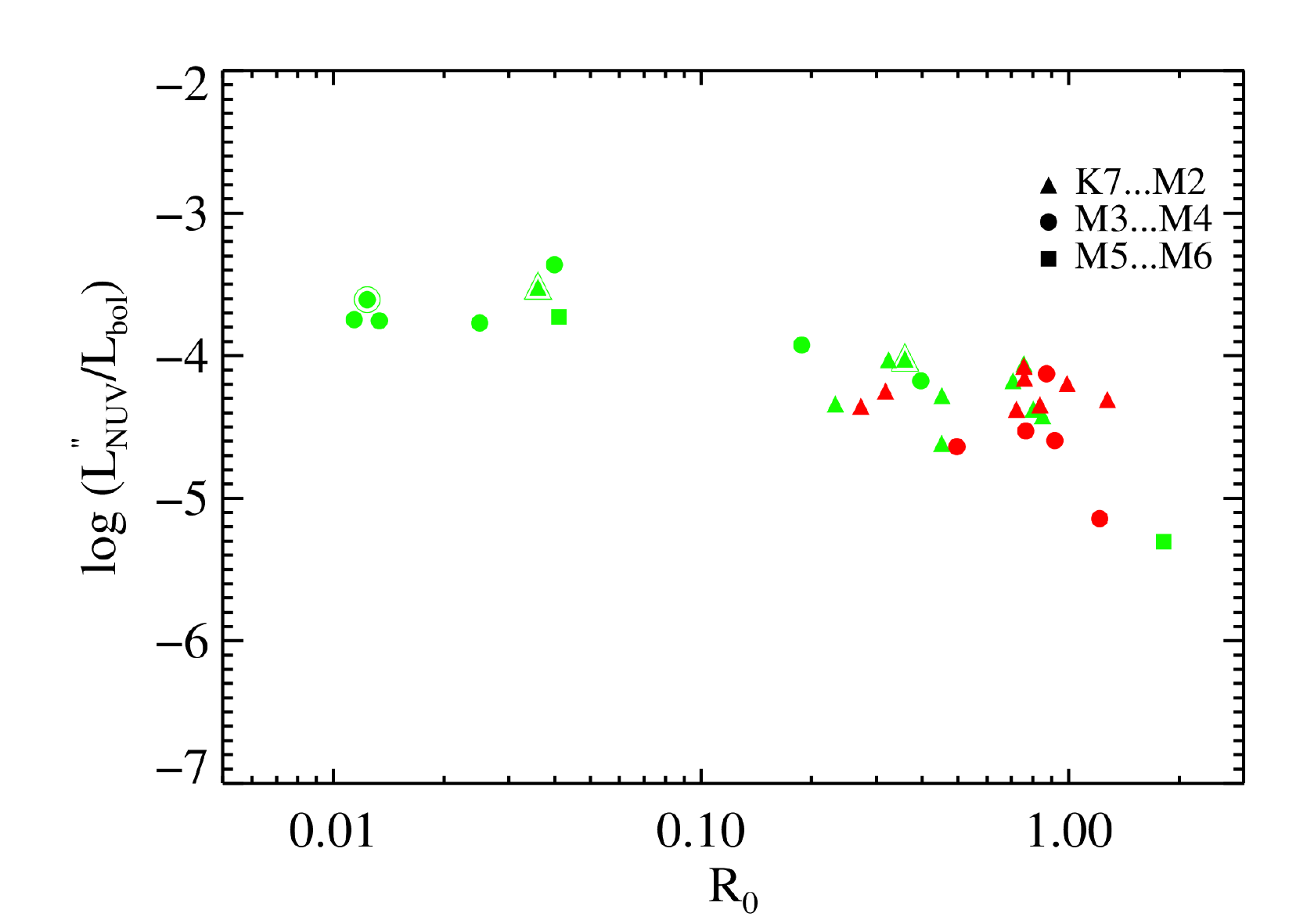}
}
}
\parbox{18cm}{
\parbox{9cm}{
\includegraphics[width=8.5cm]{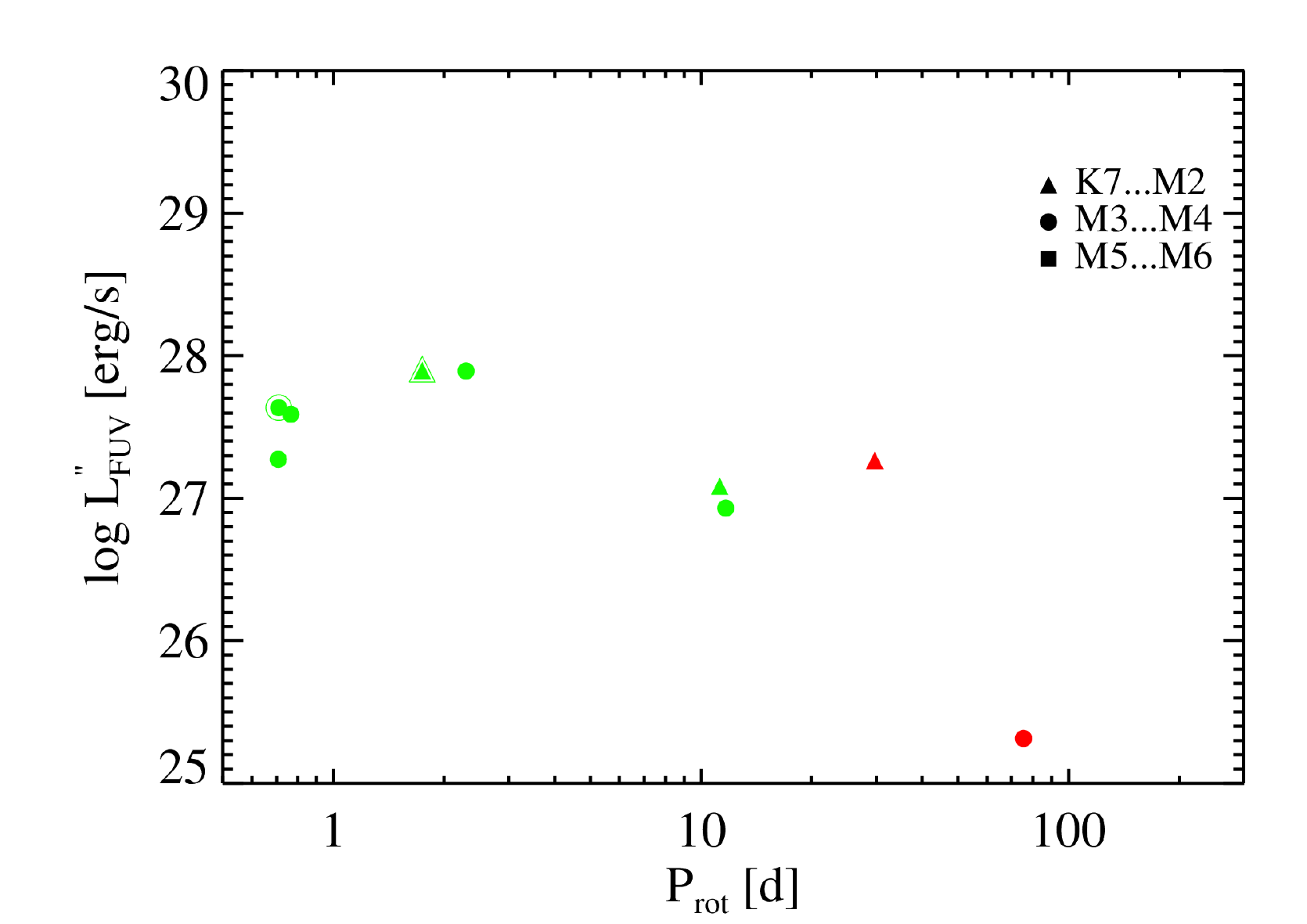}
}
\parbox{9cm}{
\includegraphics[width=8.5cm]{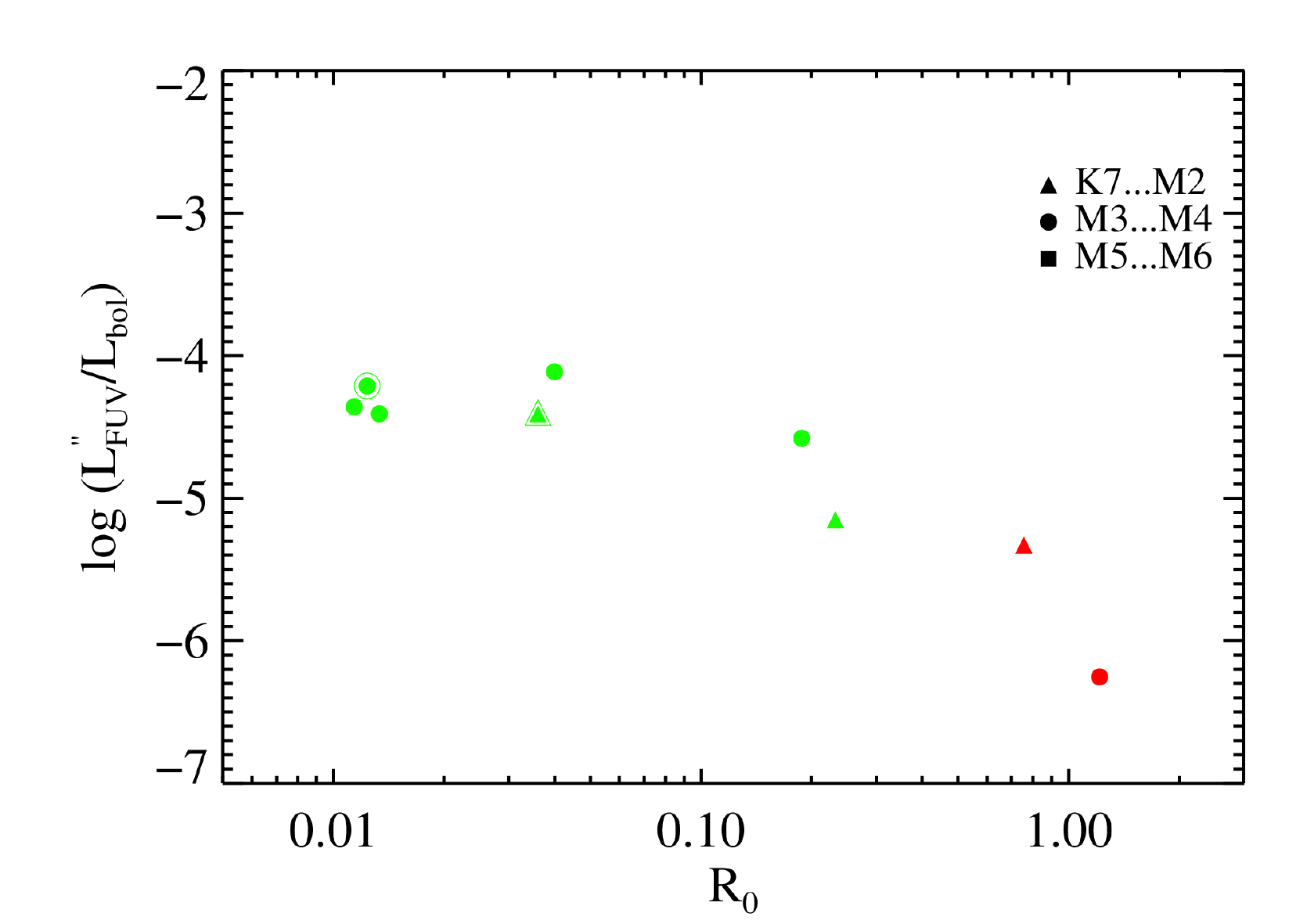}
}
}
\parbox{18cm}{
\parbox{9cm}{
\includegraphics[width=8.5cm]{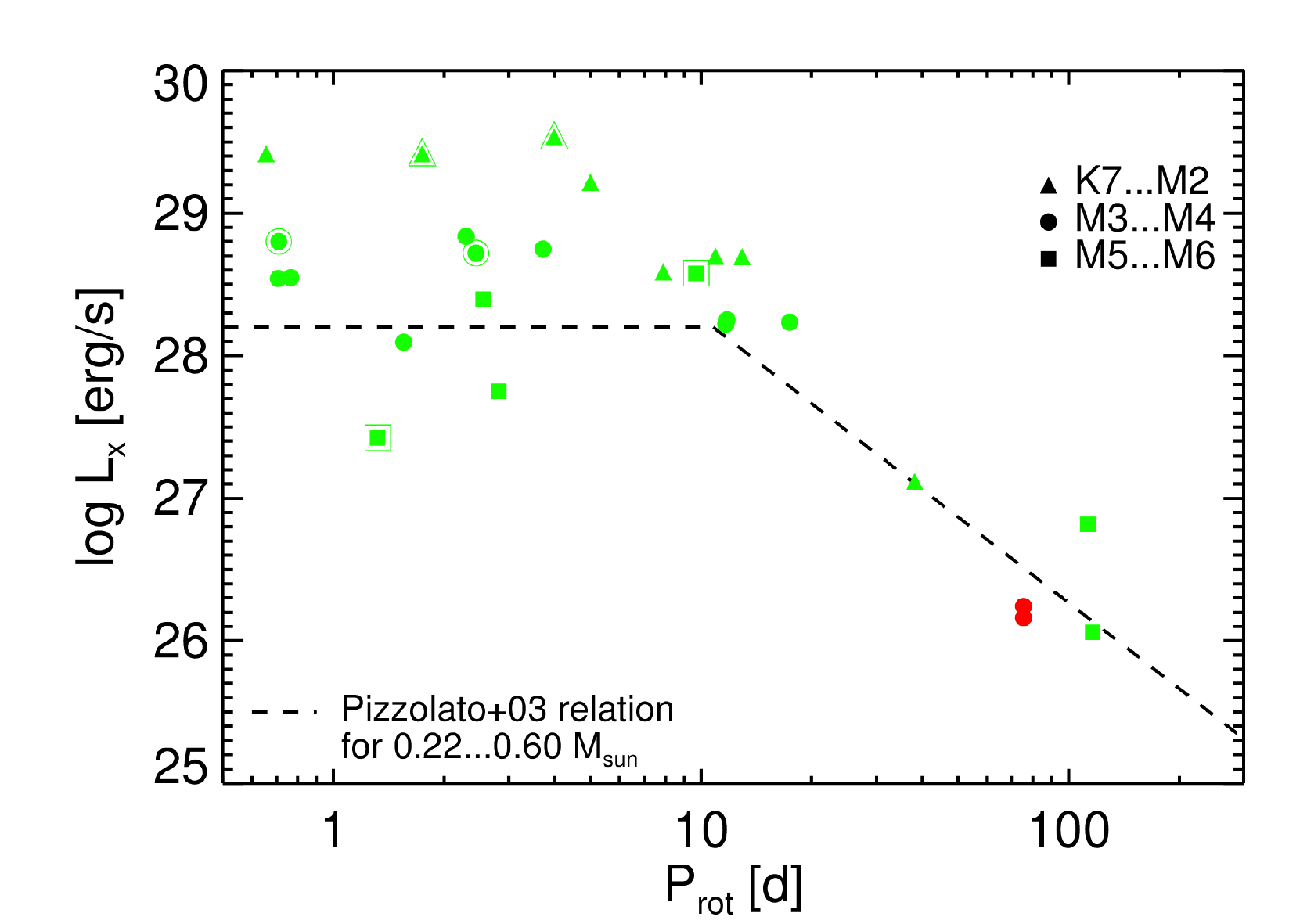}
}
\parbox{9cm}{
\includegraphics[width=8.5cm]{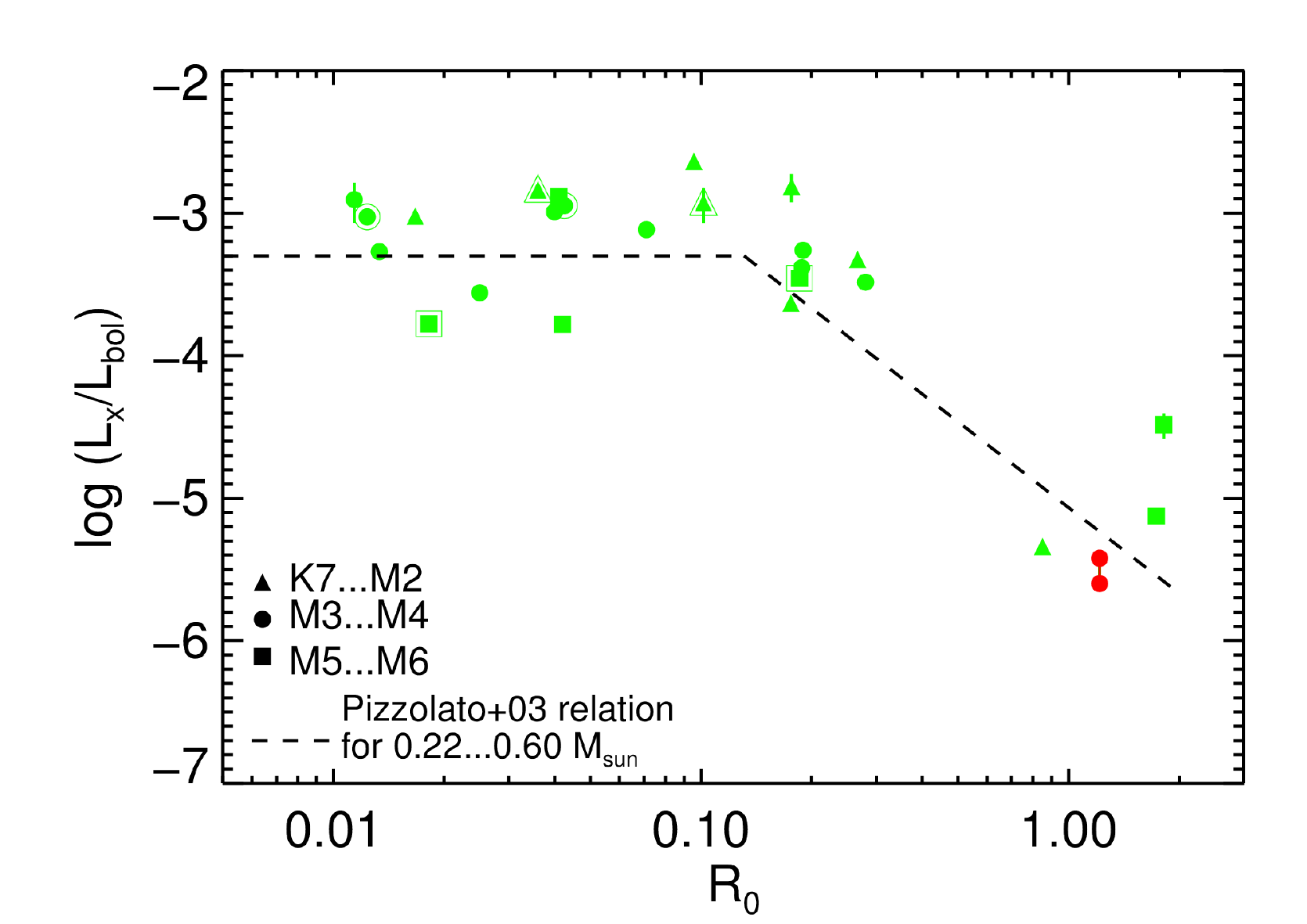}
}
}
\caption{K2 rotation periods combined with archival NUV (top), FUV (middle) and 
X-ray (bottom) data for the campaign C0...C4 Superblink M stars. In the left panels,
luminosities vs rotation period, in the right panels activity indices vs Rossby number. 
Periods flagged `?' are shown in red, unresolved binaries are represented
with large annuli.} 
\label{fig:rotact}
\end{center}
\end{figure*}

The activity -- rotation relation is traditionally expressed using X-rays, 
Ca\,{\sc ii}\,H\&K and
H$\alpha$ emission as activity indicators. Measurements of these diagnostics have historically
been easiest to achieve \citep{Pallavicini81.1, Noyes84.0}. 
Yet, as described in Sect.~\ref{sect:intro}, the dependence between magnetic activity and rotation 
 has remained poorly constrained for M stars. In Fig.~\ref{fig:rotact} we present 
an updated view using the X-ray data extracted from the archives 
and the newly derived rotation periods from K2. 
We also add here, to our knowledge for the first time for field M stars, UV emission
as diagnostic of chromospheric activity in conjunction with photometric 
rotation periods. 

All but two of the $26$ K2 Superblink stars 
with X-ray detection have reliable rotation period measurement (flag `Y').
The first exception is EPIC-206019392 for which we 
find through sine-fitting a period of $\sim 75$\,d. While there are no doubts on a periodic
spot-modulation, the slight deviations of its lightcurve
from a sinusoidal make the value for the period uncertain (therefore flagged `?' in 
Table~\ref{tab:rot}). 
For the other case, EPIC-210500368, we can not identify a dominating period, yet
the lightcurve shows a long-term trend superposed on a variability with a time-scale of
$\sim 10$\,d. 
Among the NUV detections we could establish the rotation period for $78$\,\% ($32 / 41$),
and $46$\,\% ($19 / 41$) of them have a `reliable' period. 
Nine of $11$ FUV detected stars have a period measurement, of which $7$ are flagged
`reliable'. 

The parameters which best describe
the connection between activity and rotation are still a matter of debate 
\citep{Reiners14.0}. We provide here plots
for luminosity versus rotation period (left panels of Fig.~\ref{fig:rotact}) 
and for activity index $L_{\rm i}/L_{\rm bol}$ with $i = NUV, FUV, X$ versus 
Rossby number (right panels).
First, it is clear that there is a decrease of the activity levels in all 
three diagnostics (NUV, FUV, X-rays) for the slowest rotators. While the sample of M stars 
with FUV detection and rotation period measurement is still very small, a division in
a saturated and a correlated regime, historically termed the ``linear" regime,  
can be seen in the relations involving NUV and X-ray emission.
The X-ray -- rotation relation is still poorly populated for slow rotators, and the 
turn-over point and the slope of the decaying part of the relation can not be well 
constrained with the current sample. Interestingly, for the NUV emission the situation
is reversed, in a sense that more stars with NUV detection are found among slow rotators. 
In terms of luminosity NUV saturation
seems to hold up to periods of $\sim 40$\,d, way beyond the critical period of $\sim 10$\,d
identified to represent a transition in the behavior of optical activity indicators
extracted from the K2 lightcurves (see Sect.~\ref{subsect:results_amplitude} 
and~\ref{subsect:results_flares}). On the other hand, the 
$L_{\rm NUV}^\prime/L_{\rm bol}$ values are slightly decreased with respect to the levels of 
the fastest rotators, and the active stars around a $\sim 30...40$\,d period are all 
late-K to early-M stars.
We also caution that a large fraction of the slowly rotating NUV detected stars have 
periods that we flagged as less reliable (red symbols in Fig.~\ref{fig:rotact}). 

In order to highlight eventual differences emerging at the fully convective 
transition, we divide the stars in Fig.~\ref{fig:rotact} into three spectral type 
groups represented by different plotting symbols. As far as the X-ray emission is 
concerned,
the two order of magnitude scatter in the saturated part of the $L_{\rm x}$ vs $P_{\rm rot}$
relation is clearly determined by the spectral type distribution, with cooler stars
having lower X-ray luminosities for given period. This is a consequence of the 
mass dependence of X-ray luminosity, and was already seen by \cite{Pizzolato03.1} 
for coarser bins of stellar mass representing a spectral type range from G to M. 
We have overplotted in the bottom
panels of Fig.~\ref{fig:rotact} the relation derived by \cite{Pizzolato03.1} for
their lowest mass bin, $M = 0.22 ... 0.60\,M_\odot$ (corresponding to 
spectral type earlier than M2). It must be noted that in \cite{Pizzolato03.1} the linear
regime was populated by only two stars of their sample and the saturated regime was
dominated by upper limits to $P_{\rm rot}$ which were estimated from $v \sin{i}$ 
measurements. Therefore, even our still limited K2 sample constitutes a significant
step forward in constraining the X-ray -- rotation relation of M dwarfs. 

We determine the saturation level for all X-ray
detections with $P_{\rm rot} < 10$\,d in the three spectral type bins 
K7...M2, M3...M4, and M5...M6 and for the whole sample with spectral types from K7
to M6. The results are summarized in Table~\ref{tab:xraysat}.
If we select the K2 Superblink M star subsample in the same mass range studied by 
\cite{Pizzolato03.1} ($M = 0.22 ... 0.60\,M_\odot$) we derive saturation levels
of $\log{L_{\rm x,sat}} \,{\rm [erg/s]} = 28.5 \pm 0.5$ and 
$\log{(L_{\rm x,sat}/L_{\rm bol})} = -3.3 \pm 0.4$, 
within the uncertainties compatible with their results. 
We confirm results of previous studies that the saturation level for a sample
with mixed spectral types converges to a much
narrower distribution if $\log{(L_{\rm x}/L_{\rm bol})}$ is used as activity diagnostic
(see Fig.~\ref{fig:rotact} and last line in Table~\ref{tab:xraysat}). 
There is marginal evidence for the very low mass stars (SpT M5...M6) to be 
underluminous with respect to this level. 
However, this assertion is not yet statistically sound according to the spread of the 
data (see standard deviations in Table~\ref{tab:xraysat}) and two-sample tests carried
out with ASURV \citep{Feigelson85.1} indicate that the $\log{(L_{\rm x}/L_{\rm bol})}$ values
of the three spectral type subgroups may be drawn from the same parent distributions
(p-values $> 10$\,\%). 
It has been widely acknowledged that the activity levels show a drop for 
late-M dwarfs \citep[e.g.][]{West08.1,Reiners12.1}, but an investigation of whether
and how this is related to $P_{\rm rot}$ has come into reach only now with the large
number of periods that can be obtained from planet transit search projects.
Using rotation periods from the MEarth program, \cite{West15.0} showed 
that the average $L_{\rm H\alpha}/L_{\rm bol}$ ratio for fast rotators
($P_{\rm rot} < 10...20$\,d) decreases by a factor two for late-M dwarfs (SpT M5...M8)
compared to early-M dwarfs (SpT M1-M4). Whether a distinct regime exists in which
H$\alpha$ activity correlates with $P_{\rm rot}$ could not be established in that study.
The X-ray and UV detections we present in this paper also do not adequately 
sample the regime of long periods. We refrain here from fitting that
part of the rotation-activity relation because our upcoming {\em Chandra} observations 
together with the larger sample of
periods that will be available for Superblink M stars at the end of the K2 mission
will put us in a much better position to address this issue. 
%
\begin{table}
\begin{center}
\caption{X-ray saturation level for M dwarfs determined for X-ray detected stars with 
$P_{\rm rot} < 10$\,d.}\label{tab:xraysat}
\begin{tabular}{lcrr}\hline
SpT & $N_{\rm *}$ & $\log{L_{\rm x,sat}}$ & $\log{(L_{\rm x,sat}/L_{\rm bol})}$ \\ 
    & & [erg/s] & \\ \hline
K7...M2 & $5$ & $29.2 \pm 0.4$ & $-3.0 \pm 0.4$ \\
M3...M4 & $7$ & $28.6 \pm 0.3$ & $-3.1 \pm 0.2$ \\
M5...M6 & $4$ & $27.9 \pm 0.5$ & $-3.5 \pm 0.4$ \\
\hline
K7...M6 & $16$ & $28.7 \pm 0.6$ & $-3.2 \pm 0.4$ \\
\hline
\end{tabular}
\end{center}
\end{table}

\subsection{Activity and rotation of planet host stars}\label{subsect:results_planethosts}

Being bright and nearby, the K2 Superblink stars have special importance for planet
search studies. In fact, at the time of writing of this paper two of our targets 
already have confirmed planets discovered by the K2 mission. 
K2-3 is a system comprising three super-Earths confirmed through radial velocity monitoring, 
with the outer planet orbiting close to the inner edge of the habitable zone 
(EPIC-201367065 observed in campaign C\,1); see \cite{Crossfield15.0, Almenara15.0}.
K2-18 (EPIC-201912552, also observed in C\,1) 
has a $\sim 2\,R_\oplus$ planet which was estimated to receive $94 \pm 21$\,\% of the Earth's 
insolation \citep{Montet15.0}. 
Both host stars are prime targets for characterization studies of the planetary 
atmospheres through transit spectroscopy. Thus, the analysis of their stellar activity 
is a necessary step toward a global physical description of these systems. 

Another two stars from our K2 Superblink sample have planet candidates presented
by \cite{Vanderburg16.0}. These objects are not yet verified by radial velocity
measurements. 
Our analysis shows that for both systems the stellar rotation is not synchronized with 
the planet orbital period 
($P_{\rm orb}=1.8$\,d and $P_{\rm rot} = 17.9$\,d for EPIC-203099398 
and $P_{\rm orb} = 14.6$\,d and $P_{\rm rot} = 22.8$\,d for EPIC-205489894, respectively).

\section{Discussion}\label{sect:discussion}

We present here the first full flare and rotation period analysis for a statistical
sample of K2 lightcurves. Our target list of bright and nearby M dwarfs represents 
a benchmark sample for exoplanet studies and will be thoroughly characterized
by Gaia in the near future. Knowledge of the magnetic activity of these stars is
of paramount importance given the potential impact it has on exoplanets. 
At the moment a planet is detected, the high-energy 
emission of any given K2 target and its variability 
becomes a prime interest \citep[see e.g.][for a recent
example]{Schlieder16.0}. With the study presented here and future analogous work on 
the remaining K2 campaigns we anticipate such concerns. 

Our primary aim here is to understand the stellar dynamo
and angular momentum evolution at the low-mass end of the stellar sequence through
a study of relations between magnetic activity and rotation. 
We characterize activity with a multi-wavelength approach involving archival X-ray and UV
observations as well as parameters extracted directly from the K2 lightcurves which
describe spot amplitudes and flares. This way
we provide a stratified picture of magnetic activity from the corona over the 
chromosphere down to the photosphere. To our knowledge this is the first time that the 
link with photometrically determined rotation periods is made for a well-defined
sample of M stars over such a broad range of activity diagnostics. 
Yet, as of today, only about $25$\,\% of the M dwarfs with K2 rotation periods have a
meaningful X-ray measurement, and this percentage is even lower for the NUV and FUV bands.
Dedicated X-ray and UV follow-up of these objects can provide the ultimate constraints
on the M dwarf rotation-activity relation. 

Visual inspection of the K2 lightcurves shows that there is not a single non-variable
star in this sample of $134$ M dwarfs. We can constrain rotation periods in $73$\,\%
of them.   
The distribution of rotation periods we find for our sample is in 
general agreement with studies from the main Kepler mission with much larger but less 
well-characterized M star samples \citep{McQuillan13.0, McQuillan14.0}. Contrary to
these studies we find long periods up to $\sim 100$\,d, thanks to our complementary use
of direct sine-fitting as period detection method next to ACF and LS periodograms. We detect 
such long periods only in the lowest mass stars ($M \leq 0.4\,M_\odot$). 
In this respect, our results resemble those obtained by \cite{Irwin11.0, Newton16.0}
based on the MEarth program where sine-fitting yielded many long periods. However, unlike
that project our sample includes also early-M type stars 
and, therefore, it has allowed us to establish that there is a dearth of 
long period detections in early-M dwarfs. Until corroborated by a larger sample,  
we can only speculate whether this is due to the evolution of spin-down history across
the stellar mass sequence 
or whether it results from a change in spot pattern and related changes 
in the detection capabilities for the associated periods.

The low cadence of the K2 lightcurves allows us to detect only the flares with duration
$\geq 1$\,h, and due to this sparse sampling we refrain from a detailed analysis of flare 
statistics. Yet, we find an unprecedented link between flares and stellar rotation.
The distribution of flare amplitudes and flare frequencies shows a clear transition
at $P_{\rm rot} \sim 10$\,d. 
The large flares seen in stars rotating faster than this boundary 
are absent in slow rotators although there is no detection bias against them. 
The smaller flares on slow rotators have no counterparts in the fast rotators 
but such events {-- if present -- would likely be} undetectable. 
We find the same bimodality 
between fast and slow rotators 
in the noise level ($S_{\rm flat}$) of the residual lightcurves after the
rotational signal, flares and other `outliers' are subtracted off: The residual 
variability seen in the fast rotators is significantly and systematically larger than
in the slow rotators with a dividing line at $P_{\rm rot} \sim 10$\,d.  
These new findings can now be added to the rotation-dependence of the spot cycle 
amplitude ($R_{\rm per}$) 
already known from the above-mentioned Kepler studies: 
A cut exists at the same period of 
$\sim 10$\,d with faster rotating stars showing larger amplitudes of the rotation cycle. 
These similarities lead us to 
speculate that the `noise' (i.e. the high values of $S_{\rm flat}$)  
seen in the fast rotators is produced by smaller or fast-changing 
spots or by micro-flares that cause seemingly random variations. 

The observed dichotomy in photometric activity levels between
fast and slow rotators points to a rotation-dependent rapid transition 
in the magnetic properties of the photospheres in M dwarfs. 
In fact, an analogous sharp transition is observed in some 
numerical dynamo models at $R_{\rm 0,l} \sim 0.1$, where $R_{\rm 0,l}$ is the 
`local Rossby number' \citep[e.g.][]{Schrinner12.0}. Assuming this 
theoretical Rossby number corresponds with its empirical definition 
(see Sect.~\ref{subsect:results_amplitude}), this corresponds roughly to our observed
critical period of $\sim 10$\,d. In the simulations, for $R_{\rm 0,l} > 0.1$ (slow
rotators) the dipolar component of the dynamo collapses giving way to a multipolar
dynamo regime. \cite{Gastine13.0} have compared these predictions to the magnetic field
structure inferred from ZDI of M dwarfs. Such observations are 
time-consuming and they require substantial modelling effort, 
and the samples tend to be biased towards fast rotators. When interpreted in terms
of the above-mentioned models, our results suggest 
photometric rotation and activity measures as a new window for observational
studies of dynamo flavors in M dwarfs. However, it must be questioned whether these
diagnostics, which represent activity on the stellar surface, are sensitive to the large-scale
component of the magnetic field. 
 
The transition seen in star spots and white-light flares also   
corresponds approximately to the period where previous studies
of the X-ray - rotation relation have placed the transition from the `saturated' to the
`linear' regime \citep[e.g.][]{Pizzolato03.1}. Different explanations 
have been put forth for this finding 
involving the filling factor for active regions, the size of 
coronal loops or the dynamo mechanism. Observationally,  
those studies have so far shown clear rotation-activity trends 
only for higher-mass stars. We extend the X-ray -- rotation relation here to 
well-studied M stars. With our data set we can, for the first time, refine the study
of X-ray emission from field M dwarfs in the saturated regime (fast rotation)  
in bins spanning spectral subclasses and we find a continuous decrease
of the saturation level $L_{\rm x}$ towards later spectral type which can 
be understood in terms of the mass dependence of X-ray luminosity. 
The tentative evidence that the saturated stars in the coolest mass bin 
(spectral types M5...M6) have lower $L_{\rm x}/L_{\rm bol}$ than 
the K7...M4 type stars is not statistically solid yet. 
If confirmed on a larger sample this might represent a 
change at the fully convective transition, whether due to magnetic field strength or 
structure, or its coupling to rotation (i.e. the stellar dynamo). It is by now well
established that there is a sharp drop of X-ray and H$\alpha$ activity at late-M 
spectral types \citep[$\sim$ M7...M8; e.g.][]{Cook14.0, West08.1} but for mid-M spectral types,  
so far, X-ray studies have not been resolved in both $P_{\rm rot}$ and spectral type space 
together. 
If, e.g., late-M stars remain saturated up to longer periods, the decrease of the saturation
level may go unnoticed in samples mixing the whole rotational distribution. 

We add in this study the first assessment of a link between rotation and chromospheric
UV emission in M stars. Similar to the archival X-ray data, 
the UV data (from the GALEX mission)
covers only a fraction of the K2 sample. A curious wealth of stars with high UV
emission levels and long periods is seen that seems to be in contrast with the findings
regarding all other activity indicators discussed in this work. 

Finally, our archive search for evidence of multiplicity in our targets 
raises an interesting point about the possible influence of multiplicity
on rotation and activity levels. We find   
a high incidence of binarity in the group of fast rotators below the critical
period at which magnetic activity apparently transitions to a lower level. 
The difference between the binary fraction of fast and slow rotators 
is statistically significant. Given the rather
large binary separations (of tens to hundreds of AU) this is puzzling because no tidal 
interaction is expected for such wide systems. 
Nevertheless, we can speculate about a possible causal connection between 
binarity and rotation level. It is well established that wide companions accelerate
the evolution of pre-main sequence disks \citep[e.g.][]{Kraus12.0}. 
Shorter disk lifetimes translate
into a shorter period of star-disk interaction and, hence, one may expect higher initial 
rotation rates on the main sequence for binary stars \citep{Herbst05.0}. As a result,
it may take binaries longer to spin down. Alternatively, we could be seeing the 
mass-dependence of magnetic braking. With our low-number statistics we can not draw
any firm conclusions. Note, however, that a relation between fast rotation and binarity,  
independent
of stellar mass, was also found in a recent K2 study of the Hyades \citep{Douglas16.0}.

\section{Summary and outlook}\label{sect:summary}

From a joint rotation and multi-wavelength activity and variability 
study of nearby M dwarfs observed in K2 campaigns C0 to C4 we infer a critical period
of $\sim 10$\,d at which photometric star spot and flare activity undergoes a 
dramatic change. This transition is coincident with the break separating saturated
from `linear' regime seen in traditional studies of the rotation-activity
relation probing higher atmospheric layers (e.g. the corona through X-rays or 
the chromosphere through H$\alpha$ emission). We present here an updated view of the 
X-ray - rotation relation for M dwarfs. The sample
analysed in this work has strongly increased the known number of long-period M dwarfs in 
the X-ray -- rotation relation. Nevertheless, at present there is not enough sensitive 
data 
in the X-ray archives to constrain the X-ray -- rotation relation for
periods beyond $\sim 10$\,d. A key questions is now whether the coronal emission of 
M dwarfs displays a break-point analogous to the optical photometric activity tracers 
or whether there is a continuous decrease of activity as seen in FGK stars. 
This problem will be addressed in the near
future with upcoming {\em Chandra} observations in which we sample the
whole observed K2 rotation period distribution. 
We will also further examine
the UV -- rotation relation in the larger M dwarf sample that will be available
at the end of the K2 mission. 
Moreover, in that larger sample we intend to search for a possible mass 
dependence of the rotation-activity relation within the M spectral sequence.  
A systematic assessment of multiplicity
for these nearby M stars with Gaia will also be useful for examining the influence of a 
companion star on rotation and activity levels.  

The observed dichotomy between fast and slow rotators in terms of their magnetic
activity level might have interesting consequences for habitability of 
planets near M stars being fried by flares and high-energy radiation 
until they have spun down to around $10$\,d. The time-scale for this process is
as yet poorly constrained but certainly on the order of Gyrs, and it becomes longer
the lower the mass of the star \citep{West08.1}. 
\cite{Segura10.0} found in models based on AD\,Leo that UV flares do not strongly 
affect planet chemistry but the accumulated effect of the exposure to 
strong flaring over most of the planet's lifetime has not been studied so far.

\appendix

\section{Phase-folded lightcurves}\label{sect:appendix_folded}

In the online materials we present the phase-folded lightcurves for all periodic stars 
in two figures, one for periods flagged `Y' (Fig.~A1) 
and another one for periods flagged `?' (Fig.~A2). 
For each star the lightcurve was folded with the `adopted' period, i.e. either the
Lomb-Scargle period (LS), the auto-correlation period (ACF) or the period from the
sine-fit (SINE); see Sect.~\ref{subsect:k2_analysis_period} for details. 


\section{Search for binarity}\label{sect:appendix_bin}

We search all K2 Superblink stars for archival evidence of binarity. 
We proceed in several steps. First, we perform a visual inspection of
POSS1$\_$RED and POSS2$\_$RED photographic plates by using the online
Digitized Sky Survey (DSS) and the interactive tools of Aladin. Epochs of
each pair of plates are separated by up to $\sim$40 years, with the most recent
plates obtained in the 1990s. 
Comparison of the two epochs can help in identifying possible blends 
in the K2 photometry. Specifically, we examine if the targets significantly approached
other stars due to their proper motion. Then, we search for photometric and
astrometric information of each possible contaminant by matching the UCAC4
and 2MASS catalogs in Vizier. Taking into account that the K2 pixel scale
is $\sim 4^{\prime\prime}/{\rm pixel}$, for those cases with possible blends we check 
the K2 imagettes and the photometric mask produced and used by A.Vanderburg
in the reduction of the K2 data \footnote{https://www.cfa.harvard.edu/$\sim$avanderb/k2.html}  
to estimate visually the
occurrence of blending and its significance. For each target, the
inspected imagette represents the sum of all the single imagettes recorded by
K2 during a campaign. From our experience such merged imagettes are usually affected by 
the shift on the sensor of the
photometric centroid due to pointing drift of the telescope.
The photometric masks are wide enough to take into account the drift
of the centroids, making any quantitative analysis of blending with other astrophysical
objects rather difficult and beyond the scope of this work.
We also search the Washington Double Star catalog (WDS) for information
about binarity including sub-arcsec separations, which can not be detected
simply by visual inspection of the photographic plates or matching with
other catalogues. For binaries in the WDS we adopt the visual magnitude
difference between the components indicated in the catalog, when available
and other photometric measurements were missing. 

With this approach, we find evidence for a companion for $25$ stars. However, 
many of the secondaries have a $J$ magnitude which is more than $4$\,mag fainter 
than our target.
These secondaries contribute at most a few percent to the flux of the system. They are
unlikely to be responsible for the observed rotational signal, and we do not consider
them any further. We list the remaining potential companions in 
Table~\ref{tab:multiplicitytable}. These objects have either a $J$ magnitude difference
of $< 4$\,mag with respect to the corresponding K2 target, or a small separation
according to the WDS catalog without known photometry, or both. 
Next to an identifier for the putative companion
(col.3) we provide the binary separation (col.4), the epoch to which it refers (col.5),
the $J$ magnitude or a magnitude difference between the two components according to 
the WDS (col.6-7), 
and flags indicating how we identfy it (through visual inspection of photographic
plates, as entry in the WDS, or in the K2 imagette; cols.8-10). In a final `Notes' column
and in footnotes we add further explanations where needed.

The most important fact to note concerns the binary Gl\,852\,AB. Both stars are in our
target list (EPIC-206262223 and EPIC-206262336) but they are clearly unresolved in 
A.Vanderburg's K2 pipeline. In fact, the lightcurves of both stars are identical because 
the aperture 
comprises an elongated object, clearly representing the two stars of the
$8^{\prime\prime}$ binary. 
We also add a special note here on EPIC-204927969. Our inspection of the K2 imagette
shows that the aperture used by Vanderburg includes other objects but our 
reconstruction of the lightcurve without the contaminated pixels proved that the 
rotational modulation is due to the target. 
Other possible contaminations to be taken serious regard the companions 
that have $J < 10$\,mag. There are two such objects listed in 
Table~\ref{tab:multiplicitytable}. Another three K2 Superblink stars have companions
with $J < 12.5$\,mag which might contribute somewhat to the variability in the lightcurve.  
Further three multiples are presented in the literature, one spectroscopic binary 
and two close visual binaries. 
For the remaining objects in Table~\ref{tab:multiplicitytable} we find no photometric 
measurements, and they are likely faint and may not influence the K2 lightcurves. 
All stars in Table~\ref{tab:multiplicitytable} are flagged on the figures involving
periods and activity measures from K2 data. 

\begin{sidewaystable*}
\caption{Companions possibly contaminating the K2 photometry}
\label{tab:multiplicitytable}
\begin{tabular}{lccrrrccccp{8cm}} \hline
EPIC ID & Campaign & Companion & Sep$^*$ & Epoch & \multicolumn{1}{c}{$J$}   & \multicolumn{1}{c}{$\Delta$}  & \multicolumn{3}{c}{Evidence for binarity} & \multicolumn{1}{c}{Notes} \\ 
        &          &           & [$^{\prime\prime}$] & & [mag]           & [mag]                         & Plates & WDS & K2 & \\ \hline
     202059188   & C0 &        WDS\,06102+2234 & $  1.90$ & $  2012$ & $   ...$ & & x & x & & 1954 vs 1997 plates: in 1997 the star was fully superposed with another object  \\
     201909533 & C1 &       WDS\,11519+0731  & SB & $  2013$ & $   ...$ & ... & & x & &  Triple star studied by \cite{Bowler15.0}; primary is a spectroscopic binary with nearly equal mass components; it forms a common proper motion pair with an object of spectral type M8V and $\Delta H = 5.4$\,mag \\
    202571062  & C2 &        WDS\,16240-2911  & $  6.20$ & $  2009$ & $  9.60$ & & & x & &  Alternative name UCAC4 305-091749  \\
     203124214 & C2 &       WDS\,16254-2710  & $  2.80$ & $  2000$ & $   ...$  & $0.5$ & & x & x & Magnitude difference from WDS; no reference found    \\
     204976998 & C2 &     UCAC4 351-084361  & $  7.60$ & $  2000$ & $ 11.25$ & & &   & x &   \\
     205467732 & C2 &      WDS\,16268-1724  & $  0.50$ & $  2010$ & $   ...$ & $0.4$ & & x &   & Magnitude difference from WDS; no reference found    \\
     205952383 & C3 & 2MASS\,22362748-16172 & $  7.00$ & $  2000$ & $ 12.24$ & & &   & x &   \\
     206007536 & C3 &     UCAC4\,377-172716 & $  6.40$ & $  2000$ & $ 10.71$ & & &   & x &   \\
     206208968 & C3 &      WDS\,22334-0937  & $  1.50$ & $  2010$ & $   ...$ & $0$($J, K$) & & x &   & Binary resolved by \protect\cite{McCarthy01.0} who report separation and magnitude difference \\
     206262223$^\dagger$ & C3 &     UCAC4\,406-139520 & $  8.00$ & $  2012$ & $  9.46$ & & &   & x &  Companion of EPIC-206262336  \\
               & C3 &      WDS\,22173-0847  & $  0.70$ & $  2014$ & $   ...$ & $1.2$($K$) & & x &   &  Additional spectroscopic companion; separation and spectral type (M7V) from WDS; magnitude difference from \cite{Beuzit04.0}.  \\
     206262336$^\dagger$ & C3 &     UCAC4\,406-139522 & $  8.00$ & $  2012$ & $  9.02$ & & &   & x &  Companion of EPIC-206262223 (see above)   \\
               & C3 &      WDS\,22173-0847  & $  2.60$ & $  2010$ & $   ...$ & & & x &   &  Additional companion; separation from WDS  \\
     210613397 & C4 &      WDS\,03462+1710  & $ 14.3$ & $  2008$ & $  9.00$ & & & x & x & Alternative name UCAC4 536-007198. Companion is brighter and of earlier spectral type (K4/5).  \\
     210651981 & C4 &      WDS\,04285+1742  & $  1.60$ & $  2004$ & $   ...$ & $1.3$($J$) & &x &  & Binary resolved by \protect\cite{Guenther05.0} who report separation and magnitude difference \\
\hline
\multicolumn{11}{l}{$^*$ ``SB" indicates spectroscopic binary; $^\dagger$ See text in Appendix~\ref{sect:appendix_bin} on this quadrupel system.} \\
\end{tabular}
\end{sidewaystable*}

\section*{Acknowledgments}

We would especially like to thank A. Vanderburg for his public release of the
analysed K2 lightcurves, upon which much of the present work is based. 
We thank the anonymous referee for a very careful reading of our manuscript. 
This research has made use of the VizieR catalogue access tool, and the 
``Aladin sky atlas", both developed at CDS, Strasbourg Observatory, France. 

\bibliographystyle{mn2e_fix} 
\bibliography{K2rotactV3}

\label{lastpage}

\end{document}